\documentclass[useAMS,usenatbib,usegraphicx]{mn2e}
\usepackage{amsmath} 
\usepackage{bm}
\usepackage{amssymb}
\usepackage{hyperref}
\usepackage{color}

\newcommand\Eqn[1]     {Eq.\,(\ref{#1})}
\newcommand\Eqns[2]    {Eqs\,(\ref{#1}) and~(\ref{#2})}

\newcommand{\be}{\begin{equation}}
\newcommand{\ee}{\end{equation}}
\newcommand{\ba}{\begin{eqnarray}}
\newcommand{\ea}{\end{eqnarray}}

\newcommand\Fig[1]     {Fig.\,\ref{#1}}

\def\bk{{\bf k}}
\def\nn{\nonumber}

\def\h{\rm h}

\def\bx{{\bf x}}
\def\by{{\bf y}}
\def\bk{{\bf k}}

\def\bq{{\bf q}}
\def\bp{{\bf p}}
\def\bs{{\bf s}}

\def\dy{{\rm d}^3\!{\bf y}}

\def\dq{{\rm d}^3{\bf q}}
\def\dpp{{\rm d}^3{\bf p}}
\def\dss{{\rm d}^3{\bf s}}
\def\dw{{\rm d}^3\!{\bf w}}

\def\de{\delta}

\def\cyc{\rm cyc}

\def\Msol{h^{-1}M_{\odot}}

\def\Mpc{\, h^{-1}{\rm Mpc}}
\def\Mpccube{\, h^{-3} \, {\rm Mpc}^3}

\def\Gpccube{\, h^{-3} \, {\rm Gpc}^3}
\def\kMpc{\, h \, {\rm Mpc}^{-1}}

\def\vx{{\bf x}}

\def\vq{{\bf q}}

\def\vw{{\bf w}}
\def\vk{{\bf k}}

\def\mm{\rm mm}
\def\hh{\rm hh}
\def\hm{\rm hm}
\def\mmm{\rm mmm}
\def\hmm{\rm hmm}
\def\hhm{\rm hhm}
\def\hhh{\rm hhh}

\def\brr{\begin{array}}
\def\err{\end{array}}

\def\be{\begin{equation}}
\def\ee{\end{equation}}
\def\bea{\begin{eqnarray}}
\def\eea{\end{eqnarray}}
\def\ba{\begin{eqnarray}}
\def\ea{\end{eqnarray}}

\title[A new method to measure galaxy bias]{A new method to measure galaxy bias}

\author[J. E. Pollack, R. E. Smith, \& C. Porciani]
{Jennifer E. Pollack$^{1}$\thanks{E-mail: jpollack@astro.uni-bonn.de}, 
Robert E. Smith$^{2,3}$\thanks{res@mpa-garching.mpg.de}, \&
Cristiano Porciani$^{1}$\thanks{porciani@astro.uni-bonn.de}\\
$^{1}$Argelander Institut f\"ur Astronomie der Universit\"at Bonn, Auf dem
H\"ugel 71, D-53121 Bonn, Germany\\
$^{2}$Max-Planck Institute for Astrophysics, 
Karl-Schwarzschild-Str.1, Postfach 1523, 85740 Garching, Germany\\
$^{3}$Astronomy Centre, Department of Physics and Astronomy,
University of Sussex, Brighton BN1 9QH}

\begin{document}

\maketitle


\begin{abstract} 
We present a new approach for modelling galaxy/halo bias that utilizes the full non-linear information contained in the moments of the matter density field, which we derive using a set of numerical simulations. Although our method is general, we perform a case study based on the local Eulerian bias scheme truncated to second-order.  Using 200 $N$-body simulations covering a total comoving volume of $675\Gpccube$, we measure several $2$- and $3$-point statistics of the halo distribution to unprecedented accuracy. We use the bias model to fit the halo-halo power spectrum, the halo-matter cross spectrum and the corresponding three bispectra for wavenumbers in the range $0.04\lesssim k \lesssim 0.12 \kMpc$. We find the constraints on the bias parameters obtained using the full non-linear information differ significantly from those derived using standard perturbation theory at leading order.  Hence, neglecting the full non-linear information leads to biased results for this particular scale range.  We also test the validity of the second-order Eulerian local biasing scheme by comparing the parameter constraints derived from different statistics. Analysis of the halo-matter cross-correlation coefficients defined for the 2- and 3-point statistics reveals further inconsistencies contained in the second-order Eulerian bias scheme, suggesting it is too simple a model to describe halo bias with high accuracy.
\end{abstract}




\section{Introduction}\label{sec:introduction}

The clustering statistics of the galaxy distribution contain a wealth
of information about the cosmological model.  However, in the absence
of a robust theory for galaxy formation, extracting this information
can only be achieved in part. In practice, to do this requires us to
assume a specific phenomenological relationship between the density
field of galaxies and that of the underlying matter, more commonly
referred to as galaxy bias. Whilst still incomplete, our leading
theories of galaxy formation, do provide a great deal of insight about
the distribution of galaxies. For instance they predict that 
galaxies should only reside in 
dark-matter haloes and be strongly associated with the distribution of
sub-structures \citep[for a detailed review of galaxy formation
  see][]{Moetal2010}.  This greatly simplifies our ability to
construct a phenomenological model for the galaxy distribution on large scales:
it should be closely related to a weighted average of the 
dark-matter-halo overdensities \citep[e.g.][]{Smithetal2007}.

There are a number of detailed analytical approaches for
characterizing the bias of dark-matter haloes with respect to the
mass distribution \citep[for a recent review see][]{Porciani2013}. However, it has yet to be determined which model provides the most accurate
description of galaxy bias. In the simplest method, the local Eulerian bias model (hereafter LEB), one assumes that the overdensities of the biased tracers can be written as some function of the matter-density field at the same location. If both densities are smoothed over the patch scale $R$, then the biased field may be written as a Taylor-series expansion \citep{FryGaztanaga1993}. If one considers sufficiently large patches, then high-order corrections are guaranteed to be small and the series may be truncated after a finite number of terms. 

Halo-clustering predictions of the LEB expressed in terms of standard perturbation theory \citep[hereafter SPT, for a review see][]{Bernardeauetal2002}
have been examined in numerous works
\citep{Scoccimarroetal2001,Smithetal2007,GuoJing2009,RothPorciani2011,ManeraGaztanaga2011,Pollacketal2012,ChanScoccimarro2012}.
One of the results to emerge from these studies is that, 
when the model is applied to halo counts within finite volumes of linear
size $R$, 
the coefficients of the bias expansion show a running with the ``cell'' size. 
However, halo-clustering statistics such as the $n$-point correlation functions 
(or the corresponding $n$-spectra) do not contain any smoothing scale and 
should not depend on $R$.  
There has been much debate in the literature on how to reconcile these
seemingly contrasting results \citep[see][for a concise summary]{Porciani2013}.

This has led some to discuss an ``effective'' or ``renormalized'' bias 
approach where
the scale-dependence of the bias coefficients is compensated by 
the contribution of small-scale perturbations in the matter density
\citep{Heavensetal1998,McDonald2006a,Schmidtetal2012}. Whilst such a
scheme may be plausible \citep{JeongKomatsu2008,Smithetal2009}, the
development of a unique renormalization method is still ongoing, especially
for dynamically evolved configurations in Eulerian space.
On the other hand, it was recently proposed by
\citet{ChanScoccimarro2012} that 
the bias parameters obtained counting halos within cells of size $R$ 
are only relevant for describing perturbations of wavenumber $k\simeq 0.8/R$ 
in the halo distribution.
While there is no challenge to their argument
when analyzing power-spectrum data, it does present a complication
when using higher-order statistics such as the bispectrum. In order to
interpret the galaxy bispectrum one would be required to compute bias
coefficients separately for each configuration of wavevectors.
This approach appears somewhat cumbersome to implement.

Currently, most observational analyses of galaxy clustering assume that
galaxy bias can be described by the truncated LEB and that the statistical properties of the non-linear matter density field can be modelled using SPT.
To leading order in the perturbations, 
this requires only one bias parameter for 2-point statistics
of the tracers and two parameters for 3-point statistics.
Present-day galaxy surveys, however, do not cover enough comoving volume to
accurately sample the spatial scales at which tree-level results
provide an accurate description of galaxy clustering. The presence of
rare large-scale structures, for instance, significantly alters the 
measurements of three-point statistics \citep[e.g.][]{Nicholetal2006}.
On smaller scales, where data are more robust, 
dynamical non-linearities pose a serious challenge to the models. 
Adopting the simplified LEB+SPT model may therefore generate systematic errors 
and thus influence the characterisation of the bias or the estimation of the 
cosmological parameters.

The LEB truncated to second order is the standard workhorse for studying
three-point statistics of galaxy clustering. Its predictions to leading 
perturbative order have been used to interpret measurements from
the two-degree field galaxy redshift survey \citep{Verdeetal2002,JingBorner2004,
Wangetal04,Gaztanagaetal05}, the Sloan Digital Sky Survey
\citep{Kayoetal2004,Hikageetal2005,PanSzapudi2005,Kulkarnietal2007,Nishimichietal2007,Marin2011,McBrideetal2011a,McBrideetal2011b,Guoetal13}, and the 
WiggleZ Dark Energy Survey \citep{Marinetal2013}.
In our previous study \citep*{Pollacketal2012}, we demonstrated that, in order 
to robustly model three-point statistics with the LEB, one 
must necessarily have an accurate model for the clustering statistics of the 
non-linear matter density on the relevant scales. This is imperative to recover
the correct values of the bias parameters in controlled numerical experiments.
Therefore, it is not surprising that past investigations based on the LEB+SPT 
model reached inconsistent conclusions. For example, studying
the galaxy bispectrum on scales
$0.1<k<0.5\kMpc$, \citet{Verdeetal2002} concluded that 2dF galaxies are 
unbiased tracers of the mass distribution. 
On the other hand, using the complete 2dF sample, 
\citet{Gaztanagaetal05} found strong evidence for non-linear biasing 
from the analysis of the three-point correlation function with
triangle configurations that probe separations between
9 and $36 \Mpc$ \citep[see also][]{JingBorner2004,Wangetal04}.

In this paper, we build upon our past experience and present a general method 
to model the clustering of biased tracers of the mass distribution
on mildly non-linear scales $k<0.1 \kMpc$.
This is key to extend studies of galaxy clustering to smaller spatial 
separations where observational data are less uncertain.
Our method relies on using N-body simulations to measure the relevant statistics
for the clustering of the underlying mass distribution. 
Related approaches have been presented by \citet{Sigadetal2000} and
\citet{SzapudiPan2004} for galaxy counts in cells  
\citep[see also][for an application to correlation functions]{PanSzapudi2005}.
We apply our general framework to the modelling of 
$n$-point clustering statistics of non-linear, Eulerian, locally
biased tracers. In our framework, bias parameters run with the patch scale
$R$. We address the running of the bias by treating the filter scale
as a nuisance parameter to be marginalized over. The major advantage
of our scheme is that we exactly recover the matter poly-spectra used
in the bias model at every order. The only truncation necessary in the
model is the choice as to what level to truncate the bias expansion, and
this may be selected by the data in a Bayesian model comparison.  We
test our modelling framework up to quadratic order 
in the local bias expansion
(as commonly done in recent observational studies), 
for the power- and bi-spectra of haloes and their
cross-spectra with matter measured from a large ensemble (200
realizations) of measurements from a series of large $\Lambda$CDM
$N$-body simulations. This ensemble of simulations resolves the
halos that should host luminous red galaxies 
over a total comoving volume of $675\Gpccube$,
and so provides us with a very stringent statistical test ground for our
model.

The sections are organized as follows. In \S\ref{sec:models} we
set our mathematical notation and introduce the LEB. The numerical
simulations used in this work are briefly described in \S\ref{sec:sims}
and used in \S\ref{sec:density} to measure several statistical quantities
for the matter and halo distributions.
In \S\ref{sec:biassec} we use Bayesian statistics to estimate the free 
parameters of the LEB and describe our main results.
Finally, in
\S\ref{sec:discussion} and \ref{sec:conclusion} 
we further discuss our findings and present our
conclusions.


\section{A new framework for modelling 
the clustering of biased tracers}\label{sec:models}

\subsection{General formalism}
Consider some discrete tracers of the large-scale structure
(dark-matter haloes or galaxies) with mean density $\bar{n}_{\rm h}$ and
physical density $\rho_{\rm h}(\bx)=[1+\delta_{\rm h}(\bx)]\,\bar{n}_{\rm h}$.
We want to relate this random field to the underlying distribution of
matter with local density $\rho(\bx)=[1+\delta(\bx)]\,\bar{\rho}$.
If we assume that the density contrast of the tracers averaged over some patch 
of linear size $R$, $\delta_{\rm h}(\bx|R)$, 
is locally related to the density of matter in the same patch, 
then we may write
\be 
\delta_{\rm h}(\bx|R)={\mathcal F}\left[\delta(\bx|R)\right]
\label{localbias}
\ee
where ${\mathcal F}$ denotes a generic function $\mathbb{R} \to \mathbb{R}$ and
the symbols
\be \delta_{\alpha}(\bx|R)\equiv\int  \dy 
\,W(|\bx-\by|,R)\,\delta_{\alpha}(\by) \ee
(where $\alpha$ stands for haloes or matter)
denote smoothed overdensity fields, $W$ being a rotation-invariant filter
function with size $R$.

Since we are dealing with smooth mathematical functions we may Taylor
expand Eq. (\ref{localbias}) to obtain \citep{FryGaztanaga1993}:
\be
\delta_{\rm h}(\bx|R) = \sum_{n=1}^{\infty} 
\frac{b_{n}}{n!}\left[ \delta^n(\bx|R)-\langle\delta^n(\bx|R)\rangle\right] \ ,
\label{eq:Eulbias}
\ee
where the terms $b_{n}$ are the Eulerian bias coefficients of
order $n$, which depend on both the smoothing scale and the exact 
definition of the tracers (e.g. halo mass, etc.).
Note that the subtraction of the terms $\langle\delta^n(\bx|R)\rangle$
at each order ensures that $\langle\delta_{\rm h}(\bx|R)\rangle=0$,
where $\langle\dots\rangle$ denote an ensemble average. On Fourier
transforming the above relation one finds, for $|\bk|\neq 0$,
\be 
\tilde{\delta}_{\h}(\bk|R) = \sum_{n=1}^{\infty}\frac{b_{n}}{n!}\,\Delta^{(n)}(\bk|R)
\label{eq:bk}\ee
where $\Delta^{(n)}(\bk|R)\equiv \widetilde{\delta^n}(\bk|R)$
can be written as
\be
\Delta^{(n)}(\bk|R)\equiv 
(2 \pi)^3 \int 
\delta^{D}(\bk-\bq_{1\dots n})\,
\prod_{i=1}^{n} \,\tilde{\delta}(\bq_i|R)\,\frac{\dq_i}{(2\pi)^3}\;.\
\label{eq:smDelk}
\ee
In the last expression
$\delta^{D}(\bk)$ denotes the Dirac-delta distribution and 
we have made use of the compact notation
$\bq_{1\dots n}=\bq_1+\dots+\bq_n$ and $\tilde{\delta}(\bq|R)\equiv
\tilde{\delta}(\bq)\,\widetilde{W}(qR)$.

We now define the power spectrum of the biased tracers
and their cross-spectrum with the matter in terms of the correlators:
\be
\langle\tilde{\delta}_{\alpha}(\bk_1|R)\,\tilde{\delta}_{\beta}(\bk_2|R)\rangle\equiv(2\pi)^3\,\delta^{D}(\bk_{12})\,{\mathcal P}_{\alpha\beta}(k_1) \label{eq:Pk}\;. 
\ee
Similarly, the corresponding bispectra can be defined as
\ba
&\langle\tilde{\delta}_{\alpha}(\bk_1|R)\,\tilde{\delta}_{\beta}(\bk_2|R)\,\tilde{\delta}_{\gamma}(\bk_3|R)\rangle& \!\!\!\!\!\equiv \nonumber\\
&\equiv&\!\!\!\!\!\!\!\!\!\!\!\!\!\!\!\!\!\!\!\!\!\!\!\!\!\!\!\!\!\!\!\!\!\!\!\!
\!\!\!\!\!\!
(2\pi)^3\,\delta^{D}(\bk_{123})\,{\mathcal B}_{\alpha\beta\gamma}(\bk_1,\bk_2)
\label{eq:Bk}
\ea
where we have suppressed the dependence of the bispectrum on the third
wavevector, since the Dirac-delta distribution imposes
$\bk_3=-\bk_{12}$. On inserting \Eqn{eq:bk} into \Eqn{eq:Pk}, we find:
\ba
\langle\tilde{\delta}_{\alpha}(\bk_1|R)\!\!\!\!\!\!\!\!&&\!\!\!\!\!\!\!\!\tilde{\delta}_{\beta}(\bk_2|R)\rangle=\\
&=&\!\!\!\!\!\sum_{l,m=1}^{\infty}\frac{\Gamma^{\alpha}_{l}}{l!}\frac{\Gamma^{\beta}_{m}}{m!} 
\langle\Delta^{(l)}(\bk_1|R)\,\Delta^{(m)}(\bk_2|R)\rangle \;. \nn
\ea
with $\Gamma_l^{\rm h}=b_l$ and $\Gamma_l^{\rm m}=\delta_{l1}^{\rm K}$ (for haloes and matter,
respectively) where $\delta_{ij}^{\rm K}$ denotes the Kronecker-delta function.
Similarly for Eq. (\ref{eq:Bk}) we have:
\ba
\langle\tilde{\delta}_{\alpha}(\bk_1)\,\tilde{\delta}_{\beta}(\bk_2)&&
 \!\!\!\!\!\!\!\!\!\!\!\!\!\!\!\!\!
\tilde{\delta}_{\gamma}(\bk_3)\rangle=
\sum_{l,m,n=1}^{\infty}\frac{\Gamma^{\alpha}_{l}}{l!}\frac{\Gamma^{\beta}_{m}}{m!} \frac{\Gamma^{\gamma}_{n}}{n!}\times\\
&\times&\!\!\!\!\!\langle\Delta^{(l)}(\bk_1|R)\,\Delta^{(m)}(\bk_2|R)\,\Delta^{(n)}(\bk_3|R)\rangle\;.\nn 
\ea
It is convenient to introduce the functions ${\mathcal P}_{(l,m)}$ and ${\mathcal B}_{(l,m,n)}$ such that
\be
\langle\Delta^{(l)}(\bk_1|R)\,\Delta^{(m)}(\bk_2|R)\rangle =
(2\pi)^3\,\delta^{D}(\bk_{12})\,{\mathcal P}_{(l,m)}(\bk_1)\; \, 
\ee
\noindent and
\ba
&\langle\Delta^{(l)}(\bk_1|R)\,\Delta^{(m)}(\bk_2|R)\,\Delta^{(n)}(\bk_3|R)\rangle& \!\!\!\!\!= \nonumber\\
&=&\!\!\!\!\!\!\!\!\!\!\!\!\!\!\!\!\!\!\!\!\!\!\!\!\!\!\!\!\!
\!\!\!\!\!\!\!\!\!\!\!\!\!\!\!\!\!\!\!\!\!
(2\pi)^3\,\delta^{D}(\bk_{123})\,{\mathcal B}_{(l,m,n)}(\bk_1,\bk_2)
\ea

In simple words, ${\mathcal P}_{(l,m)}$ denotes the cross power spectrum between
the smoothed random fields 
$[\delta(\vx|R)]^l-\langle [\delta(\vx|
r)]^l\rangle$ and 
$[\delta(\vx|R)]^m-\langle [\delta(\vx|R)]^m\rangle$, while ${\mathcal B}_{(l,m,n)}$ is the corresponding bispectrum.
Thus for the halo and matter power and bispectra we have:
\ba 
{\mathcal P}_{\alpha\beta}(\bk_1) & = & 
\sum_{l,m=1}^{\infty}\frac{\Gamma^{\alpha}_{l}}{l!}\frac{\Gamma^{\beta}_{m}}{m!} \,{\mathcal P}_{(l,m)}(\bk_1)\;, \\
{\mathcal B}_{\alpha\beta\gamma}(\bk_1,\bk_2) \!\!\! & = & \!\!\!\! 
\sum_{l,m,n=1}^{\infty}\frac{\Gamma^{\alpha}_{l}}{l!}\frac{\Gamma^{\beta}_{m}}{m!}  \frac{\Gamma^{\gamma}_{n}}{n!}\,
{\mathcal B}_{(l,m,n)}(\bk_1,\bk_2)\;.
\ea

The above sets of equations provide us with models for the power spectra
and the bispectra of halo counts in cells of size $R$.
However, what we really want to model is the halo 2- and 3-point functions,
$P_{\alpha\beta}$ and $B_{\alpha\beta\gamma}$.
We assume that these quantities can be approximately recovered by 
``de-smoothing'' ${\mathcal P}_{\alpha\beta}$ and ${\mathcal B}_{\alpha\beta\gamma}$
\citep{Smithetal2007,Smithetal2008b,Sefusatti2009}:
\ba
P_{\alpha\beta}(\vk_1) & = & \frac{{\mathcal P}_{\alpha\beta}
(\vk_1)}{W^2(k_1 R)} \ ;
\label{eq:Pdsm}\\
B_{\alpha\beta\gamma}(\vk_1, \vk_2, \vk_3) & = & \frac{{\mathcal B}_{\alpha\beta\gamma}
(\vk_1, \vk_2, \vk_3)}{W(k_1 R) W(k_2 R) W(k_3 R)} \ .
\label{eq:Bdsm}\
\ea
Note that when considering a model of halo bias beyond linear order this operation does not fully remove the dependence of the theory on $R$. In Section \ref{sec:biassec}, we will use the models presented in \Eqn{eq:Pdsm} and \Eqn{eq:Bdsm} to fit simulation data. Nevertheless, our choice to ``de-smooth'' the theoretical model is equivalent to analyzing counts in cell data with a smoothed model.  This is due to the fact that in Fourier-space the smoothing kernels can be treated as multiplicative factors, which means that if we factorize the expressions by dividing out the product of the window functions the relation between the model and the data still holds.  Hence, fitting counts in cells data with a smoothed model is indifferent to analyzing unsmoothed data with a ``de-smoothed'' or factorized model.

The smoothing scale must therefore be considered as a free parameter
of the model, and so it must be either determined by fitting a set of data
or marginalized over.

In \S\ref{app:one} and \S\ref{app:two} we show how the terms
${\mathcal P}_{(l,m)}$ and ${\mathcal B}_{(l,m,n)}$ are related to the
$p$-point matter spectra, where $p=l+m$ or $p=l+m+n$, respectively. In
\S\ref{app:Plmsym} we prove that the functions ${\mathcal P}_{(l,m)}$
are totally symmetric in $l$ and $m$.  For $l \ne m \ne n$, the functions ${\mathcal
  B}_{(l,m,n)}(\bk_1,\bk_2,\bk_3)$ are not in general symmetric in
$l$, $m$, and $n$, unless the wavevectors $\bk_i$ are also exchanged,
i.e.  whilst ${\mathcal B}_{(l,m,n)}(\bk_1,\bk_2,\bk_3)={\mathcal
  B}_{(m,l,n)}(\bk_2,\bk_1,\bk_3)$, ${\mathcal
  B}_{(l,m,n)}(\bk_1,\bk_2,\bk_3)\ne{\mathcal
  B}_{(m,l,n)}(\bk_1,\bk_2,\bk_3)$.
Note that in this study we choose to work with $n$-point spectra, 
${\mathcal P}_{(\alpha_1\dots\alpha_n)}$,
that are
symmetric to an exchange of their vectorial arguments, and we
accomplish this through the symmetrization operation:
\be
{\mathcal P}^{\rm (s)}_{(\alpha_1\dots\alpha_n)} =
\frac{\sum_{i_1,\dots,i_n}^{n} 
|\epsilon_{i_1\dots i_n}|\,{\mathcal P}_{(\alpha_1\dots\alpha_n)}(\bk_{i_1},\dots,\bk_{i_n})}
{\sum^{n}_{i_1,\dots,i_n} |\epsilon_{i_1\dots i_n}|} , 
\ee
where $\epsilon_{i_1\dots i_n}$ denotes the $n$-dimensional
generalization of the Levi-Civita symbol and we take its absolute
value.

In previous studies, the functions ${\mathcal P}_{(l,m)}$ and
${\mathcal B}_{(l,m,n)}$ have been modelled through the use of a
combination of perturbation theory and semi-empirical models.  In
\citet{Pollacketal2012} we recovered 
these functions exactly from an $N$-body simulation and 
demonstrated that they are essential to measure the bias parameters in an
unbiased way.
We will revisit these issues  in \S\ref{sec:density} and \S\ref{sec:biassec}. 


\subsection{Case study: biasing to second order}

As an example, let us evaluate the case when the bias is taken to
second order and all higher-order bias coefficients are vanishing.
This is a widespread assumption often used to interpret observational data
from massive redshift surveys (see \S\ref{sec:introduction} for a long list
of references).
We will consider a unique set of dark-matter haloes. 
For the case where we have multiple halo
bins (e.g. mass selected), the expressions are more cumbersome but no more
complicated. Starting with the two-point statistics, one can
formulate the halo auto- and cross-power spectra with the total mass
up to second-order in the LEB:
\ba
\mathcal{P}_{\hm}(\bk) & = &  b_1 \mathcal{P}_{(1,1)}(\bk) + 
\frac{b_2}{2} {\mathcal P}_{(2,1)}(\bk) \;, \label{eq:hm} \\
\mathcal{P}_{\hh}(\bk)  & = &  b_1^2 \mathcal{P}_{(1,1)}(\bk) +
b_1 b_2 {\mathcal P}_{(2,1)}(\bk) + 
\frac{b_2^2}{4} \mathcal{P}_{(2,2)}(\bk) \;,  \label{eq:hh} \nn \\
\ea
where from \S\ref{app:one}, we see that 
\ba
\mathcal{P}_{(2,1)}(\bk) \!\!& \equiv &\!\! \int \frac{\dq}{(2 \pi)^3} 
\,\mathcal{B}(\vq,\vk-\vq,-\vk) \;, \label{eq:P3} \\
\mathcal{P}_{(2,2)}(\bk) \!\!& \equiv & \!\!\int \frac{\dq}{(2 \pi)^3} 
\frac{\dw}{(2 \pi)^3} 
\,\mathcal{P}_{4}(\vq,\vk - \vq,\vw,-\vk-\vw)\;. 
\label{eq:P4p}
\ea
Note that the $\mathcal{P}_{(l,m)}$ functions are $(l+m-2)$-dimensional
integrals over the smoothed matter correlators of order $n=l+m$,
$\langle \tilde{\delta}(\vk_1|R)\dots
\tilde{\delta}(\vk_n|R) \rangle=(2\pi)^3\,\delta^D(\vk_{1\dots n})\,\mathcal{P}_{n}(\vk_1,\cdots,\vk_n)$. 
These include connected and disconnected terms (see
\S\ref{app:pt}).

For the three-point statistics, the symmetrized auto- halo and cross-bispectra
with respect to the matter, up to second order in the bias model, may
be written:
\ba 
{\mathcal B}^{(\rm s)}_{\hmm} \!\!\! & = & \!\! b_1 {\mathcal B}^{(\rm s)}_{(1,1,1)}   + 
\frac{b_2}{2}{\mathcal B}^{(\rm s)}_{(2,1,1)}\ ;\label{eq:shmm}\\
\mathcal{B}^{(\rm s)}_{\hhm} \!\!\! & = & \!\! b_1^2 {\mathcal B}^{(\rm s)}_{(1,1,1)} + 
b_1b_2{\mathcal B}^{(\rm s)}_{(2,1,1)} 
+\frac{b_2^2}{4}{\mathcal B}^{(\rm s)}_{(2,2,1)} \ ;\label{eq:shhm}\\
{\mathcal B}^{\rm (s)}_{\hhh} \!\!\! & = & \!\! b_1^3 {\mathcal B}^{(\rm s)}_{(1,1,1)} + 
\frac{3b_1^2b_2}{2}{\mathcal B}^{(\rm s)}_{(2,1,1)}
+\frac{3b_1b_2^2}{4}{\mathcal B}^{(\rm s)}_{(2,2,1)}+ \nn \\
&&+\,\,\frac{b_2^3}{8}{\mathcal B}^{(\rm s)}_{(2,2,2)} \ , \label{eq:shhh}
\ea
where for brevity we suppressed the dependence of the bispectra on
$(\bk_1,\bk_2,\bk_3)$. In \S\ref{app:two}, the ${\mathcal B}^{\rm (s)}_{l,m,n}$ functions
are $(l+m+n-3)$-dimensional integrals of the polyspectra of order $l+m+n$.
Specifically:
\ba 
{\mathcal B}^{\rm (s)}_{(2,1,1)} \!\! & \equiv & \!\!
\frac{1}{3}\int \frac{\dq}{(2\pi)^3}\,\mathcal{P}_{4}(\bq,\bk_1-\bq,\bk_2,\bk_3) +2\,\cyc\, \;,\label{eq:P4}\\
{\mathcal B}^{\rm (s)}_{(2,2,1)} \!\! & \equiv & \!\!
\frac{1}{3}\int \frac{\dq_1}{(2\pi)^3}  \frac{\dq_2}{(2\pi)^3}{\mathcal P}_{\rm 5}(\bq_1,\bk_1-\bq_1,\bq_2,\bk_2-\bq_2,\bk_3)  \nn \\   
& & + 2\,\cyc\, \ ;\label{eq:P5} \\
{\mathcal B}^{\rm (s)}_{(2,2,2)} \!\! & \equiv & \!\!
\int \frac{\dq_1}{(2\pi)^3} \dots \frac{\dq_3}{(2\pi)^3}  \nn \\
& & \times\,\, {\mathcal P}_{6}(\bq_1,\bk_1-\bq_1,\bq_2,\bk_2-\bq_2,\bq_3,\bk_3-\bq_3) \ . \nn \\
 \label{eq:P6} 
\ea

In \S\ref{sec:density} we show how one may estimate
$\mathcal{P}_{(l,m)}$ and ${\mathcal B}^{\rm (s)}_{(l,m,n)}$ directly from
an $N$-body simulation.

\section{N-body Simulations}\label{sec:sims}

In order to test the LEB and also to determine the covariance
matrices of the various spectra we have simulated 200 realizations of
a flat $\Lambda$CDM cosmological model.  The specific cosmological
parameters that we have adopted are:
$\{\sigma_8=0.8,\Omega_m=0.25,\Omega_b=0.04,h=0.7,n_s=1.0\} $ where:
$\sigma_8$ is the variance of linear mass fluctuations in top-hat spheres of
radius $R=8 \Mpc$; $\Omega_m$ and $\Omega_b$ are the matter and baryon
density parameters; $h$ is the dimensionless Hubble parameter in units
of 100 km s$^{-1}$ Mpc$^{-1}$; and $n$
is the power-law index of the primordial density power spectrum. Our
adopted values were inspired by the results from the WMAP experiment
\citep{Komatsuetal2009}.

All of the $N$-body simulations were run using the publicly available
Tree-PM code {\tt GADGET-2} \citep{Springel2005}.  This code was used
to follow with high accuracy the non-linear evolution under
gravity of \mbox{$N=750^3$} equal mass particles in a periodic comoving cube of
length \mbox{$L=1500\Mpc$}, giving a total sample volume of 
\mbox{$V=675\Gpccube$}. Newtonian two-body forces were softened below
scales \mbox{$l_{\rm soft}=60\, {\rm kpc}\,h^{-1}$}. The transfer
function for the simulations was generated using the publicly
available {\tt cmbfast} code \citep{SeljakZaldarriaga1996}, with high
sampling of the spatial frequencies on large scales.  Initial
conditions were laid down at redshift $z=49$ using the serial version
of the publicly available {\tt 2LPT} code \citep{Crocceetal2006}.

We use only the simulation outputs at redshift $z=0$ for analysis and 
identify dark matter haloes using the code {\tt BFoF}. This is a 
Friends-of-Friends algorithm \citep{Davisetal1985}, where we adopted a 
linking length corresponding to $b=0.2$ times the mean inter-particle 
spacing. The minimum number of particles an object must contain to be 
considered a bound halo was set to 20. This implies a minimum halo mass 
of $M_{\rm min}=1.11 \times 10^{13} \Msol$ and a mean number density of 
$\bar{n}_{\rm h}\approx 3.7\times 10^{-4}$ $h^3$ ${\rm Mpc}^{-3}$.  Further
 details regarding this set of $N$-body simulations can be found in
\citet{Smith2009} and \citet{Smithetal2012}.


\begin{figure*}
 \centering{
   \includegraphics[width=8cm]{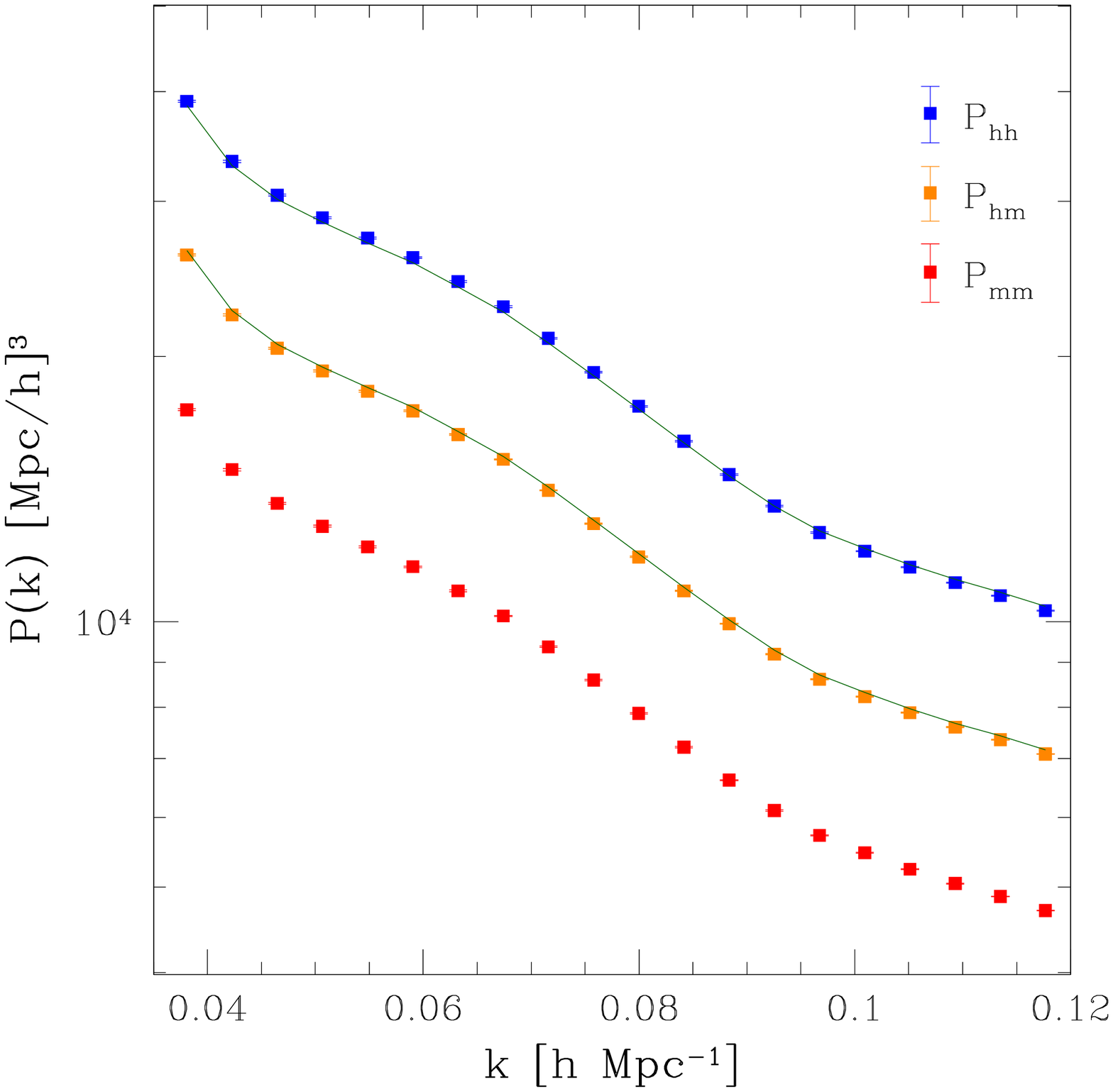}\hspace{0.2cm}
   \includegraphics[width=8cm]{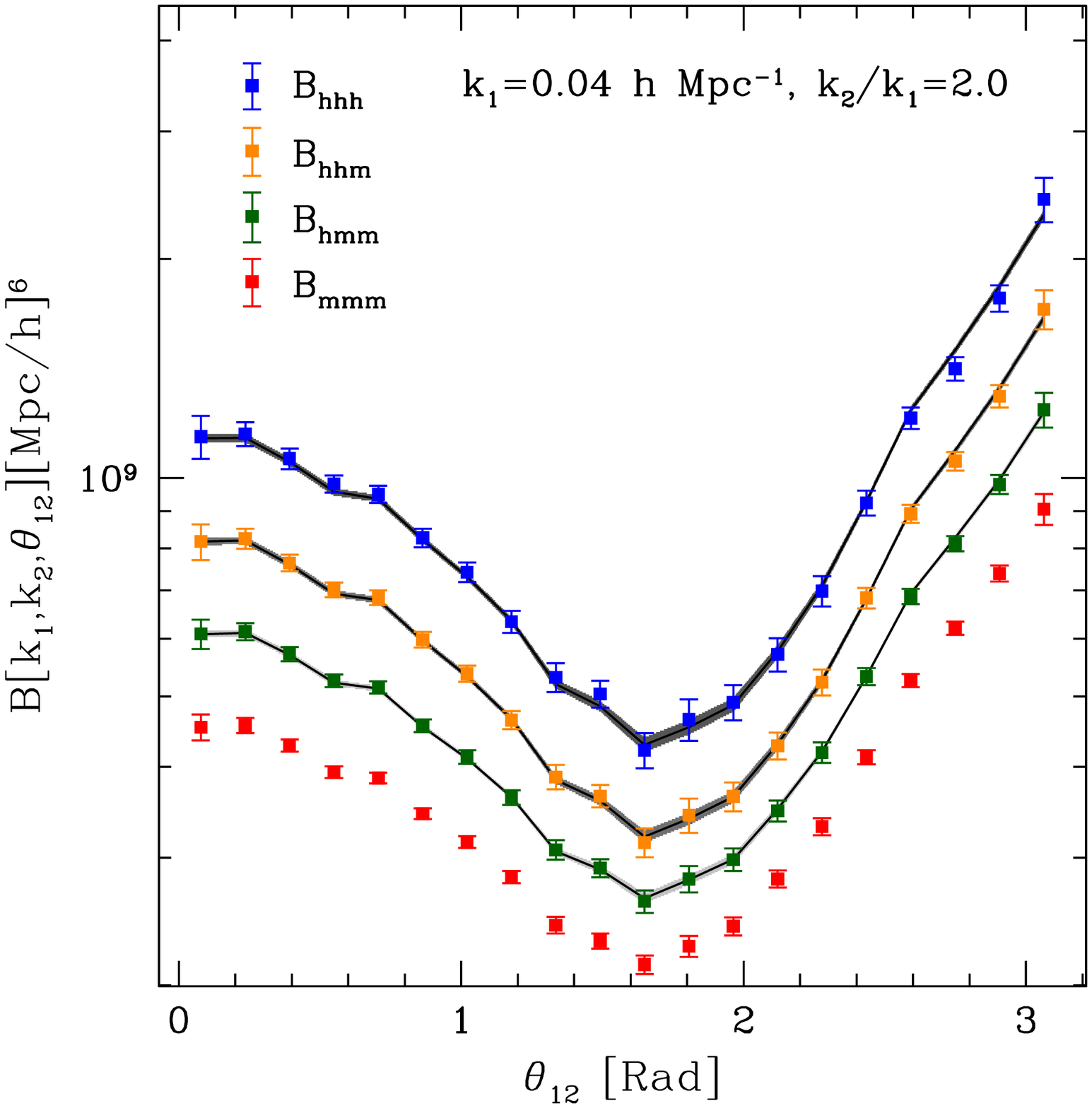}}
\caption{
Power-spectra and bispectra measurements averaged over 200 $\Lambda$CDM $N$-body
  simulations at redshift $z=0$. {\em Left}: Power spectra as a function of
  wavenumber. The blue, orange, and red symbols denote
  $P_{\hh}, P_{\hm}$ and  $P_{\mm}$, respectively. 
{\em Right}: 
Bispectra as a function of triangle configuration. The
  blue, orange, green, and red symbols represent $B_{\hhh},
  B_{\hhm},B_{\hmm},B_{\mmm}$, respectively.  In both panels, the errorbars show
the standard error on the mean. On the other hand, the
  black lines denote the posterior mean for the different statistics
obtained by fitting the second-order LEB to the simulation data.
The shaded grey areas (which are unnoticeably narrow for the power 
spectrum) indicate the predictions for the models that are located within one
rms value of the posterior distribution around the mean
(see \S\ref{ssec:bofR} for more details)}
\label{fig:PowBisp200}
\end{figure*}

\section{Estimating the spectra}\label{sec:density}

In this section we describe how we estimate all the halo and matter
polyspectra that enter the second-order LEB from the $N$-body simulations
at redshift $z=0$.


\subsection{The halo auto- and cross-power and bispectra}

To begin, the halo and matter density fields are interpolated onto a
cubical Cartesian mesh using the cloud-in-cell (CIC) algorithm. 
Throughout we use mesh sizes corresponding to $N_{\rm cell}=1024^3$. 
We then Fourier transform
these grids using the Fast Fourier Transform technique and correct each mode
for the CIC assignment.
The three
power spectra $P_{\rm mm}$, $P_{\rm hm}$, $P_{\rm hh}$, and the four
bispectra, $B_{\rm mmm}$, $B^{(\rm s)}_{\rm hmm}$, $B^{(\rm
  s)}_{\rm hhm}$, $B^{(\rm
  s)}_{\rm hhh}$, are then estimated using
the expressions:
\ba 
\hat{P}^{\rm d}_{\alpha\beta}(\bk_1) \!\! & = & \!\!
\frac{L^3}{N(k_i)}\sum_{i}^{N(k_i)}\delta_{\alpha}(\bk_i)\delta_{\beta}(-\bk_i) \;,\label{est1}\\
\hat{B}^{\rm d}_{\alpha\beta\gamma}(\bk_1,\bk_2,\theta_{12}) \!\! & = & \!\!
\frac{1}{3}\frac{L^6}{N_{\rm tri}}\sum^{N_{\rm tri}}_{\epsilon(\bk_i,\bk_j)}
\delta_{\alpha}(\bk_i)\delta_{\beta}(\bk_j)\times \nn \\
& & \times\, \delta_{\gamma}(-\bk_i-\bk_j)+2\,{\rm cyc}\;,
\label{est2}\ea
where $N(k_i)$ is the number of Fourier modes in a narrow shell centred on
$k_1$, $\epsilon(\bk_i,\bk_j)$ represents the pair of vectors which lie in thin shells centred on $k_1$ and $k_2$, whose angular separation lies in the angular bin centred on $\theta_{12}$, 
and $N_{\rm tri}\equiv N_{\rm tri}(k_i,k_j,\theta_{ij})$ is the total number of
triangles with this configuration in Fourier space. 
The superscript ``d'' denotes that these are spectra of a discrete distribution of points (i.e. haloes) and
must be corrected for shot noise. The forms of the Poissonian shot-noise
corrections we adopt were presented in \citet{Pollacketal2012}:
\ba
\hat{P}^{\rm shot}_{\alpha\beta}(\bk_1) \!\! & = & \!\! \frac{\delta^{\rm K}_{\alpha\beta}}{\bar{n}_{\alpha}} \label{eq:powshot}\\
\hat{B}^{\rm shot}_{\alpha\beta\gamma}(\bk_1,\bk_2) \!\! & = & 
\frac{1}{3}\frac{\delta^{\rm K}_{\alpha\beta}}{\bar{n}_{\alpha}}
\left[P_{\beta\gamma}(\bk_1)+2\,{\rm cyc}\right]\nn \\
& & 
+\frac{1}{3}\frac{\delta^{\rm K}_{\beta\gamma}}{\bar{n}_{\beta}}
\left[P_{\gamma\alpha}(\bk_1)+2\,{\rm cyc}\right]\nn \\
& & 
+\frac{1}{3}\frac{\delta^{\rm K}_{\gamma\alpha}}{\bar{n}_{\gamma}}
\left[P_{\alpha\beta}(\bk_1)+2\,{\rm cyc}\right]+
\frac{\delta^{\rm K}_{\alpha\beta}\delta^{\rm K}_{\alpha\gamma}}{\bar{n}_{\alpha}^2}\label{eq:bishot}
\ea
\noindent where $\bar{n}_{\alpha}$ denotes the mean number density of either the 
matter particles or the halo population.

Figure~\ref{fig:PowBisp200} presents the various power- and bi-spectra
averaged over the 200 realizations with the corresponding standard errors on 
the mean. All spectra were
corrected for shot noise using \Eqns{eq:powshot}{eq:bishot}.  
The bispectra were measured for triangle
configurations with fixed lengths $k_1=0.04\kMpc$ and $k_2=2k_1$, but with
varying angle $\theta_{12}$.  We adopt the convention $\theta_{12}=0$ for $\bk_1$
and $\bk_2$ parallel. In order to use the same range of wavenumbers,
the power spectra were measured over the scale range
$0.04<k<0.12$ $\kMpc$.  

\begin{figure*}
 \centering{
   \includegraphics[width=6cm]{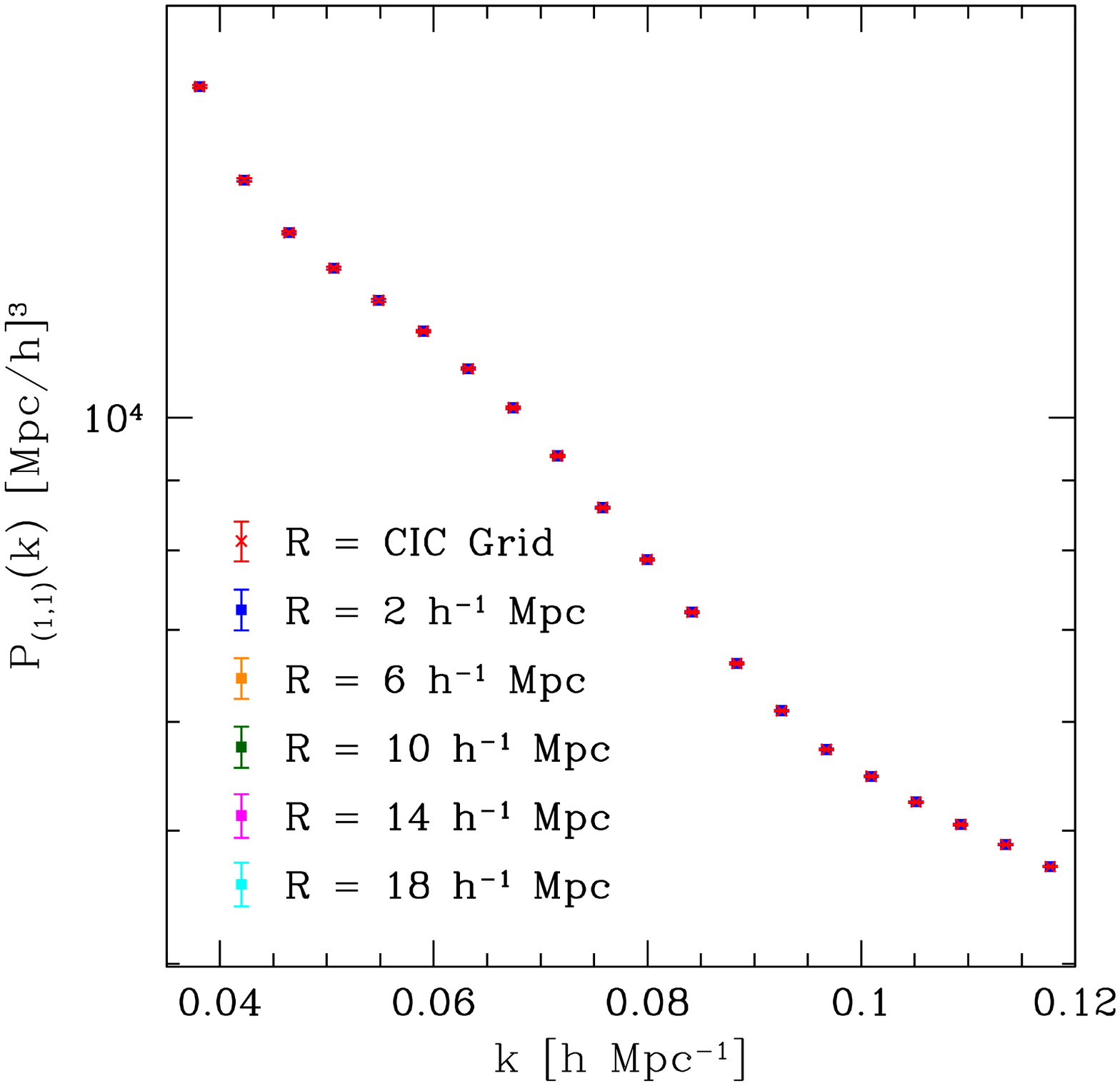}
   \includegraphics[width=6cm]{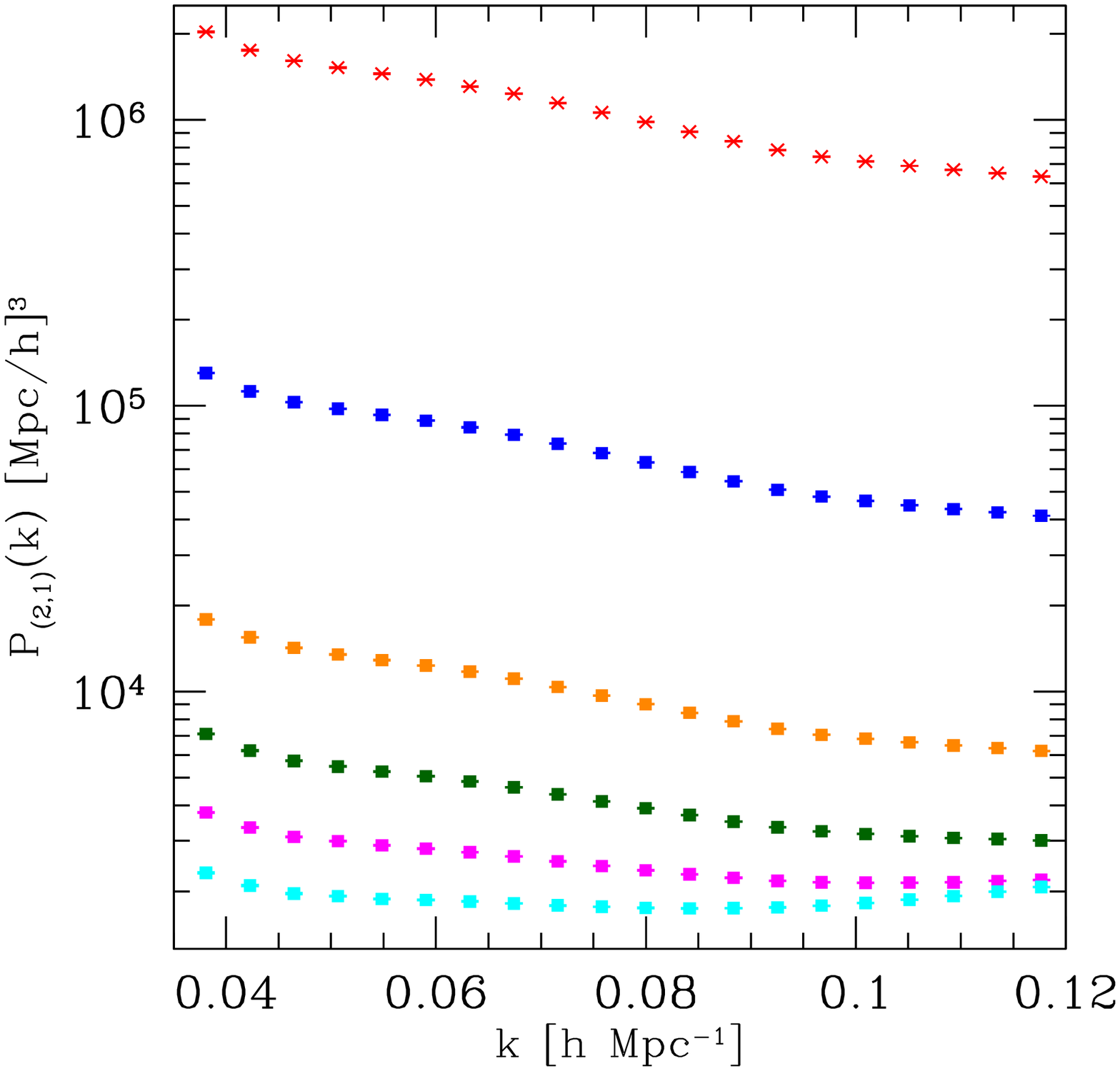}
   \includegraphics[width=6cm]{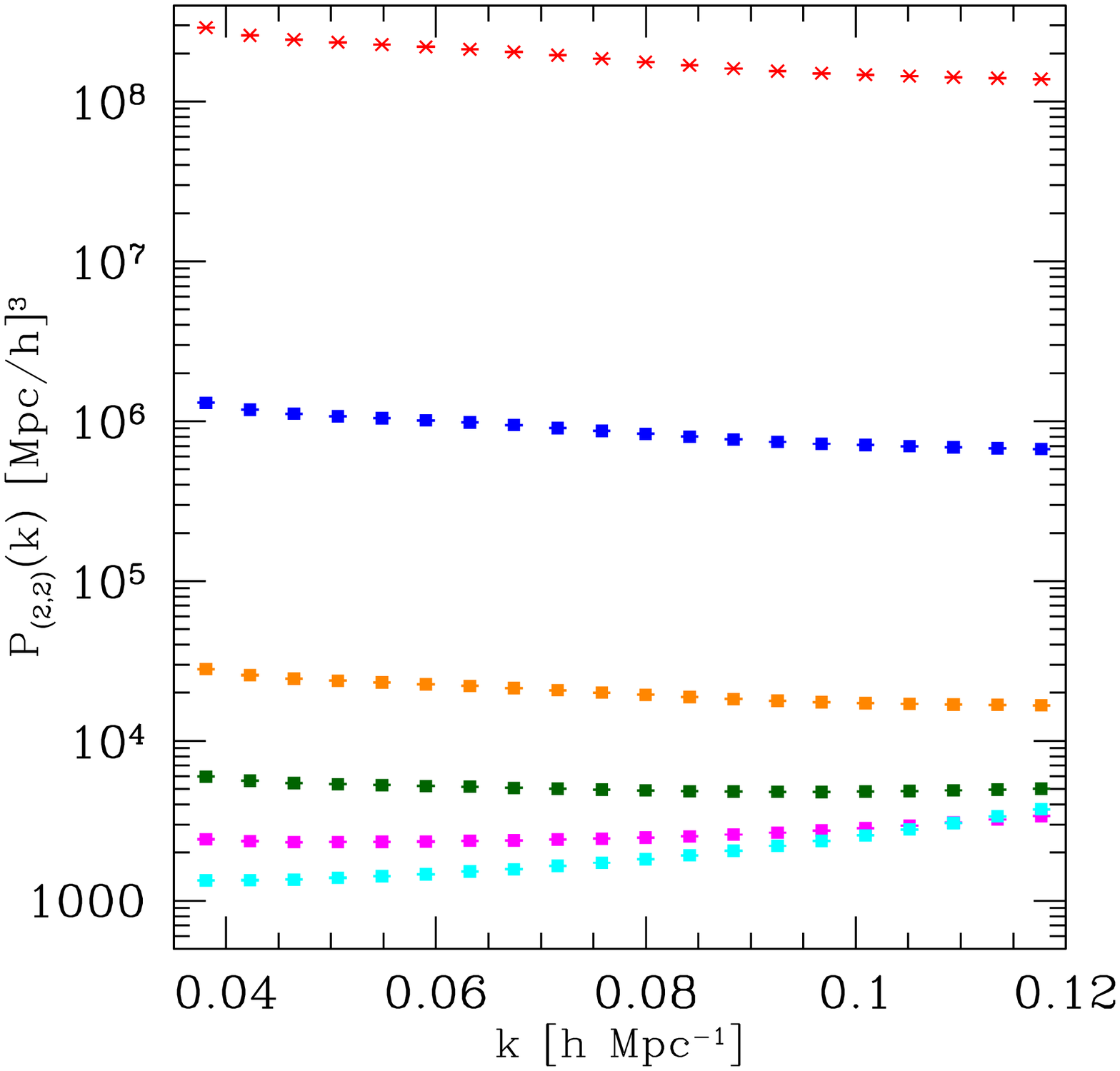}
  }
\caption{Measurements of the de-smoothed terms
  $P_{(1,1)}$, $P_{(2,1)}$, and $P_{(2,2)}$ averaged over 200
  $N$-body simulations. We
  show results for a number of smoothing scales within the range $2\leq R\leq18
\Mpc$ in comparison with our basic CIC grid 
(see the main text for more details).  The errorbars denote the
  standard error on the mean.}
\label{fig:dsmPowpterms}
\end{figure*}


\subsection{Estimating $P_{(l,m)}$ and $B^{(\rm s)}_{(l,m,n)}$}
\label{sec:estmodels}

The polyspectra $P_{(l,m)}$ and
$B^{(\rm s)}_{(l,m,n)}$ that enter the expressions for the halo
power- and bi-spectra in the LEB
are affected by the non-linear evolution of the
matter fluctuations. While these terms are usually approximated with 
perturbative techniques, we measure them directly from our
$N$-body simulations. We do this as follows.
First, we correct each Fourier mode of the mass-density field 
for convolution with the CIC grid. Then we multiply the result
by a Gaussian smoothing function
\mbox{$W(kR)=\exp\left[-(kR)^2/2\right]$} 
and inverse transform back to real space.
Next, we generate the fields $\delta^l(\bx|R)$ 
for the relevant values of $l$ and re-transform them into Fourier space.
We then deconvolve these fields for the original smoothing, which means 
simply multiplying each Fourier mode by \mbox{$W^{-1}(kR)$}.
Finally, the required $P_{(l,m)}$ and $B^{(\rm s)}_{(l,m,n)}$ terms, defined in terms of $\Delta^{(l)}(\vk|R)$ (see \Eqn{eq:smDelk}), can be estimated as follows 
\be 
\hat{P}_{(l,m)}(\bk_1) =
\frac{L^3}{N(k_i)}\sum_{i}^{N(k_i)}\Delta^{(l)}(\bk_i|R)\Delta^{(m)}(-\bk_i|R) \;, 
 \label{PlmEst1}
\ee
\noindent and
\ba
&\hat{B}^{(\rm s)}_{(l,m,n)}(\bk_1,\bk_2,\theta_{12})&\!\!\!\!=\frac{1}{3}\frac{L^6}{N_{\rm tri}}\sum^{N_{\rm tri}}_{\epsilon(\bk_i,\bk_j)}\Delta^{(l)}(\bk_i|R)\times \nn \\
&&\!\!\!\!\!\!\!\!\!\!\!\!\!\!\!\!\!\!\!\!\!\!\!\!\!\!\!\!\!\!\!\!
 \times\,\Delta^{(m)}(\bk_j|R) \Delta^{(n)}(-\bk_i-\bk_j|R)+2\,{\rm cyc}\;.
\label{BlmnEst2}\ea
%

\begin{figure*}
 \centering{
   \includegraphics[width=8cm]{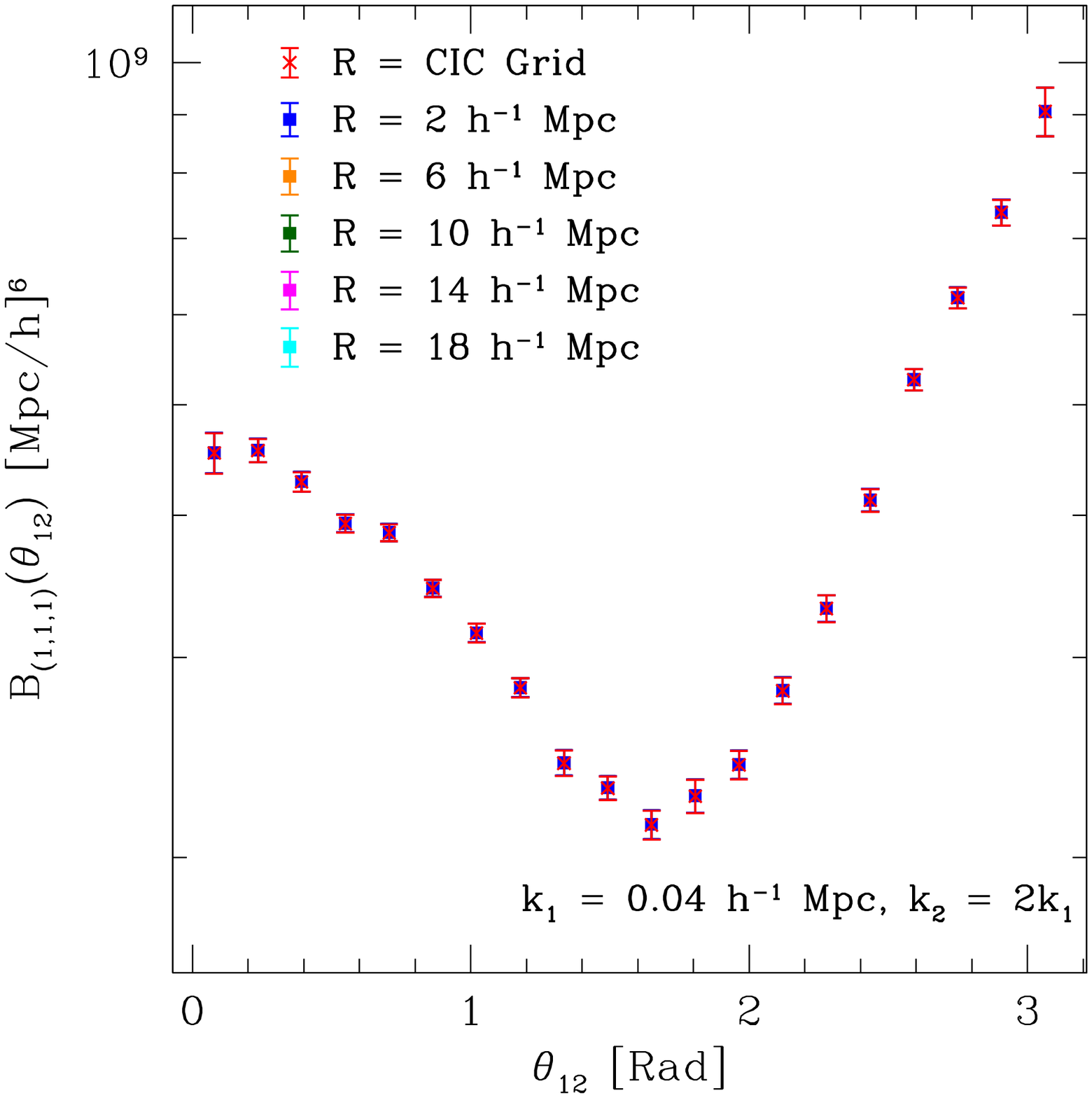}\hspace{-5pt}
   \includegraphics[width=8cm]{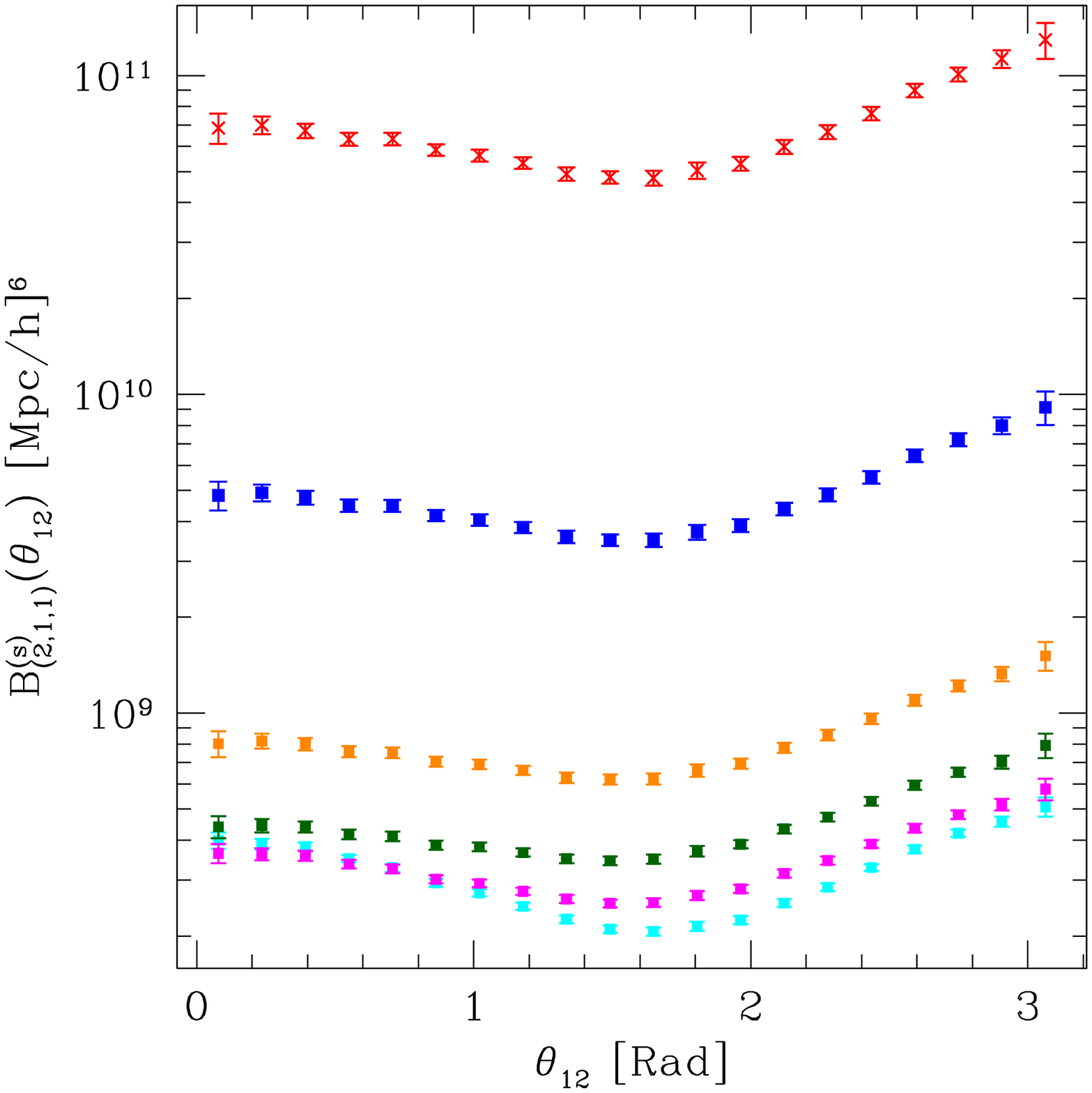}}\vspace{-10pt}
 \centering{
   \includegraphics[width=8cm]{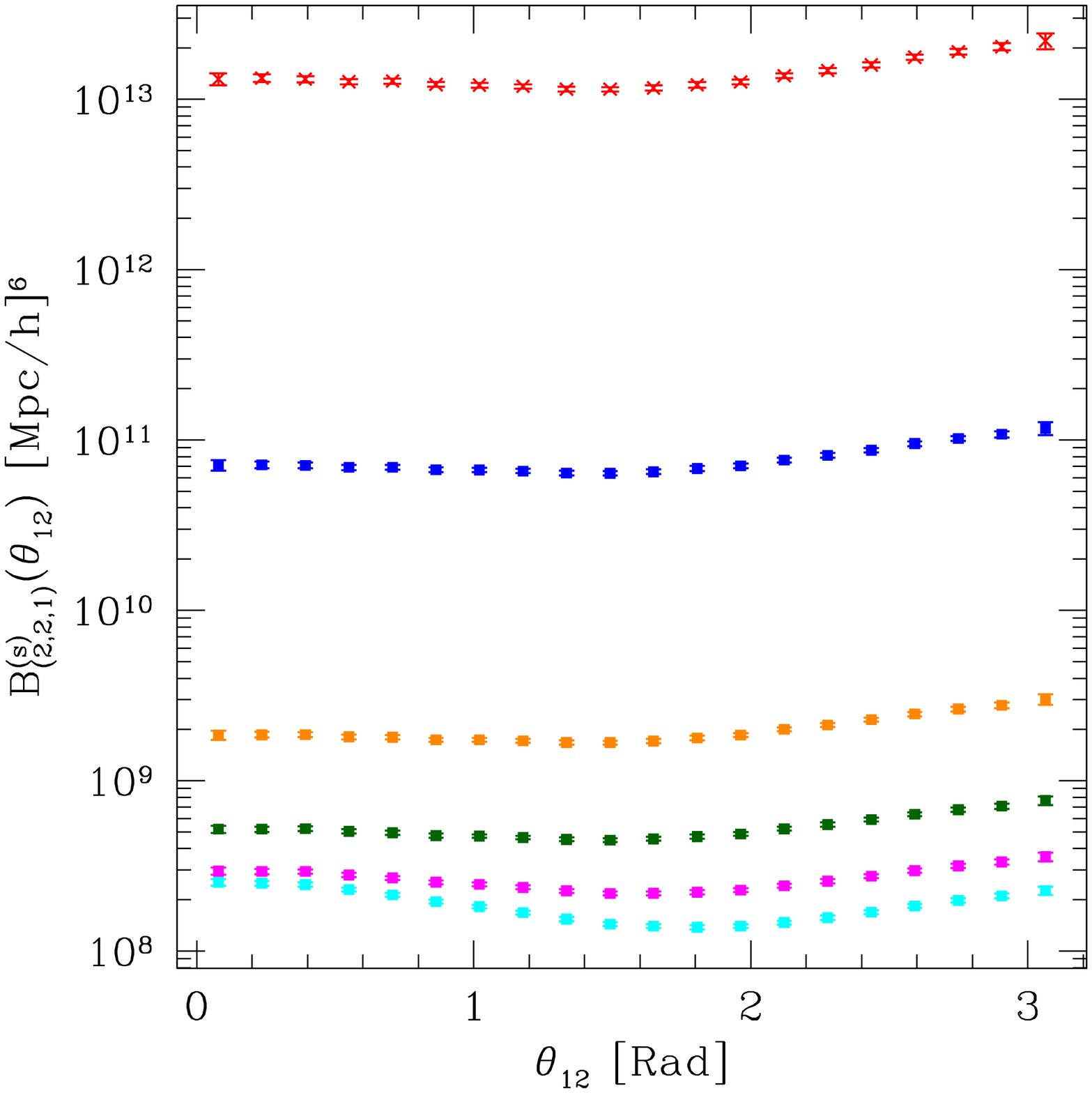}\hspace{-5pt}
   \includegraphics[width=8cm]{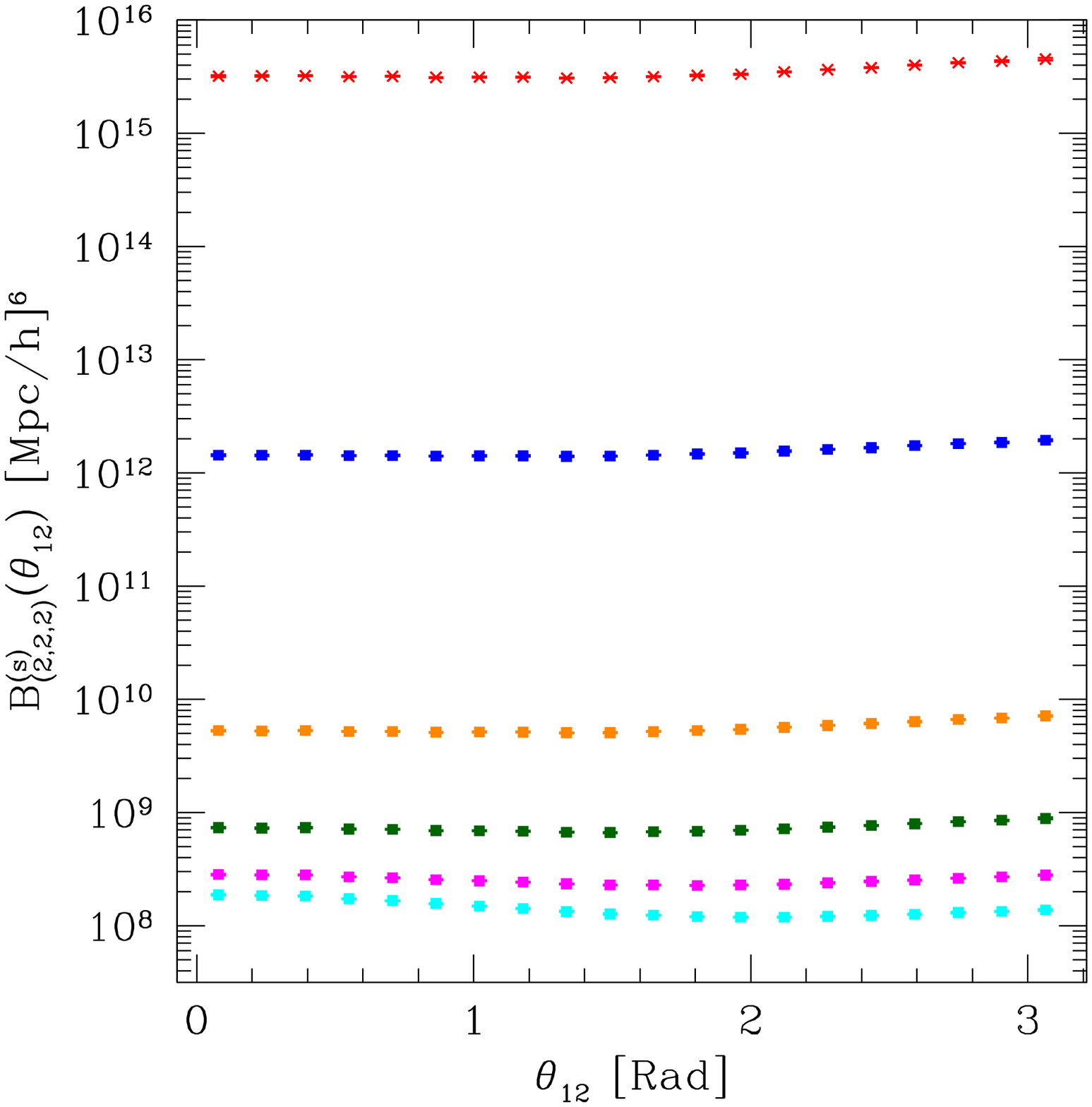}}
\caption{Same as for Figure \ref{fig:dsmPowpterms} but for the bispectrum terms
  $B_{(1,1,1)}$, $B^{(\rm s)}_{(2,1,1)}$, $B^{(\rm s)}_{(2,2,1)}$, and $B^{(\rm s)}_{(2,2,2)}$.}
\label{fig:dsmpterms}
\end{figure*}

We note that the functions $P_{(l,m)}$ and $B^{(\rm s)}_{(l,m,n)}$ slowly vary with $R$ 
and so can be smoothly interpolated.
Based on this knowledge, we measure the spectral functions
over the range: $R = \left[2,\,18\right]\Mpc$, in increments of
$\triangle R = 2 $ $\Mpc$, but including an additional measurement at $R = 7$ $\Mpc$. The lower limit was
adopted because we do not wish to smooth below the Lagrangian size of
haloes, which for our sample is of the order of $\sim 3-4\,\Mpc$. The upper
bound of $R=18\Mpc$ we justify by noting that we do not want the
largest $k$-mode entering our computations of the halo power- and bi-spectra
to be too heavily smoothed. 

Before inspecting the functions $P_{(l,m)}$ and $B^{(\rm s)}_{(l,m,n)}$, we first report the level of non-linearity
present in the smoothed matter-density field, $\de(\vx|R)$. 
We quantify this by measuring the variance of the
density perturbations, $\sigma^2(R)$ and the fraction
of cells where the density contrast exceeds unity, $f$,
as a function of the filter scale (see Table
\ref{tab:vardelta}). Our results show that
$\sigma^2(R)<1$ for $R\gtrsim 4 \Mpc$. We
therefore expect the quadratic bias model to be a poor description for smaller
values of $R$. However, we note that the fraction of the cells
with $\delta\ge1$ is $f\lesssim 0.1$ for all of the filter scales
considered. Furthermore, in our previous work \citep{Pollacketal2012}, we evaluated the scatter plots of $\delta_h$ versus $\delta$ measured from our $N$-body simulations for different smoothing radii.  We found that expressing $\delta_{\h}$ as a polynomial function at second-order  in $\delta$ can describe reasonably well the mean trend of the scatter.

%
\begin{table}
\centering
\caption{Level of non-linearity in the smoothed mass-density field at 
redhift $z=0$. Column
  1: filter scale, $R$; Column 2: variance of density fluctuations,
  $\sigma^2(R)$; Column 3: volume fraction with $|\delta(R)|>1$, $f$.}
\label{tab:vardelta}
\begin{tabular}{|c|c|c|} \hline
$R \, [\Mpc]$ & $\sigma^2(R)$  & $100\times f$\\ \hline 
 2  & 2.44 & 10.0 \\
 4  & 0.71 & 8.4 \\
 6  & 0.38 & 6.0 \\ 
 7  & 0.28 & 4.9 \\
 8  & 0.22 & 3.8 \\
 10 & 0.15 & 2.2 \\ 
 12 & 0.10 & 1.1 \\
 14 & 0.08 & 0.5 \\
 16 & 0.06 & 0.2 \\
 18 & 0.05 & 0.1 \\
\hline
\end{tabular}
\end{table}


\begin{figure*}
 \centering{
 \includegraphics[width=8cm]{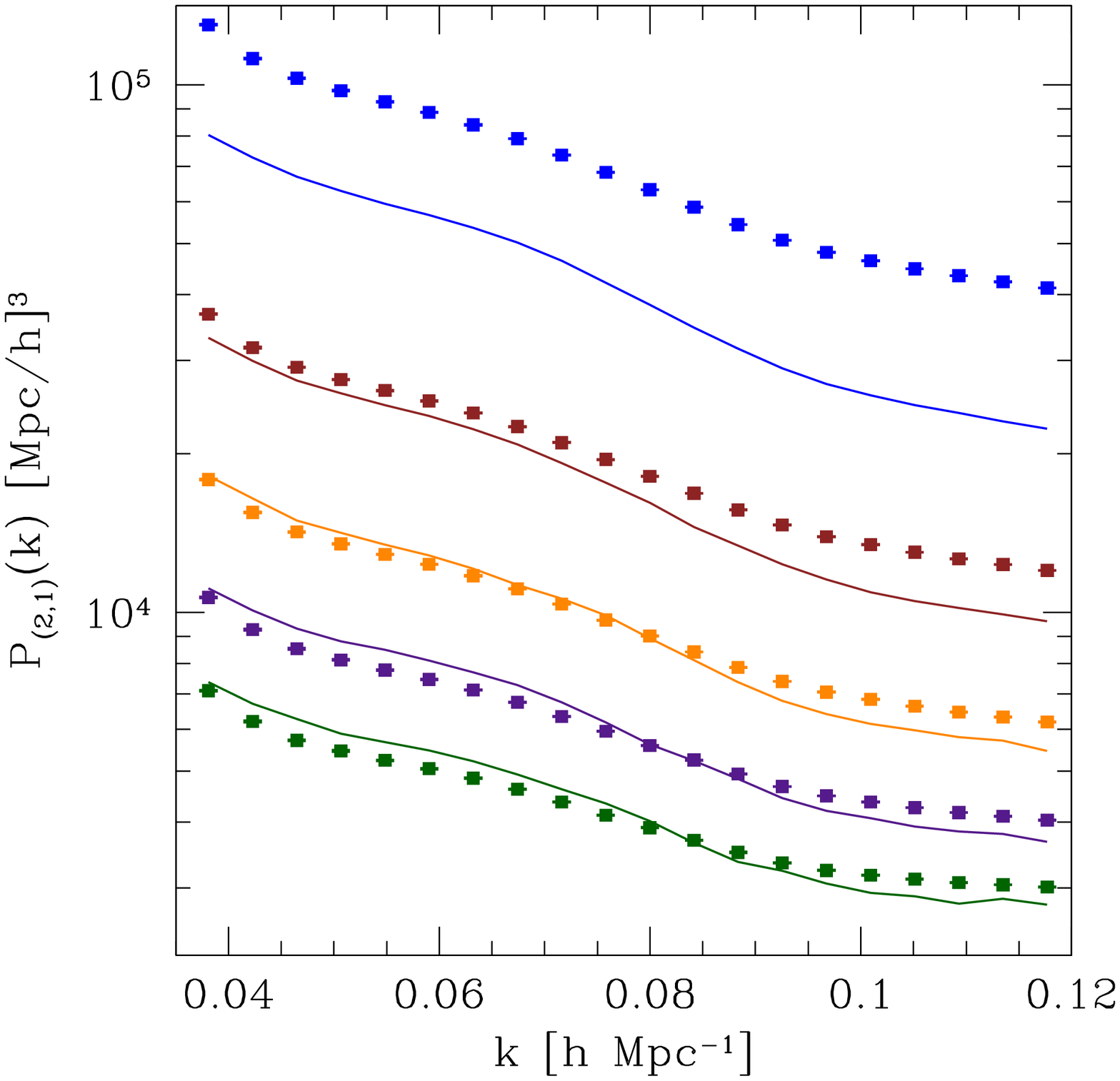}\hspace{0.2cm}
   \includegraphics[width=8cm]{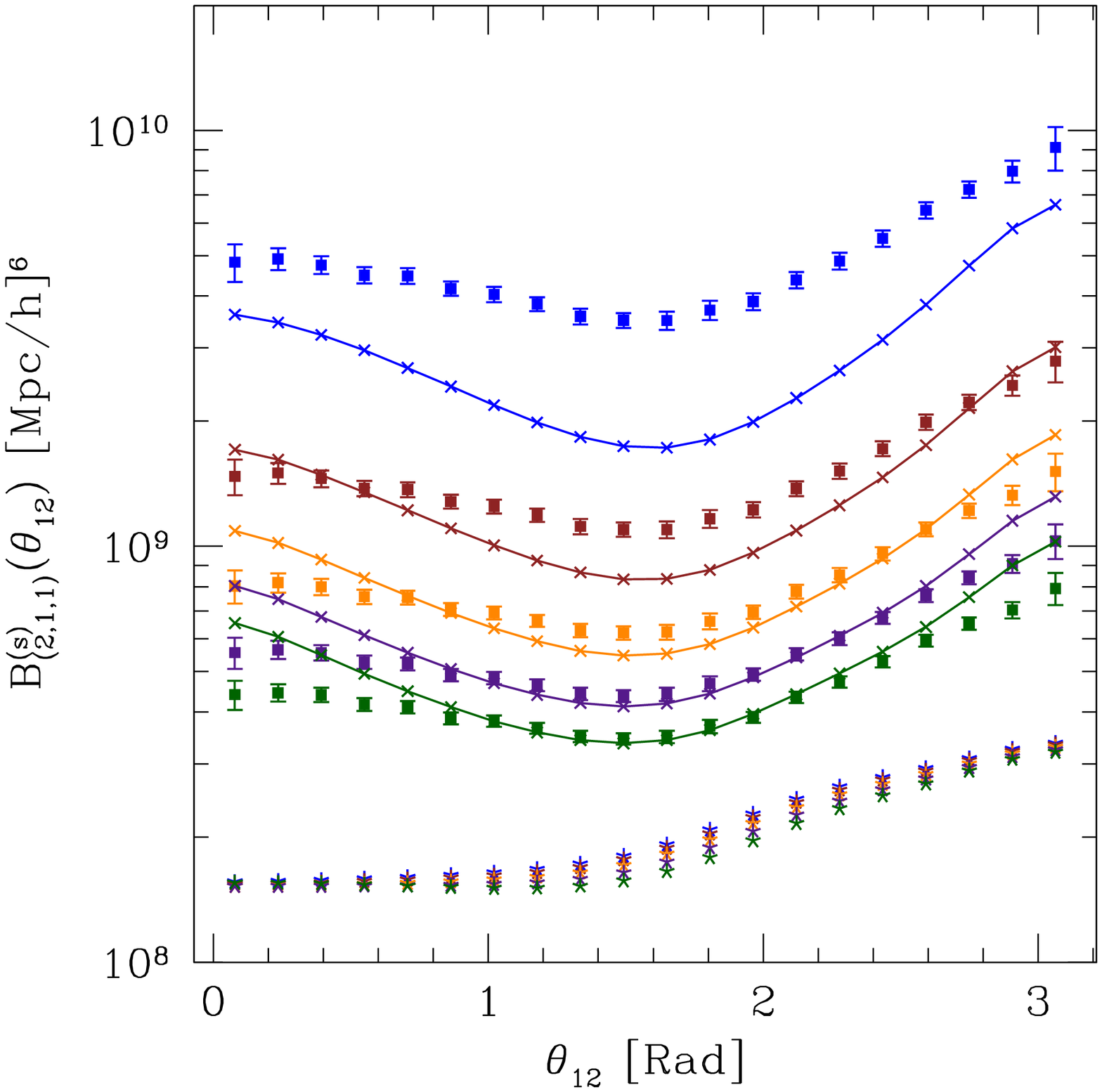}}
 \caption{Left: Measurements of the $P_{(2,1)}$ term from the 
simulations (points with errorbars) are compared with the analytical 
predictions from leading-order SPT (solid lines) for different filter radii 
(from top to bottom: $R = 2, 4, 6, 8, 10 \Mpc$).
Right:
Same as in the left panel but for $B^{(\rm s)}_{(2,1,1)}$.
The star-shaped points represent the contribution to $B^{(\rm s)}_{(2,1,1)}$ from the 
disconnected parts of the fourth-order correlators at tree level in SPT 
(i.e. cyclical products of the linear power spectrum).}
\label{fig:P4dsm}
\end{figure*}

\subsection{Results for $P_{(l,m)}$ and $B_{(l,m,n)}$}
\label{sec:methods}

\noindent 
Figures~\ref{fig:dsmPowpterms} and ~\ref{fig:dsmpterms} show the results for the ensemble-averaged
de-smoothed power and bispectra, $P_{(l,m)}$ and $B_{(l,m,n)}$, respectively. 
Focusing on the power spectrum, the panels show (from left to right) the matter power spectrum $P=P_{(1,1)}=P_{\mm}$ followed by the terms $P_{(2,1)}$, and $P_{(2,2)}$.

For the bispectra, the panels show:
the matter bispectrum $B=B_{(1,1,1)}=B_{\mmm}$ (top left), $B_{(2,1,1)}^{\rm (s)}$ 
(top right),
$B_{(2,2,1)}^{\rm (s)}$ (bottom left), and $B_{(2,2,2)}^{\rm (s)}$ (bottom right). We have restricted
the triangle configurations to those which enter the auto- and
cross- halo bispectra shown in Figure \ref{fig:PowBisp200}. 
Each panel shows six sets of points with errorbars
which denote the results obtained for different smoothing scales.
The red crosses denote the resulting
polyspectra when no Gaussian smoothing (and de-smoothing) is applied on top of 
the CIC assignment. 

Comparing the different panels reveals how the amplitudes of the
de-smoothed quantities vary. Obviously, 
for the matter power and bispectra, $P_{(1,1)}$ and $B_{(1,1,1)}$,
all of the spectra overlap with the CIC result as the smoothing
and the de-smoothing procedures perfectly cancel each other out.
However, for the remaining $P_{(l,m)}$ and $B_{(l,m,n)}$ functions, the
de-smoothed quantities vary with the scale $R$. In particular, as $R$ decreases,
the overall amplitude of the spectra increases due to the 
contributions of small-scale modes. For the largest smoothing scales, the
configuration dependence of the spectra is also modified. 
In order to gain some insight into the origin of this behaviour, let us
consider, for instance, the term
\be
P_{(2,1)}(\vk)=\int \frac{\dq}{(2\pi)^3}\,B(\vq,\vk-\vq,-\vk)\,\mathcal{W}(\vq,\vk-\vq)
\label{eq:underst}
\ee
where $\mathcal{W}$ is a generic weighting function defined as
\be
\mathcal{W}(\vk_1,\vk_2)=
\frac{W(k_1 R)\,W(k_2 R)}{W(|\vk_1 + \vk_2|R)} \, .
\ee
\noindent For Gaussian smoothing, the weighting function in \Eqn{eq:underst} can be re-expressed as $\mathcal{W}(\vq,\vk-\vq) =\exp{\left[-R^2\left(
q^2-kq\mu \right) \right]}$, with $\mu=\vk\cdot\vq/(kq)$ the cosine of the angle between $\vk$ and $\vq$.
The contribution to the integral from
all modes with $qR \gg 1$ is exponentially suppressed (i.e. $\mathcal{W}\ll 1$).

The contribution to the integral from
all modes with $qR \gg 1$ is exponentially suppressed (i.e. $\mathcal{W}\ll 1$).
However, at fixed $k$, $\mathcal{W}$ assumes values larger than unity 
for $\mu>0$ and $q<k\mu$ (independently of $R$)
and presents an absolute maximum for $q=k/2$ and
$\mu=1$ where it takes the value $\mathcal{W}_{\rm max}=\exp[(kR)^2/4]$. Note
that, when $kR\ll 1$, $\mathcal{W}_{\rm max}\simeq 1+(kR)^2/4\simeq 1$ so that
all configurations where $\mathcal{W}>1$ receive nearly the same weight. In this
case, the parameter $R$ regulates how quickly the function $\mathcal{W}$
drops when $q$ moves away from the region where $\mathcal{W}>1$. In other words,
$\mathcal{W}$ behaves nicely as a smoothing function.
This is not true, however, when $kR \gg 1$ and 
the value of $\mathcal{W}_{\rm max}$ grows very large.
In this case, $P_{(2,1)}$ receives dominant contributions
from a narrow shell of modes located at $q\simeq k/2$ and $\mu\lesssim 1$. 
This effect is clearly seen in Figure~\ref{fig:dsmPowpterms} for $R=18$ $\Mpc$ where the over-smoothing (i.e. the fact that
$kR$ is significantly larger than unity for $k\sim 0.1$) leads to a change
in shape for $P_{(2,1)}$ which is particularly evident for the configurations with the largest wavenumbers.  

It is interesting to investigate why, for $kR\ll 1$, the configuration
dependence of $P_{(2,1)}$ changes very little with $R$ and only the overall 
normalisation appears to depend on the smoothing scale.
If we assume that the amplitude of the bispectrum $B(\vk_1,\vk_2,-\vk_1-\vk_2)$
keeps nearly constant at all scales assuming a value $\simeq B_0$, Equation (\ref{eq:underst}) then gives
\be
P_{(2,1)}(\vk)\simeq \frac{\pi^{3/2}  \exp{((k R)^2/4)}}{ R^3}\, B_0\;.
\ee
The first term on the right-hand-side gives the $q$-space volume over which the bispectrum
is averaged to get $P_{(2,1)}$.
At fixed $k$,
this expression diverges as $R^{-3}$ when $R\to 0$ 
and exponentially as $R\to \infty$ 
while it shows broad minimum around $kR\sim 2.5$. 

Clearly, had we not
smoothed the density field, the resulting $P_{(l,m)}$ and $B_{(l,m,n)}$ would be
divergent in any $\Lambda$CDM cosmology.

\subsection{Modelling $P_{(2,1)}$ and
$B^{\rm (s)}_{(2,1,1)}$ with SPT}

In order to better understand what drives the amplitude and functional
form of the $P_{(l,m)}$ and $B^{\rm (s)}_{(l,m,n)}$ terms we have attempted to model 
their signal with SPT. For simplicity, we have focused on the lowest-order
non-trivial terms $P_{(2,1)}$ and $B^{\rm (s)}_{(2,1,1)}$.

To leading order in the perturbations, the matter bispectrum can be written
as $B(\vk_1,\vk_2,\vk_3)=2 \,F_2(\vk_1,\vk_2)\,P_{(0)}(k_1)\,P_{(0)}(k_2)+
{\rm 2 \ cyc}$ with $F_2$ the second-order SPT kernel 
(see Appendix \ref{app:pt}) and $P_{(0)}$ the linear power spectrum.
In Figure \ref{fig:P4dsm} (left panel) we show the results obtained after
inserting this expression into Equation (\ref{eq:underst}) 
in comparison with the $P_{(2,1)}$ measurements from the $N$-body simulations.
The SPT-based model displays the same scaling behaviour with $k$ and $R$ 
as the data. However, for $R>6 \Mpc$ the SPT predictions are accurate to better than $13$ per cent, which is still not at the level of precision required for future galaxy clustering datasets; the deviations become larger with smaller $R$.
It follows from the definition of the $B^{\rm (s)}_{(2,1,1)}$
term that (see Appendix \ref{app:pt} for $\mathcal{B}^{\rm (s)}_{(2,1,1)}$)
\ba
 B^{\rm (s)}_{(2,1,1)}\!\!\!\!\!\!&&\!\!\!\!\!\!\!\!(\bk_1,\bk_2,\bk_3)
= \nonumber 
\frac{2}{3}\left[
 P(\bk_2) P(\bk_3)\mathcal{W}(\vk_2,\vk_3)    + 2 \ {\rm cyc}
\right] +\\
&&+\,\,\frac{1}{3}\int \frac{\dq_1}{(2\pi)^3}  
T(\bq_1,\bk_1-\bq_1,\bk_2,\bk_3)\mathcal{W}(\vq_1,\vk_1-\vq_1) \nn \\
&& +\,\,{2\ \cyc} \; , 
\ea
where $T$ denotes the matter trispectrum (i.e. the connected
part of the $4$-point correlator).  
The SPT contribution to lowest non-vanishing order is simply:
\be
 B^{\rm (s)}_{(2,1,1)}(\bk_1,\bk_2,\bk_3)
\simeq 
\frac{2}{3}\!\left[
P_{(0)}(\bk_2) P_{(0)}(\bk_3)\mathcal{W}(\vk_2,\vk_3) + 2 \ {\rm cyc}
\right]\; \label{eq:treeP4}\! .
\ee
In the right panel of Figure~\ref{fig:P4dsm} we show that this approximation
(star-shaped points) strongly underestimates the outcome from the $N$-body simulations
(solid symbols with errorbars) and does not display the same scaling behaviour
with $k$ and $R$ as the data. A common approach performed during observational data analysis is to substitute in place of the linear power spectrum, $P_{(0)}$, shown in Eq. \ref{eq:treeP4}, the fully non-linear power spectrum, $P_{(1,1)}$. We found that performing this substitution has little effect on the resulting amplitudes, remaining roughly equivalent as the lowest nonvanishing contributions.
We then go one step further and compute the next-to-leading-order corrections
to $B^{\rm (s)}_{(2,1,1)}$ which are of 
sixth-order in terms of the linear density field.
This gives
\ba
B^{\rm (s)}_{(2,1,1)}\!\!\!\!\!\!\!\!&&\!\!\!\!\!\!\!\! (\bk_1,\bk_2,\bk_3)\simeq
\frac{2}{3}\mathcal{W}(\vk_2,\vk_3)\left[\frac{}{}
P_{(0)}(\bk_2)P_{(0)}(\bk_3 )+ \right. \nn\\
&+&\!\!\!\!\!\left. P_{(0)}(\bk_2) P_{(1\ell)}(\bk_3)\right. \nn  
+\!\! \left.
\frac{}{}P_{(1\ell)}(\bk_2)P_{(0)}(\bk_3) \right]+ 2\ {\rm cyc} \nn \\
& &\!\!\!\!\!\!\!\!\!\!\!+\,\,\frac{1}{3}\int \frac{\dq_1}{(2\pi)^3}\mathcal{W}(\vq_1,\vk_1-\vq_1)  
T_{(0)}(\bq_1,\bk_1-\bq_1,\bk_2,\bk_3) \nn \\
& & \!\!\!\!\!\!\!\!\!\!\!+\,\,{2\ \cyc} \ ,
\ea
where $P_{(1\ell)}$ denotes the first loop
correction to the power spectrum (i.e. $P\simeq P_{(0)}+
P_{(1\ell)}+\dots$) and the term $T_{\rm 0}$
represents the tree-level contribution to the connected trispectrum.
In Appendix~\ref{app:pt} we provide the expressions needed for 
evaluating all these quantities, which are de-smoothed according to \Eqn{eq:Bdsm}.
Our final results are shown in Figure~\ref{fig:P4dsm} (solid lines).
The SPT approximation shows the correct scaling with $R$, but for $R>2 \Mpc$ it tends to overpredict the amplitude for collinear (i.e. $\theta\simeq 0$ and $\theta\simeq\pi$) configurations.  For $4<R\leq 8 \Mpc$ it also underpredicts the amplitude for triangles in which $\vk_1$ and $\vk_2$ are nearly perpendicular.  However, as $R$ increases the discrepancy lessens and at $R=10 \Mpc$ SPT performs better.
This suggests that using SPT to fit galaxy bispectra in the scale range $0.04\lesssim k \lesssim 0.12 \kMpc$ 
may possibly lead to seriously biased estimates for the parameters of the LEB.

Nevertheless, whilst the analytic calculations of 
${\mathcal P}_{(2,1)}$ and
${\mathcal B}^{\rm (s)}_{(2,1,1)}$
are feasible, computing higher-order terms 
becomes increasingly challenging. 
However, estimating these quantities from simulations
is no more demanding than measuring the low-order terms and so our
approach offers a distinct advantage over the classical SPT calculations.

\section{Estimation of halo bias}\label{sec:biassec}


\subsection{Bayesian parameter estimation}\label{ssec:biasest}

The second-order LEB contains three parameters: 
\mbox{$\bm\theta\equiv\{b_1,b_2,R\}$}.
In this section, we use Bayesian statistics to determine their  
values that best represent 
the halo power and bispectra extracted from our simulations.
For simplicity, we assume that the cosmological parameters are perfectly known
and that the measurement errors are Gaussian distributed, i.e.
\ba 
\label{eq:likelihood}
{\mathcal L}({\bf x}|\bm\theta) \!\!\!\!\!&=&\!\!\!\!\! (2\pi)^{-N/2}\,|{\mathbfss{C}}|^{-1/2}
{\rm e}^{
-\frac{1}{2}[(\bx-\mu(\bm\theta))^{\rm T}\mathbfss{C}^{-1}(\bx-\mu(\bm\theta))]}\;= \\
&=& \!\!\!\!\!(2\pi)^{-N/2}\,|{\mathbfss{C}}|^{-1/2}
{\rm e}^{-\frac{\chi^2(\bx,\bm\theta)}{2}}\;, \nn 
\ea
where ${\bf x}^{\rm T}$ is the $N$-dimensional vector containing
the power spectra or bispectra for different configurations,
$\mu(\bm\theta)$ is the model prediction and ${\mathbfss{C}}$ is the
covariance matrix. In theory ${\mathbfss{C}}$ is a model dependent
quantity, however owing to the technical challenge of estimating this
matrix and its inverse, we have decided to determine
${\mathbfss{C}}$ directly from the data. 

Equation (\ref{eq:likelihood}) gives the likelihood of the data given the model,
but what we need in order to perform parameter estimation is the
posterior probability of the model parameters given the data. 
This can be obtained using Bayes' theorem:
\be
P(\bm\theta|\bx)=\frac{\Pi(\bm\theta)\,
{\mathcal L}(\bx|\bm\theta)}{p(\bx)}\ ,
\label{eq:posterior}
\ee
where $\Pi(\bm\theta)$ is the prior probability for the model parameter while
the evidence,
\be p(\bx)\equiv\int \Pi(\bm\theta)\, {\mathcal L}(\bx|\bm\theta)\, 
d^3\,\theta \ ,\ee
simply acts as a normalizing factor and does not influence the search
for the best fit. In what follows we will always assume flat
priors on the parameters, but bounded over a finite domain which is much more
extended than the likelihood function.
Moreover, $b_1$ and $R$ will always be assumed to be positive.


\subsection{Covariance matrix estimation}\label{sssec:covar}

The sample covariance matrix
\be 
\widehat{{\mathbfss{S}}}\equiv\frac{N}{N-1}\langle\Delta{\bf x}^{\rm T}\Delta{\bf x}\rangle_N\ ; 
\hspace{0.5cm}\Delta{\bf x}\equiv {\bf x}-\langle\bx\rangle_N \;,\label{eq:cov}
\ee
where $\langle\dots\rangle_N$ denotes the arithmetic mean over $N$
independent measurements,
provides an unbiased estimator of the covariance matrix for the measurement
errors.  

However, this estimator is extremely unstable and inefficient. It 
generally provides 
matrices where the smallest eigenvalue is too small and the largest one is
too big. Very large samples are thus needed to obtain accurate estimates
of the covariance.

On using our ensemble of 200 simulations for both the power and the bispectra,
we could measure the
diagonal elements of the covariance with an accuracy of
$\sim10$ per cent. On the other hand, the off-diagonal elements 
had a much smaller absolute value and 
were scattering around zero with errors of the order of $\sim 100$ per cent.
All this suggests that the covariance should be close to diagonal as expected
for a Gaussian random field with infinitesimally narrow bins in $k$-space.

Due to these large uncertainties in the off-diagonal elements, we opted for 
implementing a shrinkage method to better estimate the covariance matrices
of our power and bispectra.
Shrinkage estimation is a variance reduction technique that shrinks
an empirical estimation of the covariance like $\widehat{{\mathbfss{S}}}$
towards a theoretical model 
for how the covariance should be, represented by a structured matrix
$\mathbfss{T}$ (the target covariance).
The shrinked estimator is given by the convex linear combination
\be
\widehat{\mathbfss C}= \lambda {\mathbfss T} + (1-\lambda) \widehat{\mathbfss S}
\ee
where $0<\lambda<1$ is the shrinkage intensity. 
This ensures the resulting covariance matrix to be positive definite even
if $\widehat{\mathbfss S}$ is singular (because it is determined from $N<{\mathrm {dim}}(\vx)$ observations).

It has been demonstrated that shrinkage techniques provide a regularized estimate of the covariance $\widehat{\mathbfss S}$ which is both more accurate and statistically efficient than either of the individual estimators $\widehat{\mathbfss{S}}$ and $\mathbfss{T}$, and they do so in a systematic way 
\citep{SchaeferStrimmer2005}. Without the need for specifying an underlying
probability distribution, \citet{LedoitWolf2003} provided a theorem that 
determines the optimal value for $\lambda$ through minimization of a quadratic
loss function such as the mean-square error of the covariance matrix.
This can be expressed in terms of the squared Frobenius norm
\ba
L(\lambda) & = & \parallel \widehat{\mathbfss{C}} - \bf{\Sigma}\parallel^2_{F} \nonumber \\
& = & \parallel\lambda \mathbfss{T} + (1 - \lambda) \widehat{\mathbfss{S}} - \bf{\Sigma}\parallel_{F} \nonumber \\
& = & \sum_{i,j=1}^{p}(\lambda t_{ij} + (1 - \lambda) s_{ij} - \sigma_{ij})^2 \ .
\label{eq:Frobnorm}
\ea
which gives a measure of the distance between the true population covariance, $\bf{\Sigma}$, and the inferred one, namely, $\widehat{\mathbfss{C}}$. 
The key is to select a suitable target, and we assume it to be a diagonal matrix
with unequal variances coinciding with the sample variances:  

\be
t_{ij} = 
\begin{cases}
{s}_{ii}, & \text{if i = j} \\
0,                & \text{if i $\ne$ j}
\end{cases}\;.
\ee

Minimizing Eq. (\ref{eq:Frobnorm}) gives the expression for the optimal 
shrinkage intensity: 
\be
\lambda_{*}  =  
\frac{\sum_{j>i} {\textrm{Var}}(s_{ij})}{\sum_{j>i}[\textrm{Var}(s_{ij})+
\sigma_{ij}^2]} =
\frac{\sum_{j>i} {\textrm{Var}}(s_{ij})}{\sum_{j>i} \textrm{E}(s_{ij}^2)} \,,
\label{lambdastar}
\ee
where ${\textrm E}(\dots)$ denotes the expectation value of a random variable.
Following \citet{SchaeferStrimmer2005}, we estimate
the sampling variance of the elements of the sample covariance using
\be
\widehat{\textrm{Var}}(s_{ij})=\frac{N}{(N-1)^3}\sum_{j=1}^N
\left(\Delta{\bf x}_j^{\rm T}\Delta{\bf x}_j-\langle\Delta{\bf x}^{\rm T}\Delta{\bf x}\rangle_N\right)^2\;.
\ee
However, while these authors approximate ${\mathrm E}(s_{ij}^2)$ in Eq. (\ref{lambdastar})
with the square of the point estimate $s_{ij}$ thus overestimating $\lambda_*$,
we adopt the square of the sample covariances $s_{ij}^2$ as a proxy for
$\sigma_{ij}^2$ \citep[e.g.][]{Kwan2011}.
In all cases, we found that the optimal shrinkage intensity was roughly 
$\hat{\lambda}_{*} \sim 0.45$ for the power spectra covariance and 
$\hat{\lambda}_{*}\sim0.23$ for the bispectra covariance, respectively.
Note that the adopted algorithm only performs shrinkage of the off-diagonal 
elements of the covariance matrix. 

\subsection{Constraining the bias parameters: $b_1$, $b_2$ and $R$}
\label{ssec:bofR}

We now determine the best-fit model parameters for 
the various power and bispectra that we have
estimated from the simulations within the scale range $0.04<k<0.12$ $\kMpc$.
We consider two second-order LEB models that differ in the 
polyspectra describing the non-linear matter distribution (see below
for the details). 
In both cases,
we map the likelihood function within a finite volume of the parameter
space that we slice into a regular Cartesian mesh.

\subsubsection{SPT tree-level model}
\label{SPTtree}
The first model uses SPT at the lowest 
non-vanishing order to approximate the $P_{(l,m)}$ and $B_{(l,m,n)}$. 
This is what is most commonly done in the literature.
For the power spectrum, the $P_{(l,m)}$ terms expressed at tree-level of SPT are:
\ba P^{\rm tree}_{(1,1)} & = & P_{(0)}(k) \;,\label{eq:PowtreeMod} \\
P^{\rm tree}_{(i,j)} & = & 0 \ {\mathrm {for}}\ i+j>2 \;,
\ea
where $P_{(0)}(k)$ denotes the linear matter power spectrum. 

Thus, the bias relation is linear and carries no dependence on the filter scale, $R$, and on $b_2$.

For the bispectrum, the evaluation of $B_{(l,m,n)}$ using only 
tree-level contributions gives:
\ba B^{\rm (s), tree}_{(1,1,1)} & = & 2\,P_{(0)}(k_1)P_{(0)}(k_2)
F_2(\bk_1,\bk_2)+2\,{\rm cyc} \;,\\ 
B^{\rm (s), tree}_{(2,1,1)} & = & \frac{2}{3}\,P_{(0)}(k_1)P_{(0)}(k_2)\mathcal{W}(\vk_1,\vk_2)+2\,{\rm cyc} \;,\\ 
B^{\rm (s), tree}_{(i,j,k)} & = & 0\ {\mathrm{for}}  \ i+j+k>4 \;,
\label{eq:BisptreeMod}
\ea
where $F_2(\bk_1,\bk_2)$ is the second-order mode-coupling kernel function from
SPT \citep[e.g.][]{Bernardeauetal2002}. 

\subsubsection{Fully non-linear model}
The second model considers the fully non-linear matter 
polyspectra extracted from the simulations.
Note that, while evaluating the LEB when varying $b_1$ and $b_2$ at fixed $R$ 
is a trivial exercise, varying $R$ would, in theory, require
recomputing all the relevant $P_{(l,m)}$ and $B_{(l,m,n)}$.
However, as we mentioned earlier in our discussion of
\Fig{fig:dsmpterms}, these functions change smoothly with $R$. We
therefore use a cubic-spline interpolation of $\log\left[P_{(l,m)}\right]$ and $\log\left[B_{(l,m,n)}\right]$
to model the $R$-dependence of the theory.  This enabled us to map the
likelihood function with arbitrary resolution.

A final comment is in order regarding the details of how the fit is performed.
There is some arbitrariness in defining what exactly are the ``observables'' 
and what is the ``model'' in the simulations.
For instance, we could have fit the outcome of each $N$-body simulation 
separately using the polyspectra extracted from the very same realization.
While being a valid test of the LEB, this method would have not had much in
common to actual galaxy redshift surveys (or even to the SPT model discussed
above), where the underlying mass 
distribution is unknown and needs to be modeled independently.
In fact, the presence of the same noise structure in the matter and halo
power and bispectra would result in overfitting.  There are a couple of alternative approaches one could follow to prevent this.  The first is to generate smooth versions of the $P_{(l,m)}$ and $B_{(l,m,n)}$ terms by averaging over the entire ensemble of simulations.  One can then use these ``theoretical models'' to simultaneously fit the halo statistics extracted from all of our 200 independent realizations. The other alternative is to subdivide the total ensemble of simulations into two subsets, where one subset would be used to construct the smooth $P_{(l,m)}$ and $B_{(l,m,n)}$ terms by averaging over the total number of simulations in the subsample and the other subset would serve as the halo statistics to be analyzed.  The partition of the ensemble of realizations into two distinct subsets ensures that the ``model'' and ``data'' are indeed independent.  Furthermore, one can exchange the roles of ``model'' and ``data'' for the two subsets and then sum the $\chi^2$s obtained from the two sets of analysis. We carried out both approaches however we only report the results from averaging over the 200 simulations as the bias model constraints compared to the partitioning approach are in extremely good agreement.

\subsection{Goodness of fit}\label{ssec:goodness}

In this Section we use the classic $\chi^2$ goodness-of-fit 
test to quantify how well the second-order LEB fit our simulated data.
We minimise the $\chi^2$ function over the parameter space using the simplex
method. The best-fit models determined this way basically coincide with those
that minimise the $\chi^2$ function in the dense grid used for our Bayesian 
analysis. Since for all power and bispectra we have always used 20 bins in $k$ or $\theta$ and the covariance matrices are full rank, the number of degrees of freedom totalled $\nu=200\cdot 20-3=3997$ for each fit.


\begin{figure}
 \centering{
   \includegraphics[height=8cm]{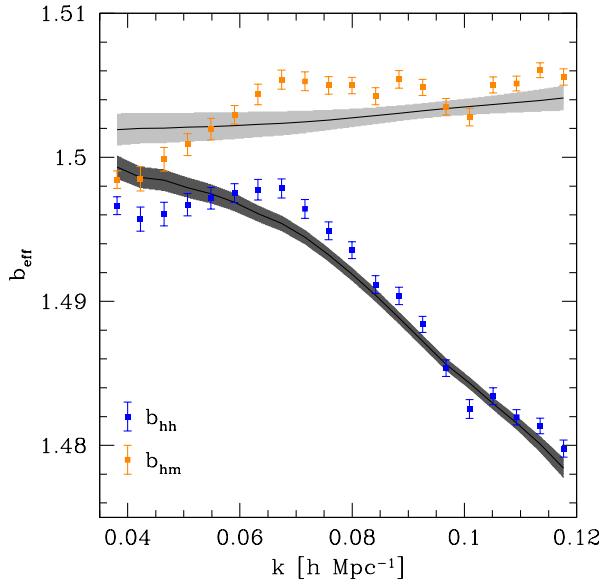}}
\caption{The effective halo-bias parameters 
$b_{\hm}=P_{\hm}/P_{\mm}$ (orange symbols) and
$b_{\hh}=(P_{\hh}/P_{\mm})^{1/2}$ (blue symbols) extracted from our simulations 
as a function of the wavenumber.
The black solid lines and shaded regions indicate the mean 
and the rms value of the effective bias obtained by averaging the predictions
of the second-order LEB over the posterior probability of the model parameters.}
\label{fig:Powbias200}
\end{figure}

\begin{figure*}
  \centering{
    {\includegraphics[width=5.7cm]{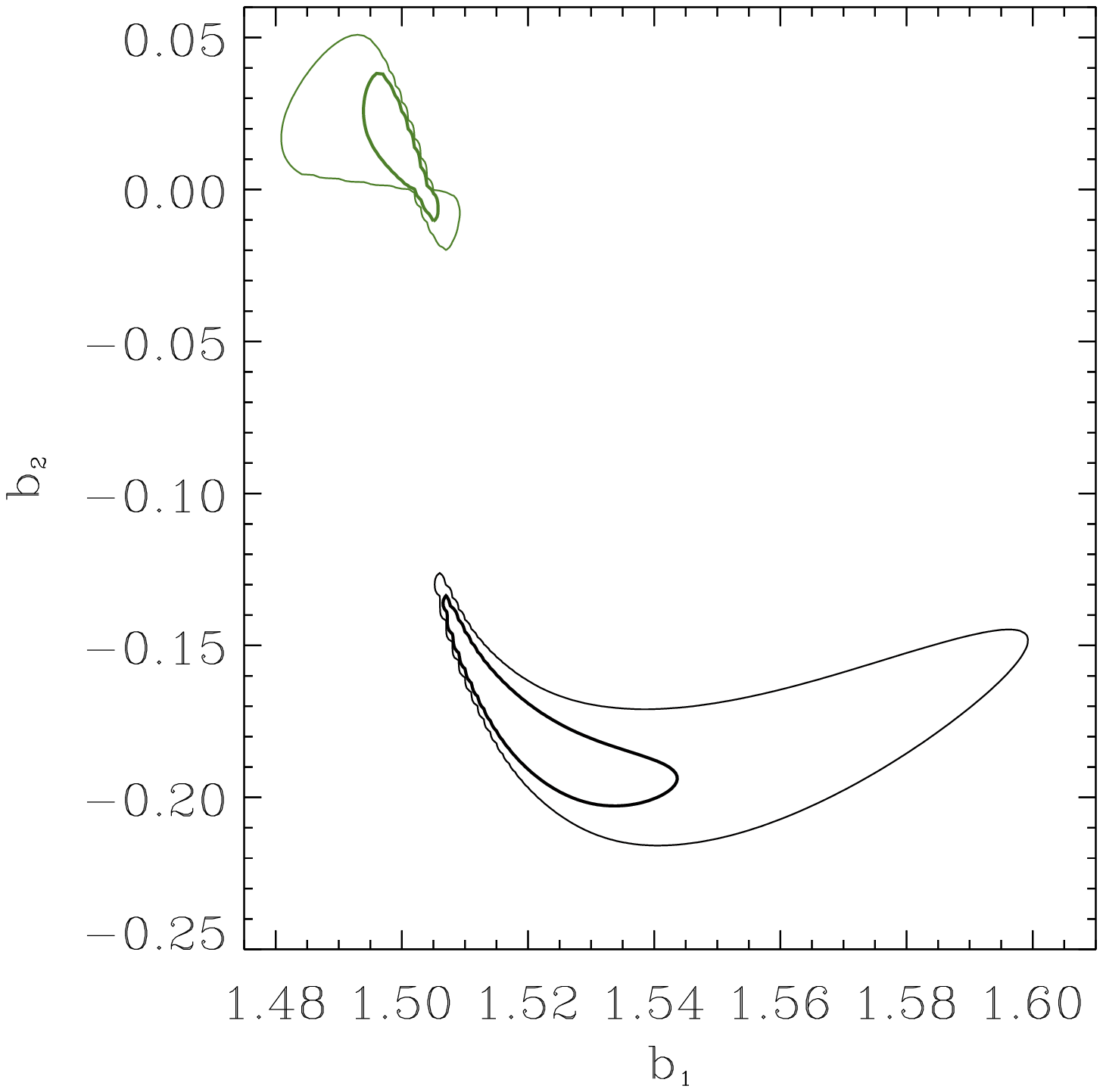}
   \includegraphics[width=5.7cm]{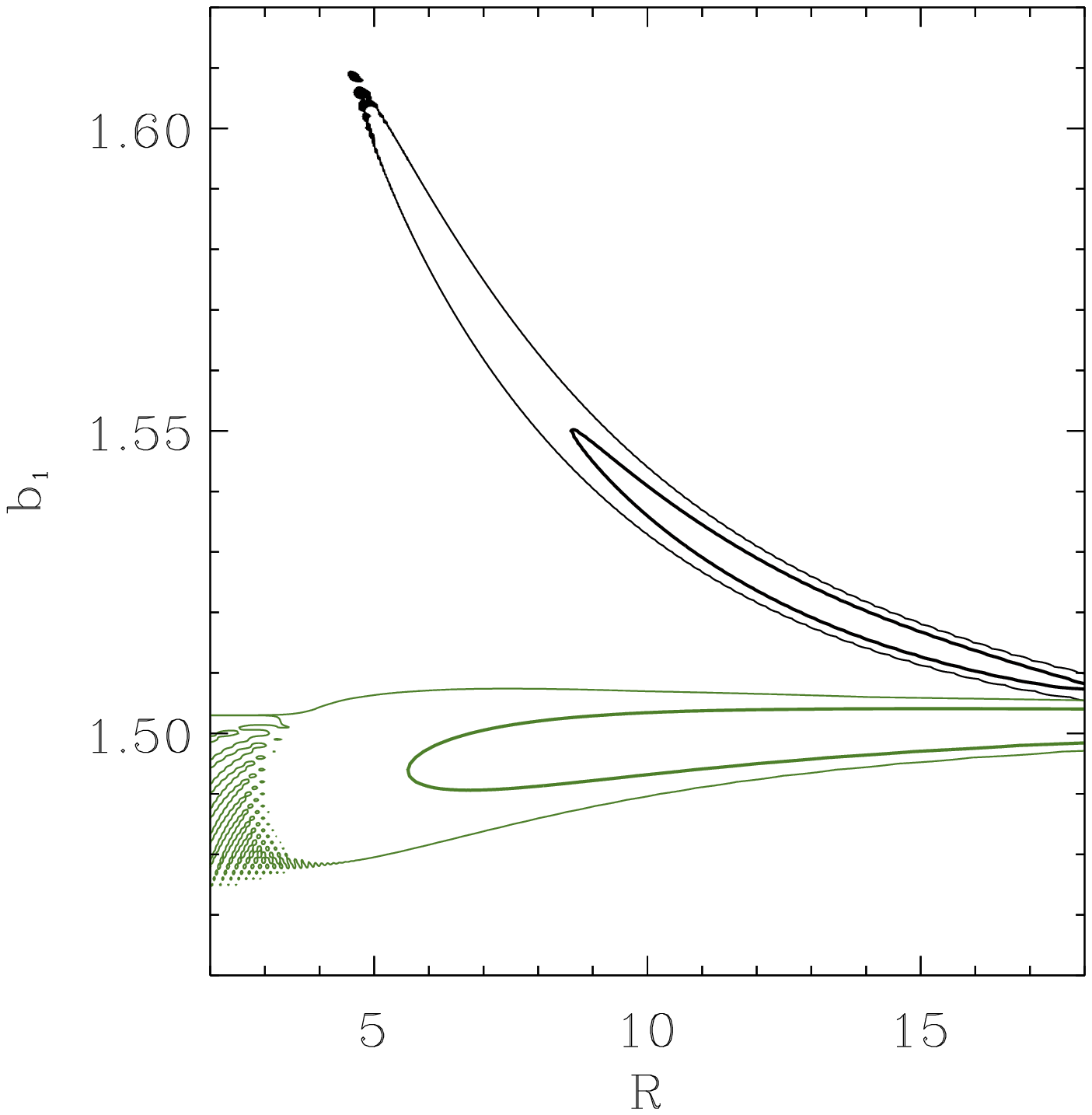}}
  \includegraphics[width=5.7cm]{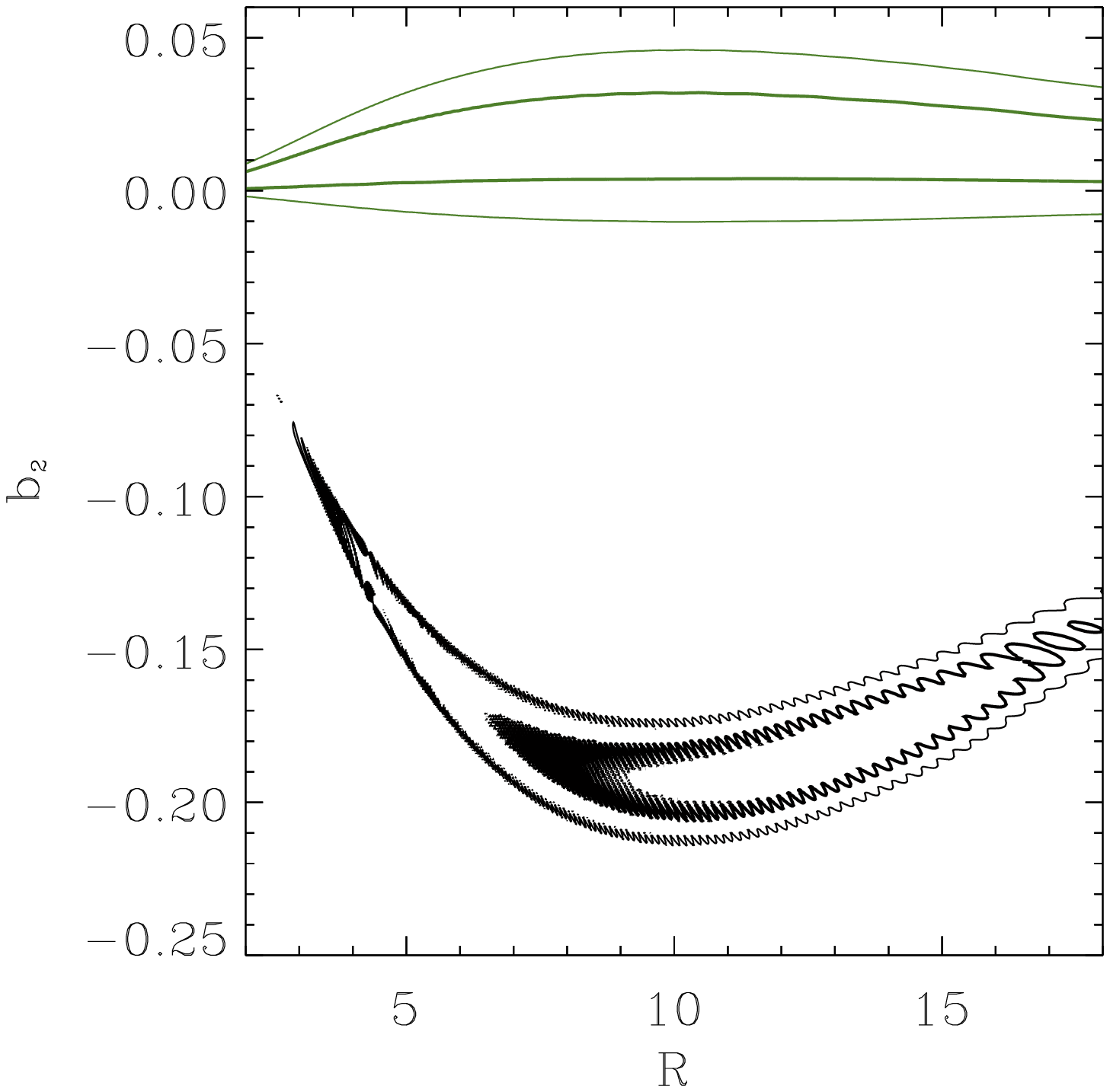}}
\caption{Joint marginal probability distribution for the parameter pairs 
$b_1$--$b_2$, $b_1$--$R$ and $b_2$--$R$ (from left to right) obtained 
using the fully non-linear model for $P_{\hh}$ (black) and $P_{\hm}$ (green).
Contours correspond to the 68.3 and 95.4 per cent credible intervals.
}
 \label{fig:PowMarg2D}
\end{figure*}

\subsubsection{Power spectra}
The tree-level SPT models for the halo power spectra provide very poor fits
to our data. The minimum $\chi^2$ value is much larger than the number of 
degrees of freedom, reaching $\chi^2_{\rm min}\simeq 7465$ for
$P_{\hm}$ and $\chi^2_{\rm min}\simeq 139,821$ for $P_{\hh}$.
These results may serve as indicators that halo biasing is non-linear 
and/or a result of the breakdown of linear SPT. To check the latter, we refit
both spectra using Equation (\ref{eq:PowtreeMod}) but after replacing
$P_{(1,1)}^{\rm tree}$ with the fully non-linear matter power spectrum $P_{(1,1)}$.
In this case, we acquire $\chi^2_{\rm min}\simeq 3903$ for $P_{\hm}$
and $\chi^2_{\rm min}\simeq 4442$ for $P_{\hh}$.   
This significant improvement to 
the tree-level results demonstrates the need to model non-linearities 
in the matter distribution very accurately.
Using the fully non-linear model with the additional free-parameters $b_2$ and
$R$ only slightly improves the goodness of fit for $P_{\hm}$, giving 
$\chi^2_{\rm min} \simeq 3901$. On the other hand, the improvement is marked for $P_{\hh}$ for which we obtain $\chi^2_{\rm min} \simeq 3915$.

It is interesting to see how the $\chi^2_{\rm min}$ value changes when $R$
is kept fixed. In this case, we find that all the fits to $P_{\hm}$ are 
equally good. However,
for $P_{\hh}$, the values of $\chi^2_{\rm min}$ undergo a sharp decrease (from 
$3944\lesssim \chi^2_{\rm min} \lesssim 3921$) 
for $2< R\lesssim 3.66$ $\Mpc$, 
then decrease slowly to the absolute minimum value at 
$R\sim 13.2$ $\Mpc$ and finally begin to slowly rise again to our cutoff scale 
of $R=18$ $\Mpc$. Hence, it appears that there is a range of preferred
smoothing scales that best fit the simulation data for $P_{\hh}$. 

\subsubsection{Bispectra}
\label{bispchi2}
Turning now to the bispectra, we find that the fully non-linear
model provides slightly better fits to the numerical data 
($\chi^2_{\rm min}\simeq 3906, 3908$ and $3913$ for $B_{\hmm}$, $B_{\hhm}$,
and $B_{\hhh}$, respectively) than the tree-level model
($\chi^2_{\rm min}\simeq 3923, 3922$ and $3925$) which, however, already
supplies $\chi^2_{\rm min}/\nu\lesssim 1$.

In all cases,
if we keep $R$ fixed and only consider 2-parameter models,
we find that the $\chi^2_{\rm min}$ value does not change much
for $2<R<13$ $\Mpc$ 
while it rapidly grows adopting larger smoothing scales.
In terms of goodness of fit, the 
non-linear model for $B_{\hmm}$ outperforms the tree-level SPT model 
for all values of $R$. On the other hand, when $B_{\hhm}$ and $B_{\hhh}$ are 
considered, the non-linear model gives a better fit only
for $R\lesssim 15$ $\Mpc$.

\subsubsection{Posterior mean}

In order to give a visual impression of the best-fit models,
in Figure \ref{fig:PowBisp200}  we show the posterior mean (black line) and 
the posterior rms error 
(shaded gray region) for the halo power and bispectra resulting from our fits 
with the fully non-linear model in comparison with the simulation data. 
In all cases, the models agree with the simulations remarkably well. 
Note that the rms error on the best-fit models for $P_{\hh}$ and $P_{\hm}$ is 
hardly visible on the scale of the plots.

\subsection{Bias from the power spectrum}

\subsubsection{Effective bias}
Due to its highly compressed ordinate,
Figure \ref{fig:PowBisp200} gives the false impression that $P_{\rm hm}$ and
$P_{\rm hh}$ are nicely described by rescaling the matter
power spectrum with constant multiplicative factors $\sim 1.5$ and $1.5^2$,
respectively.
In order to examine the bias relation more closely as a function of scale, we 
introduce two effective bias coefficients by taking different ratios of the halo
power spectra after\footnote{Very similar results are obtained if one averages the ratios instead 
of taking the ratio of the averages.} averaging them over the 200 N-body 
simulations:
$b_{\rm hm}=\langle P_{\rm hm}\rangle/\langle P_{\rm mm}\rangle$ and
$b_{\rm hh}=(\langle P_{\rm hh}\rangle/\langle P_{\rm mm}\rangle)^{1/2}$.
We present our results in Figure~\ref{fig:Powbias200}.
The solid points with errorbars represent the effective biases estimated using 
the 
shot-noise corrected quantities of both the auto- and cross halo power spectra.
We compute the $1\sigma$ uncertainties via error propagation accounting for the
covariance between the different observables. 
It can be seen that $b_{\hh}$ and $b_{\hm}$ do not perfectly match each other. On large-scales ($k < 0.06\,\kMpc$), $b_{\hh}$ keeps roughly constant while 
it shows a significant scale dependence for $k>0.06\,\kMpc$, whereas $b_{\hm}$ 
shows the opposite trend although its scale dependence
is weaker for the large scales. At $k\simeq 0.04\,\kMpc$, $b_{\hm}$ and 
$b_{\hh}$ assume very similar values. However, $b_{\hm}>b_{\hh}$ for all
wavenumbers. Our high-quality data also provide some hints for the presence 
of weak oscillatory features in the effective bias parameters on the scales
of baryonic acoustic oscillations. 

Figure~\ref{fig:Powbias200} also tests how the fully non-linear
second-order LEB model is able to reproduce
the scale-dependence of $b_{\hm}$ and $b_{\hh}$ in fine details. 
The black curves represent the posterior mean of the effective
bias coefficients and the shaded grey regions denote the corresponding 
rms value of their posterior distribution.  
Although the models are not able to reproduce all the features which are
present in the numerical data, they are in reasonable agreement with the
simulations, especially for $k>0.08\,\kMpc$.
Nevertheless, we see that for both $b_{\hm}$ and $b_{\hh}$ the power spectum 
models actually are less accurate at small $k$ (i.e. on the large scales) 
in the proximity 
of the point where the trend from constant-to-scale dependence (and vice versa)
occurs. 
On these scales, the models systematically overpredict the effective biases and
the largest discrepancy is of the order of $\sim0.3$ per cent.
%

\begin{figure*}
\centering{
    \includegraphics[bb = 25 166 577 516,height=8.5 cm]{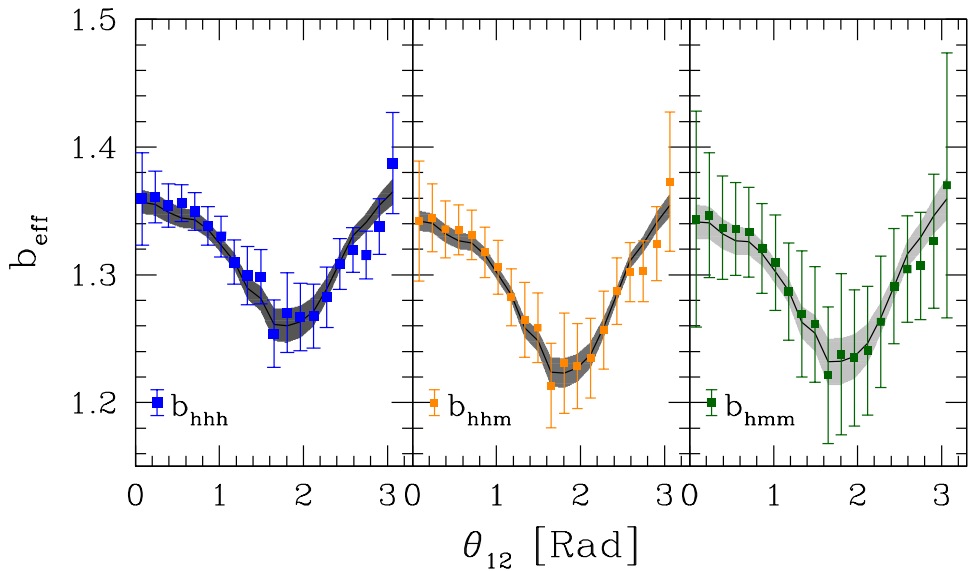}}
\caption{As in Figure \ref{fig:Powbias200} but for the effective bias parameters
$b_{\rm hmm}=B_{\rm hmm}/B_{\rm mmm}$,
$b_{\rm hhm}=(B_{\rm hhm}/B_{\rm mmm})^{1/2}$ and
$b_{\rm hhh}=(B_{\rm hhh}/ B_{\rm mmm})^{1/3}$.}
\label{fig:Bispbias200}
\end{figure*}


\subsubsection{Marginal credible regions}
Now we compare the level of the consistency between the model-parameter 
constraints deriving from the fits to the halo power spectra, 
$P_{\hh}$ and $P_{\hm}$.
Figure~\ref{fig:PowMarg2D}
shows (from left to right) the marginal posterior distributions 
for the parameter pairs $b_1$--$b_2$, $b_1$--$R$ and $b_2$--$R$ of our
fully non-linear model.  
The black and green
contours denote the $68.3\%$ and $95.4\%$ credible regions for the
parameters of the LEB obtained from analyzing 
$P_{\hh}$ and $P_{\hm}$, respectively.
The first apparent observation is that 
the contours of $P_{\hm}$ and $P_{\hh}$ span different regions of the parameter 
space: while the $P_{\hm}$ data prefer $b_1\lesssim 1.5$ and $b_2\gtrsim 0$, 
$P_{\hh}$ favours $b_1\gtrsim 1.5$ combined with $-0.15 \lesssim b_2\lesssim 
-0.2$.
In other words,
the second-order LEB model provides a succesful fit to 
$P_{\hh}$ or $P_{\hm}$ but requires two incompatible parameter sets.
Improper modelling of the shot noise in $P_{\hh}$ might be the primary cause of 
the inconsistency \citep[e.g.][]{Hamausetal2010}.
Note, however, that the best-fit values for $b_1$ and $b_2$ that we derive from 
$P_{\hh}$ are in good agreement with the predictions of theories that follow
the collapse of dark-matter halos \citep[e.g. Equation (14) and (15) in][]
{Scoccimarroetal2001}.
It is also worth mentioning that,
for Gaussian fluctuations in the matter density,
the cross-spectrum of locally-biased tracers is always exactly 
proportional to $P_{\mm}$ even though this is not apparent from the mathematical
formulation of the LEB \citep{FruscianteSheth12}.
The fact that our measurement of $P_{\hm}$ needs $b_2\simeq 0$ might 
simply suggest that a similar
relation holds true also in the presence of non-Gaussian perturbations
(at least approximately, since $b_{\hm}$ keeps nearly constant with $k$
as shown in Fig. \ref{fig:Powbias200}).

\subsection{Bias from the bispectrum}

\subsubsection{Effective bias}
To investigate the bias relation as a function of
scale using the halo bispectra,  
we define a set of coefficients by taking the following ratios: 
$b_{\rm hmm}=\langle B_{\rm hmm}\rangle_{200}/\langle B_{\rm mmm}\rangle_{200}$,
$b_{\rm hhm}=(\langle B_{\rm hhm}\rangle_{200}/\langle B_{\rm mmm}\rangle_{200})^{1/2}$,
$b_{\rm hhh}=(\langle B_{\rm hhh}\rangle_{200}/\langle B_{\rm mmm}\rangle_{200})^{1/3}$.
The results are shown in Figure \ref{fig:Bispbias200}: 
all the effective bias coefficients present a characteristic configuration 
dependence and are in agreement within their $1\sigma$ 
uncertainties (although $b_{\rm hhh}$ tends to assume slightly higher values
for all triangle configurations).
The posterior means of the effective bias coefficients from the fully non-linear
models also closely match the data as expected from the $\chi^2$ test
presented in \S\ref{bispchi2}.
All this suggests the second-order LEB provides a 
suitable description of the bias relation at the three-point level.


\begin{figure*}
  \centering{
    {\includegraphics[width=6cm]{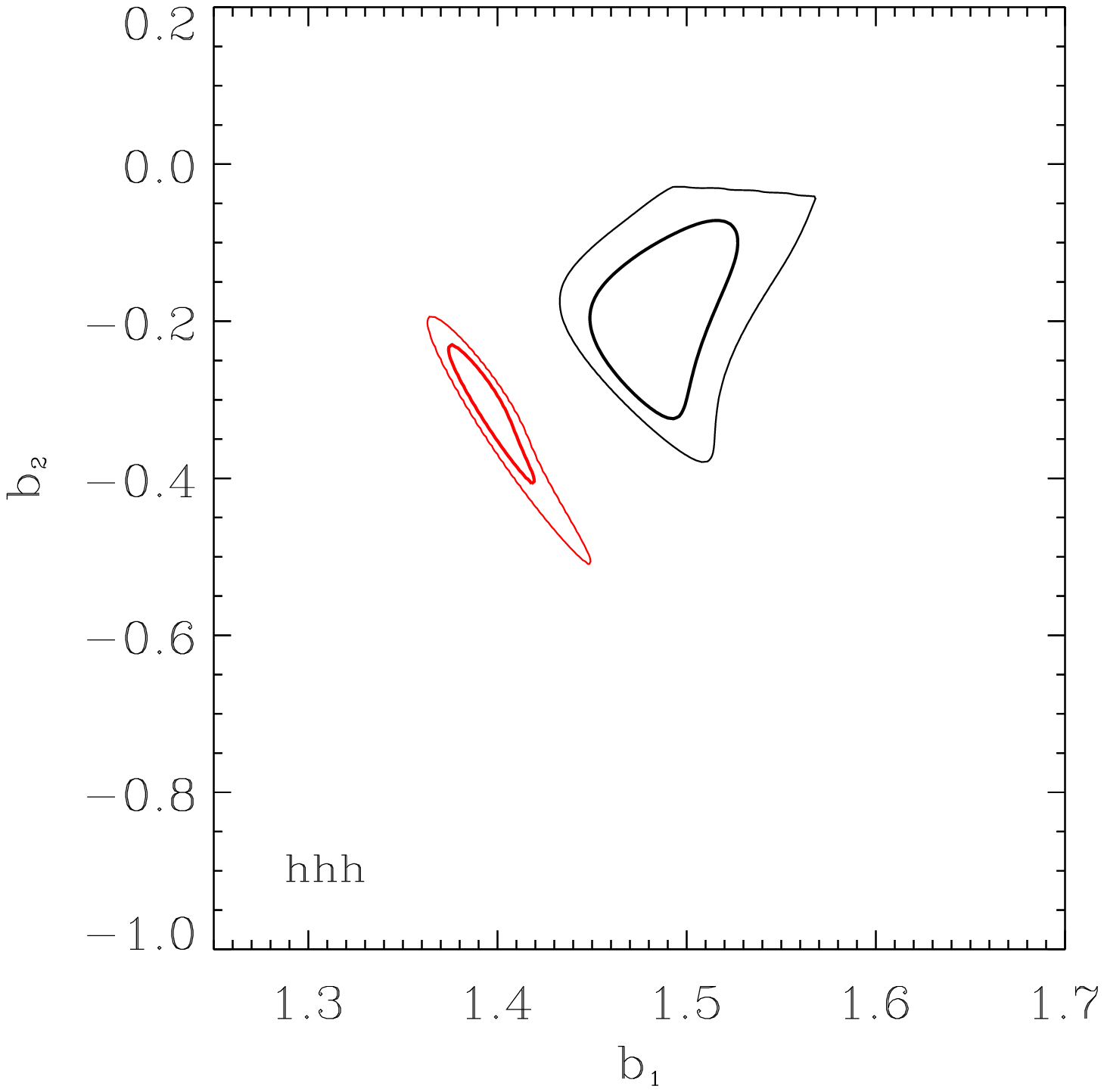}
    \includegraphics[width=6cm]{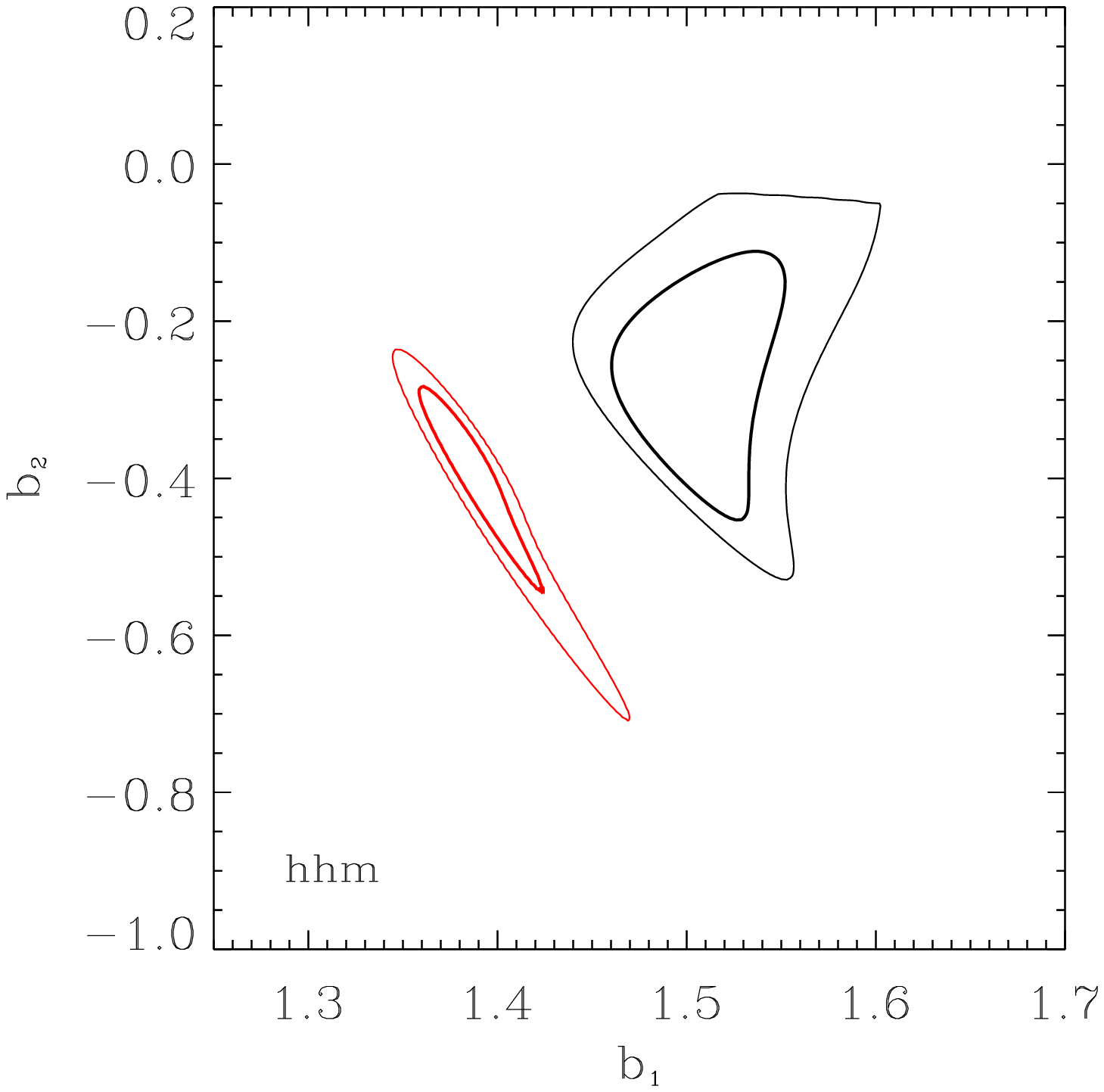}}
    \includegraphics[width=6cm]{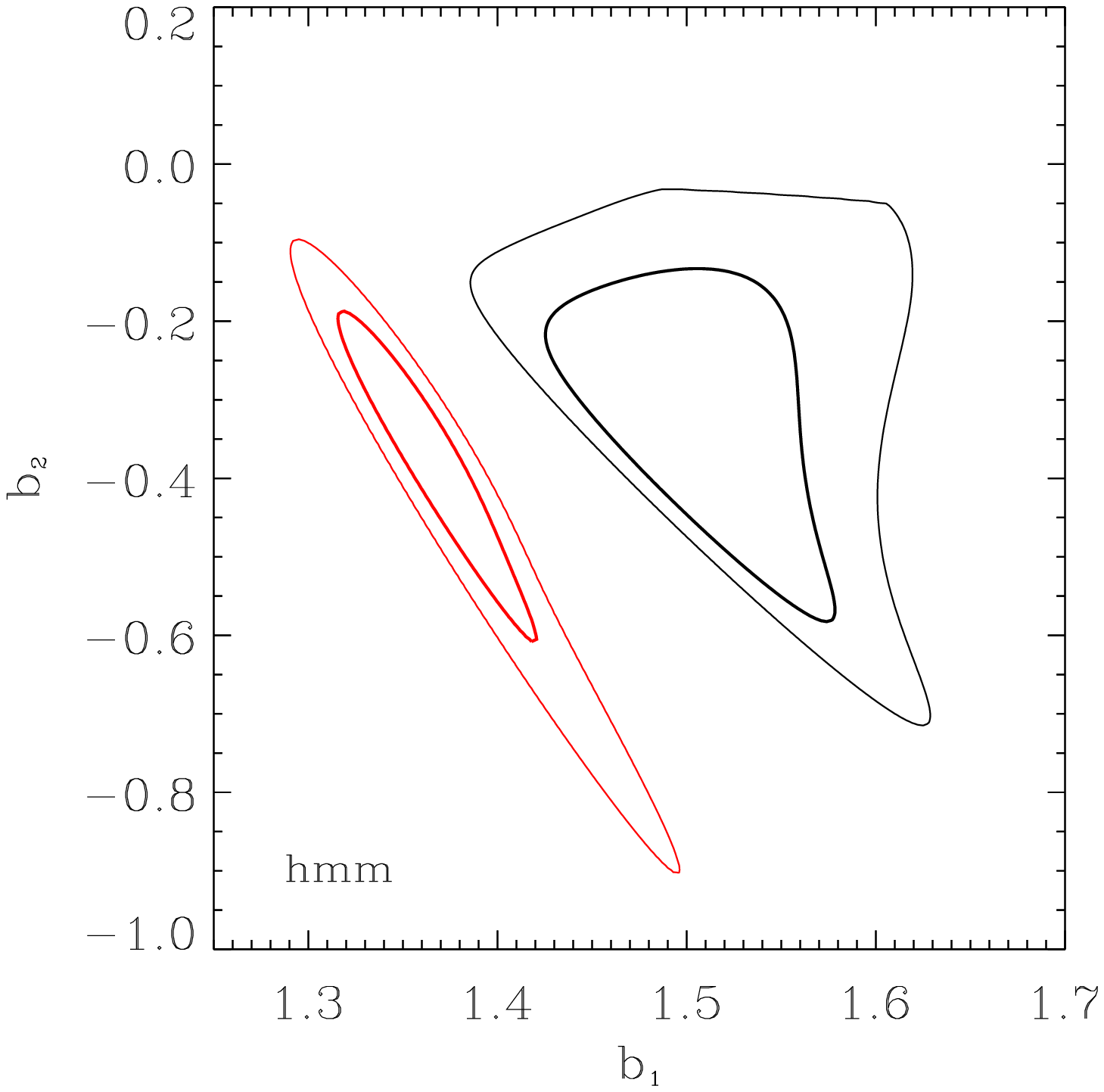}}
\centering{
    {\includegraphics[width=6cm]{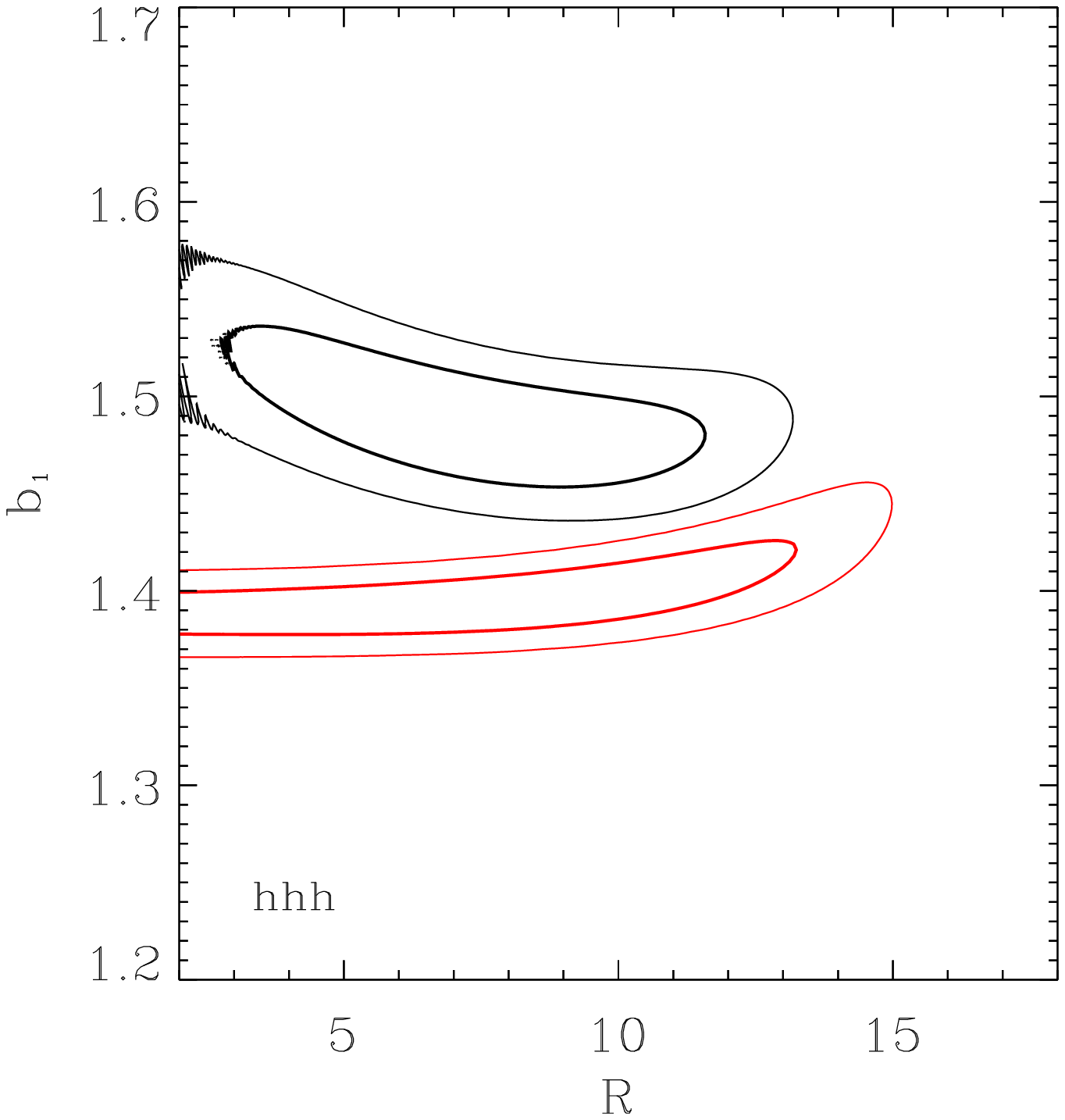}
      \includegraphics[width=6cm]{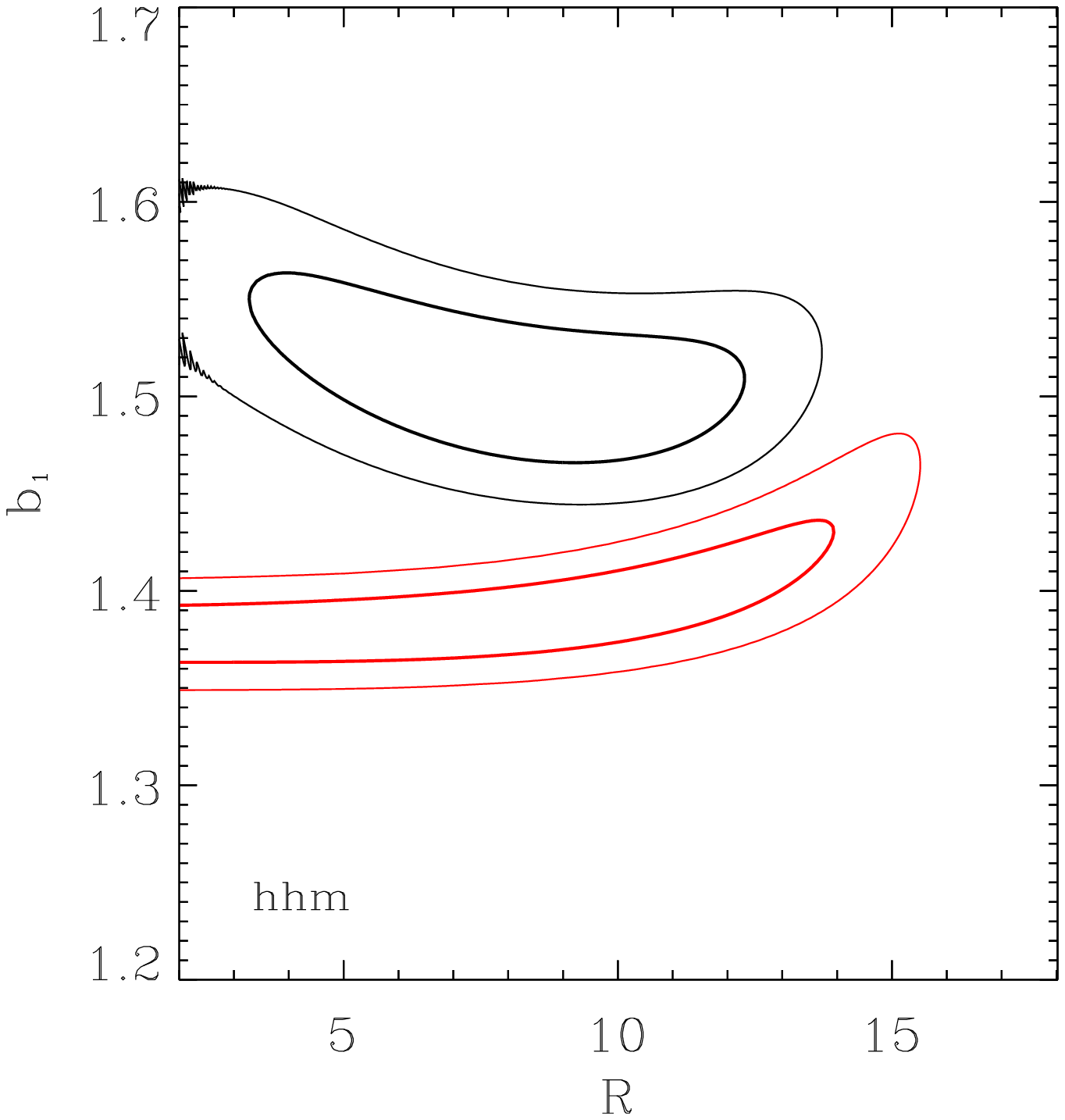}}
    \includegraphics[width=6cm]{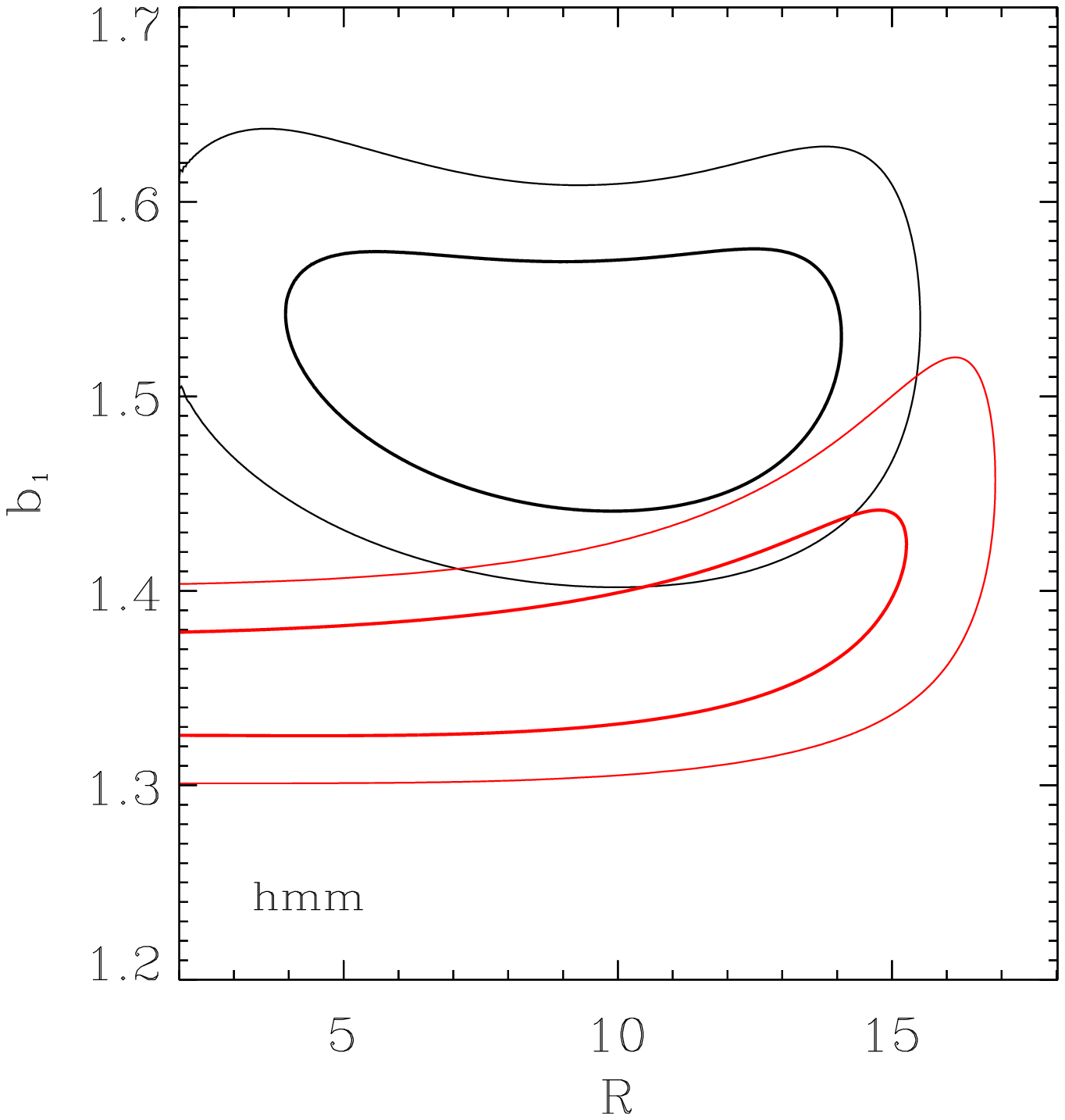}}
\centering{
    {\includegraphics[width=6cm]{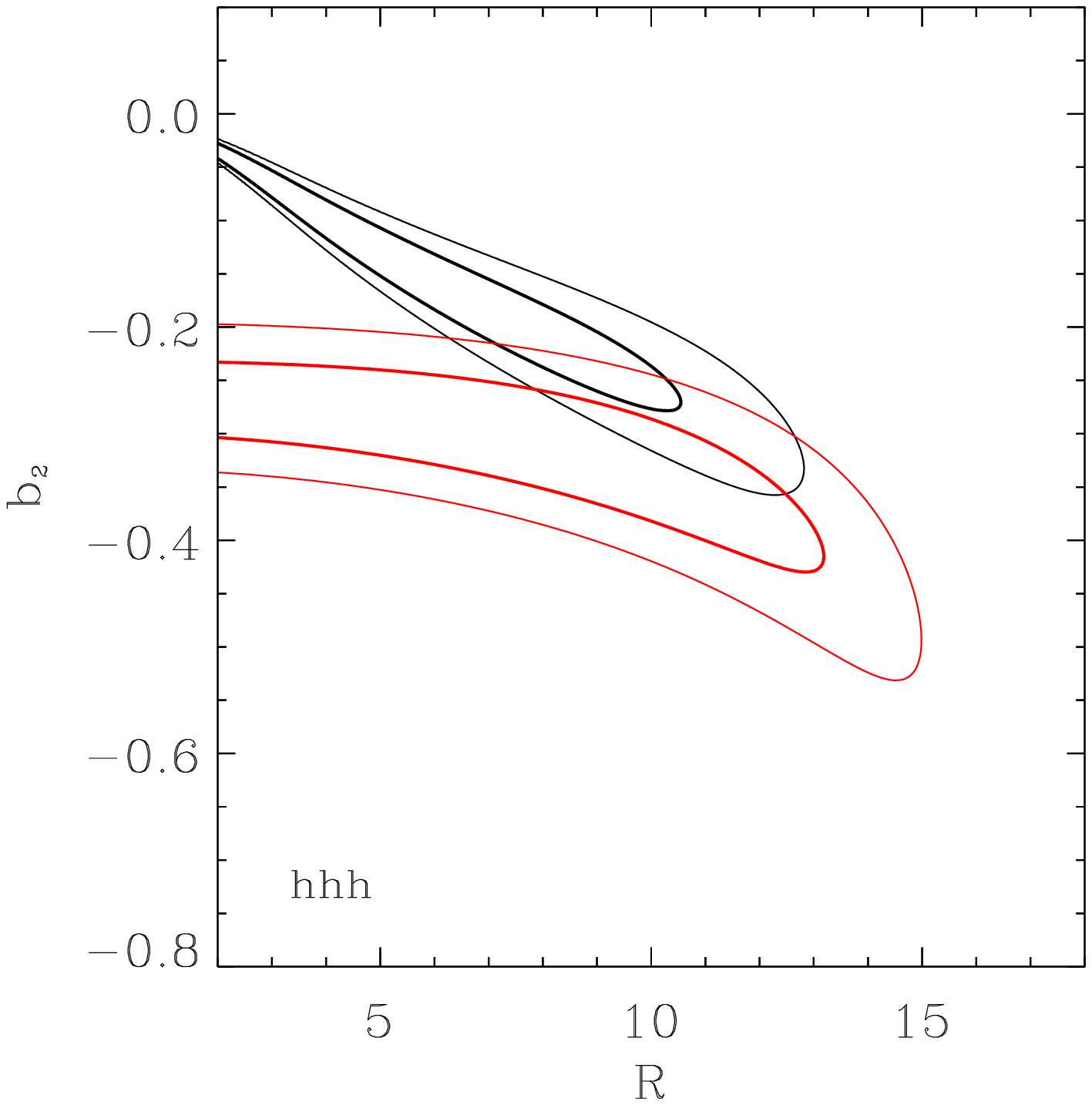}
      \includegraphics[width=6cm]{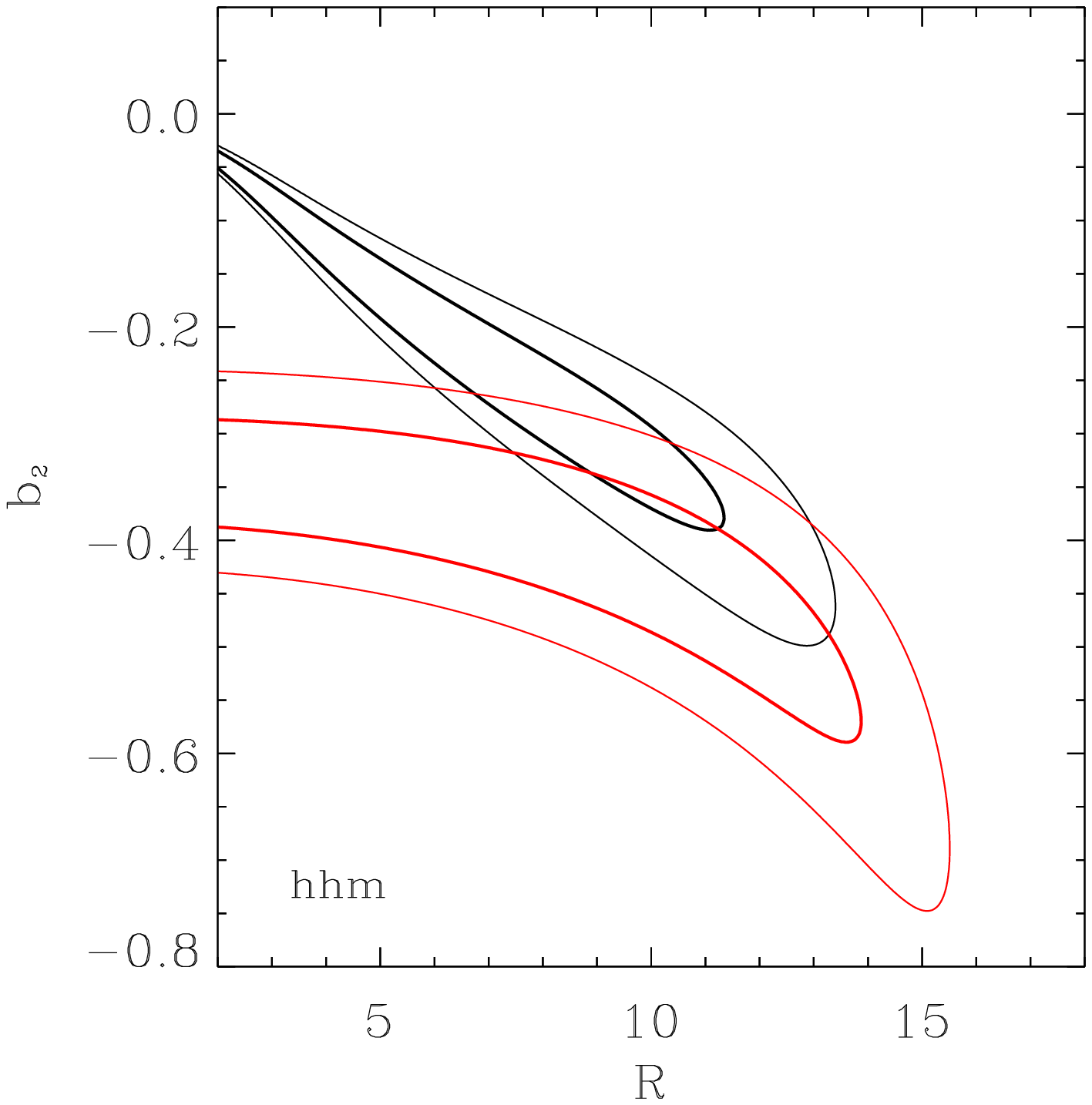}}
    \includegraphics[width=6cm]{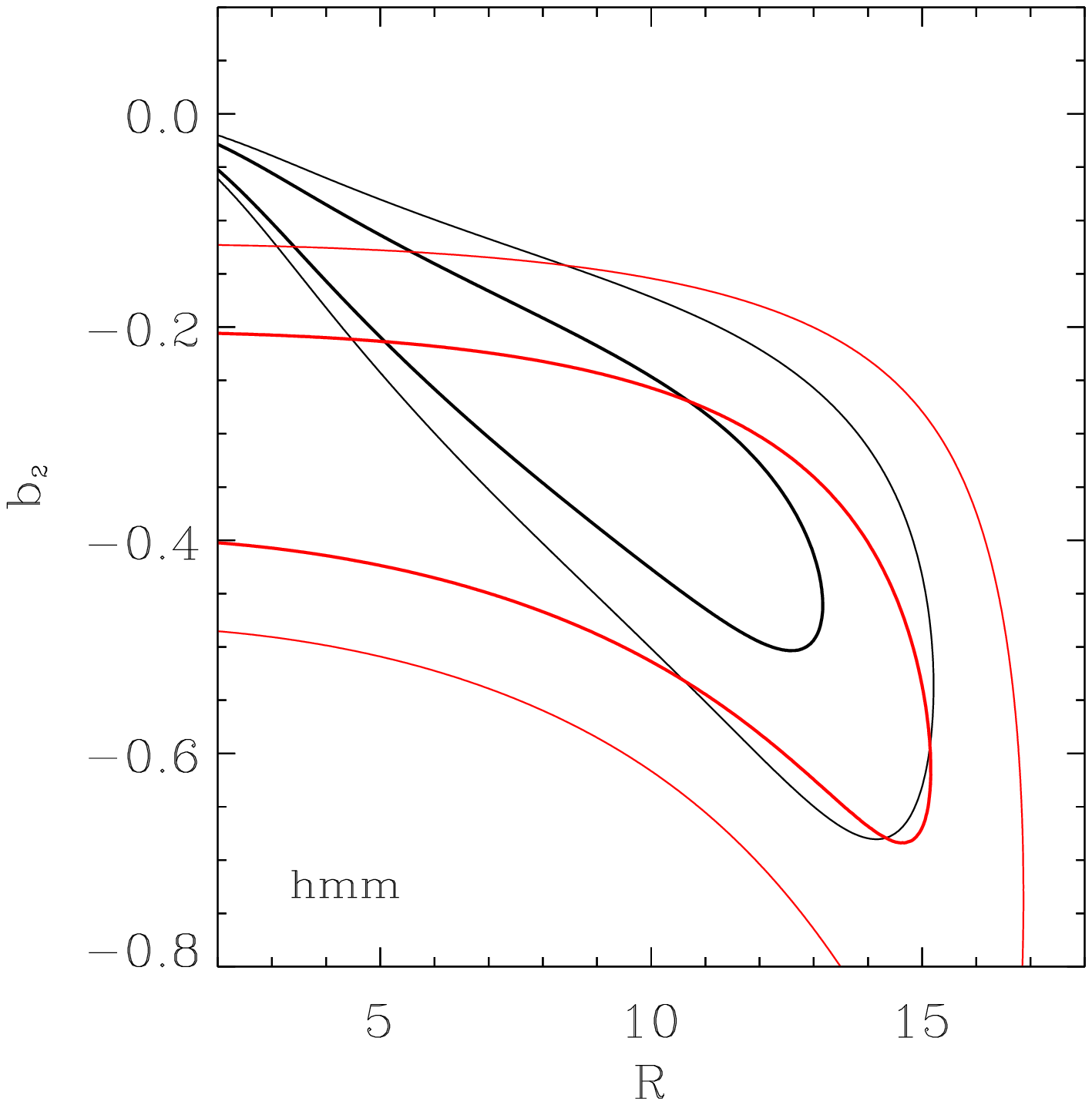}}

 \caption{Joint marginal probability distribution for the parameter pairs 
$b_1$--$b_2$ (top), $b_1$--$R$ (middle) and $b_2$--$R$ (bottom) obtained 
fitting the data for $B_{\hhh}$ (left), $B_{\hhm}$ (center) and $B_{\hmm}$ (right).
Contours correspond to the 68.3 and 95.4 per cent credible intervals and
refer to the full non-linear model (black) and to the approximation based
on tree-level SPT (red).}
 \label{fig:margb1b2}
 \end{figure*}


\subsubsection{Marginal credible regions}

We now evaluate the consistency of the model-parameter constraints for the 
halo bispectra.
Figure~\ref{fig:margb1b2} shows the marginal posterior distribution 
for the parameter pairs $b_1$--$b_2$, $b_1$--$R$ and $b_2$--$R$,
respectively. Each panel refers to a particular bispectrum, as indicated from 
the label in the bottom left corner. The black
contours denote the $68.3\%$ and $95.4\%$ credible regions for the
parameters of the fully non-linear model.
The red contours, instead, indicate the corresponding regions 
for the SPT tree-level model described in \S\ref{SPTtree}.

The first thing that may be noticed is that the estimates for 
$b_1$ and $b_2$ from the tree-level and fully non-linear models are in 
disagreement: the tree-level constraints show a systematic shift, preferring
lower $b_1$ and slightly more negative $b_2$ values.
This implies that inferences made about the
non-linearity of galaxy bias using the galaxy bispectrum and
tree-level perturbation theory will be significantly biased. Note that
this statement also applies to rather large scales $k_1\lesssim0.12\kMpc$.
If one uses triangle configurations on smaller scales
\citep[e.g.][]{Verdeetal2002} then the discrepancy becomes larger.
Therefore, our program to use N-body simulations for determining the matter 
terms in the bias relation is key to correctly estimate the bias (and thus the 
cosmological parameters) from forthcoming observational data.

The second important point to notice is that 
fits to $B_{\hhh}$, $B_{\hhm}$ and 
$B_{\hmm}$ with the fully non-linear model
give consistent constraints for $b_1$, $b_2$ and $R$.
The precision with which we are
able to determine the bias parameters increases as we go from
$B_{\rm hmm}$ to $B_{\hhh}$. This finding is consistent with our
earlier results \citep{Pollacketal2012}. 

Note that the best-fit values for $b_1$ appear relatively insensitive to 
variations in $R$. 
One has to consider rather large smoothing scales in order to see any changes 
in the best-fit models.  However, the best-fit solutions for $b_2$ are strongly
 degenerate with $R$: as $R$ decreases, $b_2$ becomes less negative and tends 
towards zero. This owes to the fact that, on changing $R$ in the
 interval $R\in[2, 13] \Mpc$, one can always find different combinations of 
$b_1$ and $b_2$ that fit the data with the same accuracy 
as previously described in \ref{ssec:goodness}.
We can more directly understand the origin of the $b_2$--$R$ degeneracy as
follows. Let us consider the bias model for $B_{\hmm}$ since
this only contains the terms ${\mathcal B}^{\rm (s)}_{(1,1,1)}$ and
${\mathcal B}^{\rm (s)}_{(2,1,1)}$.  As shown in the top panels of
\Fig{fig:P4dsm}, the de-smoothing of ${\mathcal B}^{\rm (s)}_{(1,1,1)}$
results in the matter bispectrum. However, the de-smoothing of
${\mathcal B}^{\rm (s)}_{(2,1,1)}$ results in a function that carries a
dependence on $R$.  If we take the ratio of ${\mathcal
  B}^{\rm (s)}_{(2,1,1)}$ defined for $R_{\rm a}$ with the same function but
defined for $R_{\rm b}$, then we will find something close to a constant for
$R=[4,10]\Mpc$.  Thus we can identify the degenerate combination:
\be b_2^{\rm a} =
\frac{{\mathcal B}^{\rm (s)}_{(2,1,1)}(\bk_1,\bk_2,R_{\rm b})}{{\mathcal B}^{\rm (s)}_{(2,1,1)}(\bk_1,\bk_2,R_{\rm a})}\, b_2^{\rm b}
\approx  A(R_{\rm a},R_{\rm b})\,b_2^{\rm b}\ee
where $A(R_{\rm a},R_{\rm b})$ is a function that is independent of the triangle
configuration. Hence, the value of $b_2$ is correlated with the size
of the smoothing scale, $R$.


\begin{figure*}
 \includegraphics{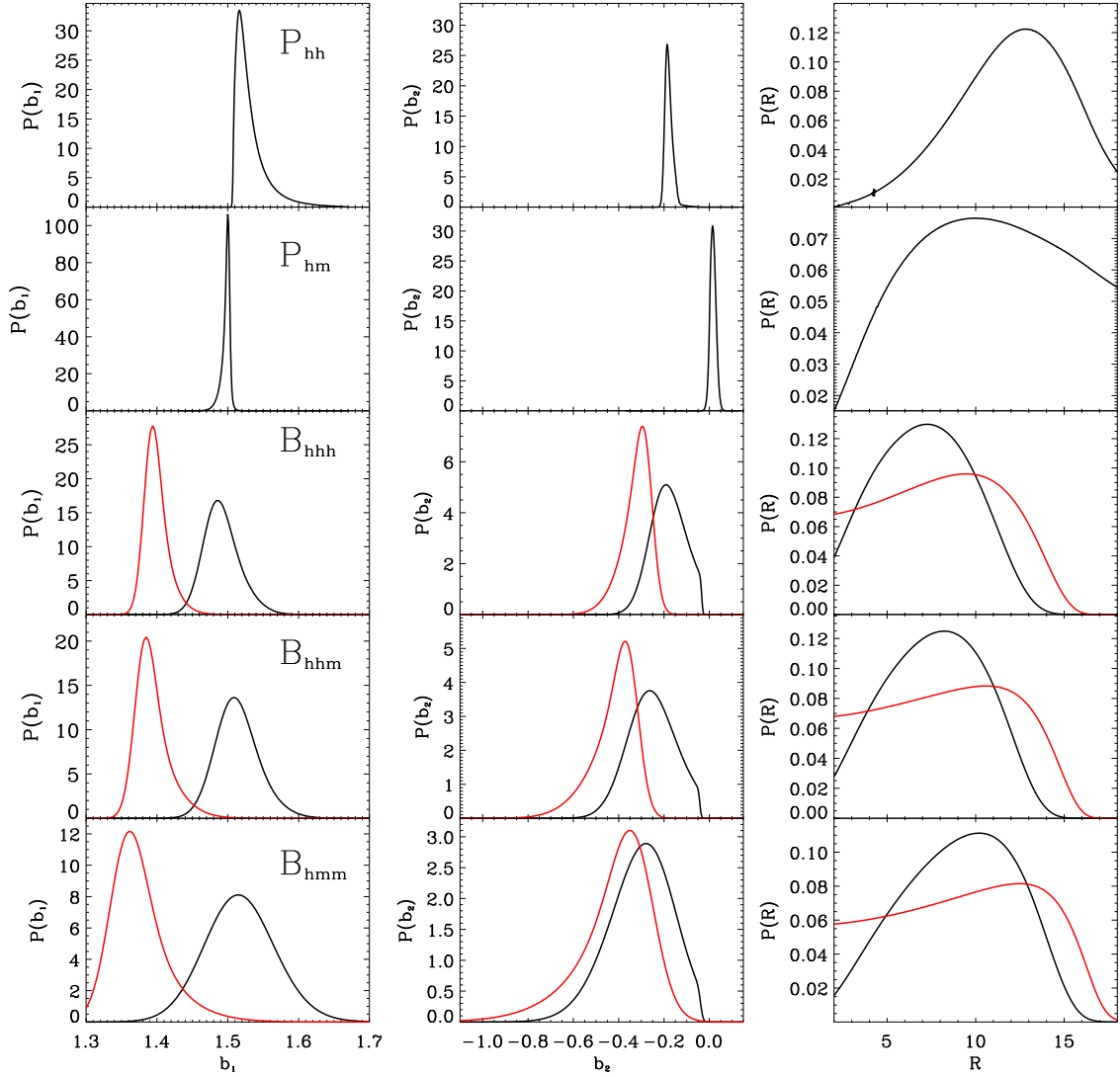}
\caption{Marginal probability distributions for the single bias parameters
$b_1$ (left), $b_2$ (center) and $R$ (right) obtained fitting various
halo statistics (from top to bottom: $P_{\hh}, P_{\hm}, B_{\hhh}, B_{\hhm}, B_{\hmm}$).
Results obtained with the full non-linear model (black) are compared with
those derived using tree-level SPT (red).
} 
\label{fig:singlemarg}
\end{figure*}


\subsection{Comparing all constraints}
\label{ssec:Pofb1b2}
\begin{table}
\centering
\caption{Posterior mean and rms error of the bias
parameters $b_1$, $b_2$ and $R$ obtained fitting various halo statistics
with the full non-linear bias model.}
\label{tab:posmeannl}
\begin{tabular}{|c|c|c|c|} \hline
 Statistic  &  $b_1 \pm \sigma_{b_1}$ & $b_2 \pm \sigma_{b_2}$ & $R \pm
\sigma_{R}$ \\ 
& & & ($\Mpc$) \\
\hline
$P_{\hh}$  & 1.53 $\pm$ 0.02 & -0.18 $\pm$ 0.02 & 12.0 $\pm$ 3.1 \\
$P_{\hm}$  & 1.48 $\pm$ 0.02 &  0.02 $\pm$ 0.01 & 10.6 $\pm$ 4.1 \\
$B_{\hhh}$ & 1.49 $\pm$ 0.03 & -0.18 $\pm$ 0.07 & 7.2 $\pm$ 2.6  \\  
$B_{\hhm}$ & 1.51 $\pm$ 0.03 & -0.26 $\pm$ 0.10 & 7.8 $\pm$ 2.8  \\
$B_{\hmm}$ & 1.52 $\pm$ 0.05 & -0.31 $\pm$ 0.14 & 9.1 $\pm$ 3.1  \\
\hline
\end{tabular}
\end{table}

In Figure~\ref{fig:singlemarg} we present the marginal posterior
probabilities for the single bias parameters extracted
from the various probes that we
have considered. The left, central and middle columns
show the results for $b_1$, $b_2$ and $R$,
respectively.
From top to bottom, the rows correspond to 
$P_{\rm hh}, P_{\rm hm}, B_{\hhh}$, $B_{\hhm}, B_{\hmm}$, respectively.
The black curves represent the results from the fully
non-linear modelling, and the red curves show the results from the
tree-level perturbation theory for the bispectra.
The corresponding mean and rms values of the marginal probabilities
for the full non-linear model are reported in Table \ref{tab:posmeannl}.

Considering the values for $b_1$ from the bispectra, we see that, as
noted earlier, the parameter constraints for the non-linear model are
consistent with one another and are significantly different from the
best-fit $b_1$ obtained from the tree-level expressions. On comparing
the bispectra results with the power-spectra results we find
reasonable consistency for the non-linear modelling, whereas for the
tree-level bispectrum model, the results disagree at high significance
\citep[see also][]{Pollacketal2012}. 
However, the marginal distributions for $b_1$ from $P_{\rm hh}$ 
and $P_{\rm hm}$ overlap very little. In fact, 
they exhibit opposite skewness although they are both narrow and
located around $b_1\simeq 1.5$.
The marginal distribution for $b_1$ computed from $P_{\hm}$
agrees remarkably well with the effective bias $b_{\hm}=1.503 \pm 0.002$. This
is because the data require $b_2\simeq 0$ in this case.
On the other hand, $b_1 > b_{\hh}=1.49\pm0.002$ in the marginal distribution extracted
from $P_{\hh}$ which requires $b_2<0$. 

Examining the results of the fully non-linear model for $b_2$, 
from the bispectra we find that the
marginal posterior distributions are fairly broad and are peaked
towards negative values ($b_2\simeq -0.2$ for $B_{\hhh}$,
$b_2\simeq -0.3$ for $B_{\hhm}$ and $B_{\hmm}$).
Overall, the various bispectra give consistent constraints. 
Note that the sharp cutoff in the marginal 
distributions at $b_2\simeq 0$ is due to the fact that our prior for $R$ 
does not consider values $R<2 \Mpc$. 
Considering the results obtained using the tree-level SPT model, 
we see that the distributions for $b_1$ and $b_2$ shift towards different values
(approximately the posterior mean of the bias parameters moves 
by $\Delta b_1\simeq \Delta b_2\simeq -0.15$). 
On comparing with the results obtained from the halo power spectra, 
we see that the marginal distribution for $b_2$ extracted from
$P_{\rm hh}$ and $P_{\hm}$ are narrowly peaked around $b_2\sim-0.18$ and 
$b_2\sim0.02$, respectively. 


We now turn to the question of whether there is a preferred smoothing
scale for the haloes we have considered. On inspecting the bispectra, we see that the marginal 
distributions for $R$ are reasonably consistent and 
display a broad peak between 5 and 12 $\Mpc$. 
The power spectra, instead, tend to prefer slighty larger values of $R$: 
$10<R<15 \Mpc$ for $P_{\hh}$ and $R>5 \Mpc$ for $P_{\hm}$,
consistent with the behaviour of the goodness of fit reported in 
\S\ref{ssec:goodness}.
In all cases, these optimal smoothing scales correspond
to a few Lagrangian radii of the halos. They are also comparable to (but
a bit smaller than) the scales
that we sample with the measurements of the power spectra and bispectra.
Note that a sphere of radius $\sim 10 \Mpc$ contains $\sim 1.5$ halos
on average so that counts in cells of this extension
are subject to sizable random fluctuations that create stochasticity
in the bias relation.

\subsection{Cross-correlation coefficients}

There are three possible explanations as to why $P_{\hm}$, $P_{\hh}$ and
the bispectra show disagreement for the full non-linear model. 
One, we may
require higher-order terms in the bias expansion, e.g. $b_3$ etc; two,
the LEB may be wrong; three, there may be uncorrelated stochasticity in the 
relation between halo overdensity
and mass overdensity. We shall now explore this latter possibility.

A number of studies have
demonstrated, using $N$-body simulations, that the relation between
$\delta_{\rm h}(\bx|R)$ against $\delta(\bx|R)$ contains
scatter, and that this scatter depends on the scale which one uses to
compute the density field
\citep[e.g.][]{DekelLahav1999,SeljakWarren2004,Smithetal2007,ManeraGaztanaga2011,RothPorciani2011,Pollacketal2012,
ChanScoccimarro2012}.
%


\begin{figure*}
 \centering{
  \includegraphics[width=8cm]{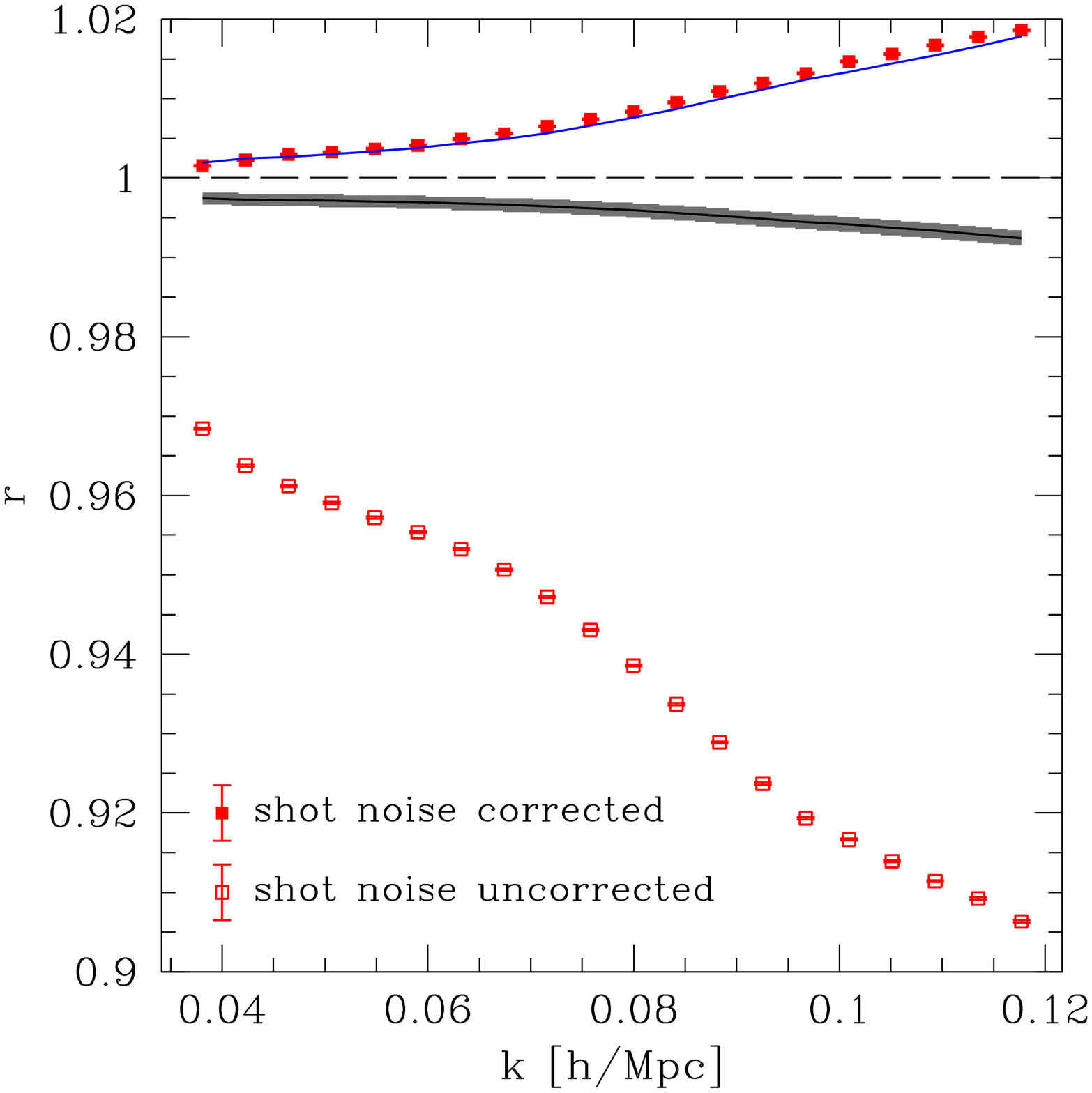}
  \includegraphics[width=8cm]{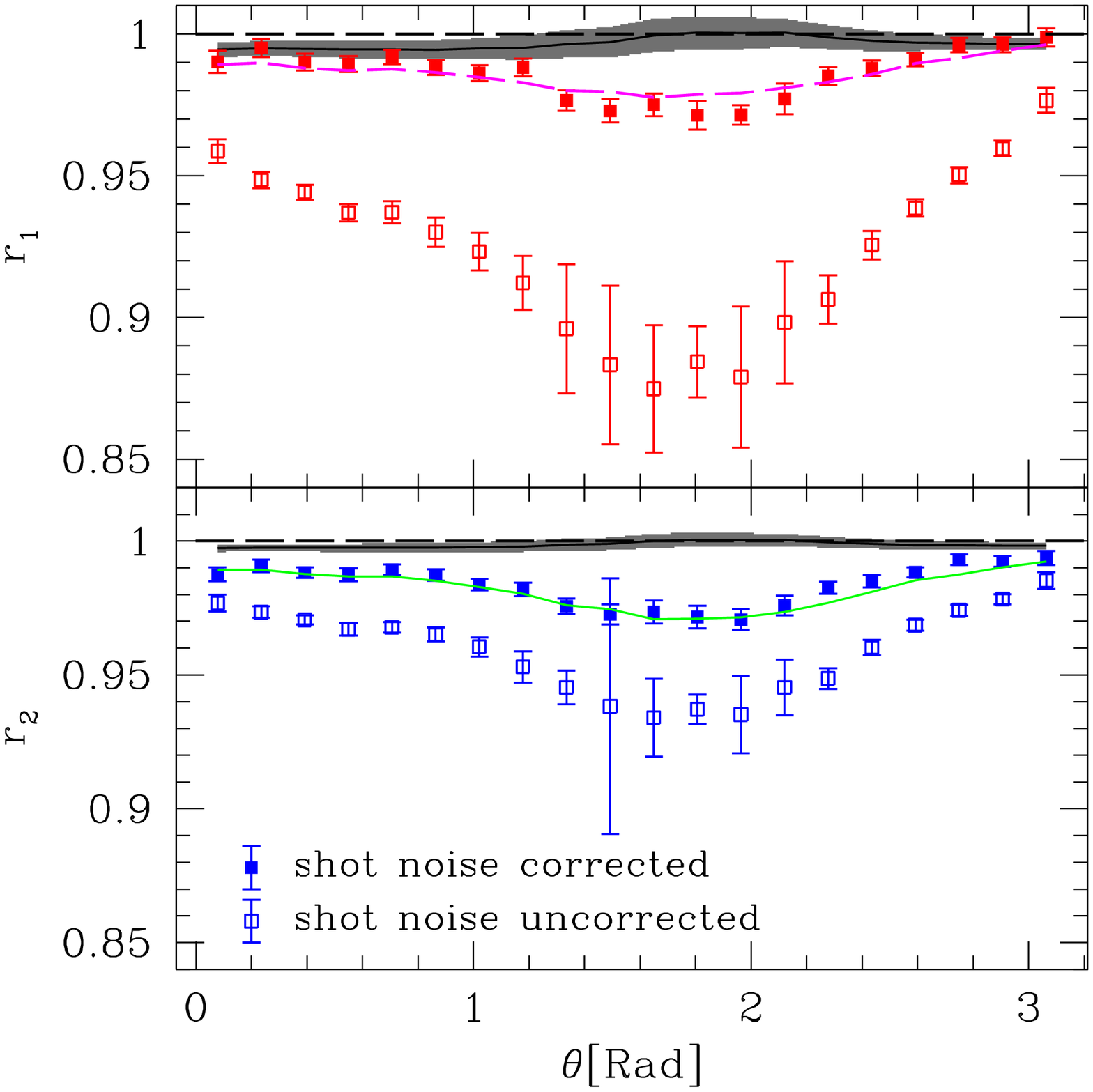}}
\caption{Left: Linear cross-correlation coefficient between the fluctuations in
the halo and matter density, $r=b_{\hm}/b_{\hh}$, for different
wavenumbers.
Closed and open symbols show the results obtained from the simulations
when $P_{\hh}$ is and is not 
corrected for shot noise, respectively. 
The black solid line and the shaded region around it indicate the mean 
and the rms value of the correlation coefficient 
obtained by averaging the predictions of the second-order LEB over the 
posterior probability of the model parameters derived from a joint fit
to $P_{\hh}$ and $P_{\hm}$.
The solid curve, instead, shows the values of $r$ that are computed using
the means for $P_{\hh}$ and $P_{\hm}$ over
the posterior distributions for the individual fits to $P_{\hh}$ and
$P_{\hm}$, respectively.
Right: As in the left panel but for the 3-point coefficients $r_1$ 
and $r_2$ defined in Equations (\ref{eq:r1}) and (\ref{eq:r2}).
In this case, 
the shaded region is obtained averaging the model over the posterior
distribution for the parameters derived from a joint fit to the
relevant bispectra, while the solid curve uses the different means from
the fits to the individual bispectra.
 }
\label{fig:rcoeff}
\end{figure*}
Another way to explore the stochasticity is through the
cross-correlation coefficient between Fourier modes. 
For two-point statistics this can be defined \citep{DekelLahav1999}:
\be r \equiv \frac{\hat{P}_{\hm}}{\sqrt{\hat{P}_{\hh}
    \hat{P}_{\mm}}} = \frac{b_{\hm}}{b_{\hh}} \ . 
\label{eq:rcoeff}
\ee
If $\delta^{\rm h}$ is a deterministic linear function of $\delta$,
then $r=\pm 1$. However, if there is uncorrelated random
noise present, i.e. $\delta^{\rm
  h}(\bx)=b\delta(\bx)+\epsilon(\bx)$, then the halo power spectrum
would be $P_{\rm hh}(k)=b^2P_{\mm}(k)+P_\epsilon(k)$, where $P_\epsilon$ denotes
the power spectrum of the noise distribution. This leads to:
\be r =\left(1+\frac{P_\epsilon}{P_{\mm}}\right)^{-1/2}<1 \ .\ee
We note that non-linearity in the bias relationship will also introduce
deviations of $r$ away from unity: consider the quadratic relation
$\delta_{\rm h}(\bx)=b_1\delta(\bx)+b_2\,\delta^2(\bx)/2$, then one
finds that the cross-correlation can be written:
\ba r & = & 
\left[1+\frac{c_2}{2}\frac{\mathcal P_{(2,1)}}{\mathcal P_{(1,1)}}\right]
\left[1+c_2\frac{\mathcal P_{(2,1)}}{\mathcal P_{(1,1)}}
+\frac{c_2^2}{4}
\frac{{\mathcal P}_{(2,2)}}{{\mathcal P}_{(1,1)}}\right]^{-1/2} \\
& \approx &  1-\frac{c_2^2}{8}
\frac{{\mathcal P}_{(2,2)}}{{\mathcal P}_{(1,1)}} \ .
\ea
where $c_2\equiv b_2/b_1$ and where the second equality follows for
the case where ${\mathcal P}_{(2,1)}\ll {\mathcal P}_{(1,1)}$ and 
${\mathcal P}_{(2,2)}\ll {\mathcal P}_{(1,1)}$.

In this case, we see
that the cross-correlation function can be either greater or less than
unity depending on the sign and magnitude of $c_2$.

Figure~\ref{fig:rcoeff} shows the cross-correlation coefficient
estimated from our ensemble of $N$-body simulations along with the
standard errors on the mean. The open symbols show the result before
we correct $P_{\rm hh}$ for shot noise, the solid symbols show the
result after the usual inverse number-density correction. We see that
before correcting for the shot noise the function is less than 1 and
decreases with scale. After the correction, $r$ is brought within a few
percent from unity and is always larger than one. 
Note that the difference from unity is very significant given the numbers of 
realizations and the comoving volume covered by our simulations.

In order to derive $r$ from the fully non-linear model,
we jointly fit the numerical data for $P_{\hm}$ and $P_{\hh}$.
We acknowledge that utilizing 200 simulations prevents us from accurately estimating a $40\times40$ covariance matrix, in particular the cross covariances between the different spectra. Therefore, we performed the joint fit in the following manner.  In order to ensure that the different spectra can be treated as independent of each other, we generated two ensembles consisting of 100 simulations each to estimate a particular spectra. We then computed a 20x20 block covariance matrix, selecting every other bin from our auto- and cross- power spectrum estimates.  The off-diagonal blocks of the covariance matrix were set equal to zero when analyzing the auto-halo and cross halo-matter power spectrum. The resulting best-fit model ($b_1\simeq 1.5, b_2\simeq -0.09, R\simeq 18$)
does not match to the data ($\chi^2_{\rm min}\simeq 2242/1997$ 
with a contribution of 1170 coming from $P_{\hm}$) 
meaning that it is impossible to simultaneously fit $P_{\hh}$ and $P_{\hm}$
with the second-order LEB.
Consequently, 
we find that the posterior mean of the cross-correlation coefficent, obtained by multiplying the likelihoods of the single fits to $P_{\hh}$ and $P_{\hm}$, is always smaller than unity and
does not provide a good description to the data (see the black line
and the shaded region in Figure~\ref{fig:rcoeff}).  
To investigate this further, we recompute $r$ using
the posterior means of $P_{\hh}$ and $P_{\hm}$ 
shown in the left panel of Figure~\ref{fig:PowBisp200}.
Inserting them in Equation (\ref{eq:rcoeff}),
we find excellent agreement with the data 
(see the blue line in  Figure~\ref{fig:PowBisp200}).  
As shown previously, the best-fit models
for $P_{\hh}$ and $P_{\hm}$ prefer different values for $b_1$ and $b_2$ when 
analyzed independently. The joint analysis of $P_{\hm}$ and $P_{\hh}$, in this 
manner, shows more clearly the inconsistency obtained when using the 
second-order LEB as a model for halo biasing.  


One can also define cross-correlation coefficients for higher-order
statistics. The second equality in \Eqn{eq:rcoeff} gives us a clear
path to make this generalization. 
From the 3-point effective bias coefficients we may form two independent ratios:
\ba
r_1 & \equiv & \frac{b_{\hmm}}{b_{\hhh}}
= \frac{B_{\hmm}}{B_{\mmm}^{2/3} B_{\hhh}^{1/3}} \label{eq:r1} \\
r_2 & \equiv & \frac{b_{\hhm}}{b_{\hhh}} 
= \frac{B_{\hhm}^{1/2}}{B_{\mmm}^{1/6} B_{\hhh}^{1/3}}  \ , \label{eq:r2}
\ea
where the dependence on the triangle configuration is understood.  Note, that a third (dependent) ratio may be also computed:
$r_3=b_{\rm hmm}/b_{\rm hhm}=r_1/r_2$. 
For a deterministic linear bias model with bias coefficient $b$, 
$r_1=1$ and $r_2={\mathrm {sgn}}(b)$. Once again, additional
stochasticity or non-linear biasing will alter the cross-correlation
coefficients.

\begin{figure}
 \centering
  \includegraphics[width=8.5cm,height=8.5cm]{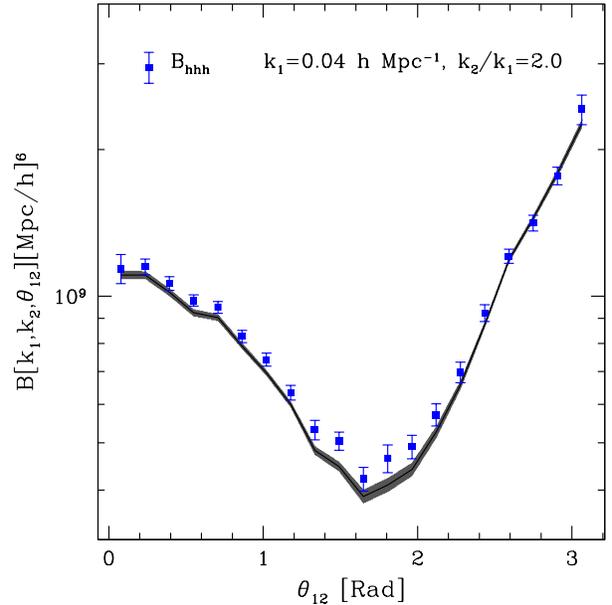}
\caption{The halo bispectrum, $B_{\hhh}$, measured from the simulations 
(solid symbols) 
is compared with the fully non-linear LEB adopting the parameters 
that best fit $B_{\hmm}$.
The line and shaded region show the mean and rms value of the LEB model 
for $B_{\hhh}$ averaged over the posterior distribution for 
$b_1, b_2$ and $R$ coming from a fit to $B_{\hmm}$. 
This shows that the parameter sets that nicely fit $B_{\hmm}$ (see Figure 
\ref{fig:PowBisp200}) 
are not able to reproduce all the features seen in $B_{\hhh}$.
\label{fig:testbisp}}
\end{figure}

In the right panel of Figure~\ref{fig:rcoeff} we present the cross-correlation 
coefficients 
$r_1$ and $r_2$ extracted from the simulations 
as a function of the triangle configuration. 
Both functions are always a few per cent below unity even 
after shot-noise subtraction.
In the same figure, we also plot the posterior mean and variance for 
the $r$ coefficients obtained from joint fits to two bispectra
(black line and shaded region) performed in the same manner as for the
power spectrum. These results are very close to unity 
and do not adequately describe the simulated data. In fact,
the joint fits prefer less negative values for $b_2$ 
than the single fits (for example, the best simultaneous fit to 
$B_{\hhh}$ and $B_{\hhm}$ gives $b_1\simeq 1.50, b_2\simeq -0.15$ and 
$R\simeq 5.5$ with $\chi^2_{\rm min}/\nu\simeq 1954/1997$). 
On the other hand, if we compute $r_1$ and $r_2$ from the individual 
posterior means for $B_{\hhh}$, $B_{\hhm}$ and $B_{\hmm}$, we get results that 
are in good agreement with the data.
This is somewhat puzzling as the fits to the various bispectra appear
to give consistent bias parameters.
However, in order to test how congruous the different fits really are,
we derive models for one bispectrum type (say $B_{\hhh}$) averaging over
the joint posterior distribution for the bias parameters derived
by fitting one of the other bispectra ($B_{\hmm}$ or $B_{\hhm}$).
An example is shown in Figure~\ref{fig:testbisp}: the fit based on $B_{\hmm}$ matches well the data for $B_{\hhh}$ for collinear triangles 
but systematically underestimates the halo bispectrum in all the other 
configurations.
It is exactly in this more precise comparison that we see the failure of the 
non-linear local bias model when analyzing the bispectra data.

\section{Discussion}\label{sec:discussion}

Our high-precision measurements of the halo-halo and halo-matter
spectra and bispectra enabled us to carry out a series of consistency tests of 
the second-order LEB.  
As seen in Figure \ref{fig:singlemarg} and in Table \ref{tab:posmeannl}, 
the marginal posterior distributions for $b_1$ and $b_2$ determined from
$P_{\hh}, B_{\hhh}, B_{\hhm}$ and $B_{\hmm}$ are all consistent with one another and, yet, are 
inconsistent with the constraints derived from the halo-matter cross spectrum.  
The primary reason is that
$P_{\hm}$ requires a positive $b_2$ that is close to 0, whereas the fits to the 
other spectra prefer a negative value for $b_2$.  
In terms of statistical significance, the stronger discrepancy is with $P_{\hh}$ as
the posterior distributions for $b_2$ extracted from all bispectra are rather broad. 
The incompatibility between the bias parameters obtained from $P_{\hh}$ and $P_{\hm}$ 
might indicate a breakdown in the modelling due to either incorrect shot-noise 
subtraction or incorrect parameterization of halo biasing. 
To better understand this issue,
it is interesting to focus for a moment onto the shot-noise free spectra $P_{\hm}$ and 
$B_{\hmm}$. Comparing their mathematical expressions given
in Equations (\ref{eq:hm}) and (\ref{eq:shmm}), we see that they have 
the same parametric form in terms of $b_1$ and $b_2$, 
it is only the non-linear matter terms which are different.
Since we directly measure these terms from the simulations and shot-noise does not play
any role here, the fact that the model-parameter constraints from $P_{\hm}$ and $B_{\hmm}$ 
are incompatible suggests that the LEB truncated to second-order is incorrect or,
at the very least, incomplete.
%
The simplest improvement would be to consider higher-order terms in the bias expansion
given in Equation (\ref{eq:Eulbias}). However, there are good reasons to believe that
more sophisticated corrections are needed.
Recent numerical work has provided strong evidence that dark-matter halos form out of linear
density peaks \citep{Ludlow-Porciani-2011}. This suggests that the halo bias with respect
to the matter fluctuations may actually be best understood as originating
in Lagrangian space \citep{Catelanetal1998,Catelanetal2000}.
However, even the simplest local Lagrangian biasing scheme 
generates a non-linear, non-local and stochastic scheme in 
Eulerian space \citep{Catelanetal1998,Catelanetal2000,Matsubara2011} 
which can be parameterized in 
terms of the invariants of the deformation tensor \citep{Catelanetal1998,
Chanetal2012,Baldaufetal2012}.  
Several terms should then be added to the bias expansion of the LEB
and this might help bring the model-parameter constraints
extracted from the different halo statistics in to better agreement.
We will revisit this issue in our future work.


\section{Conclusions}\label{sec:conclusion}
The use of galaxy clustering to extract information on the cosmological 
parameters
is currently limited to very large scales where both galaxy
biasing and the process of structure formation are expected to be linear 
and thus simple to model.  
Although more precise data are already available on smaller scales,  
they are not usually considered to avoid daunting complications in the modeling
that might introduce systematic effects in the results.  
Pursuing the goal of extending clustering studies to smaller scales,
we propose to use N-body simulations to measure the relevant 
statistics for the matter distribution that enter any biasing scheme.

While our framework is explicitly general, as an example, we apply it to the 
Eulerian local bias model truncated to quadratic order.  
This scheme represents the minimal theoretical model for studying 
three-point statistics of the galaxy distribution on large spatial separations.
Its predictions are easily computed to leading order in SPT and are commonly 
used to interpret observational results 
\citep{Verdeetal2002,JingBorner2004,
Wangetal04,Kayoetal2004,Gaztanagaetal05,Hikageetal2005,PanSzapudi2005,Kulkarnietal2007,Nishimichietal2007,Marin2011,McBrideetal2011a,McBrideetal2011b,Guoetal13,Marinetal2013}.

We use a set of 200 $N$-body simulations to study the clustering properties
of dark-matter halos and their relation to the underlying matter distribution
with unprecedented accuracy. 
Our halo catalogs cover a total 
comoving volume of $675\Gpccube$, much larger than the effective volume of the 
SDSS LRG sample ($0.26\Gpccube$), the BOSS BAO sample ($2.4\Gpccube$) and the 
planned spectroscopic survey of the Euclid satellite ($19.7\Gpccube$).
We consider halos with mass $M>1.11 \times 10^{13} \Msol$ 
corresponding to a number density
of $3.7\times 10^{-4}\Mpccube$ so that the effective volume  
(i.e. the actual volume weighted by the factor $\bar{n}\,P_{\hh}$)
roughly coincides with the total volume for the wavenumbers analyzed here
($0.04\lesssim k \lesssim 0.12 \kMpc$) that match the observable scales of 
current and future surveys.
All this allows us to measure the halo power spectrum to sub-percent accuracy 
(better than 0.3 per cent at $k\simeq 0.04 \kMpc$) and the halo bispectrum
to a few per cent accuracy.

We make a twofold use of our simulations: to measure the moments of the non-linear
matter distribution on several scales (and compare them against SPT predictions)
and to test how well the LEB truncated to quadratic order 
fits several statistics of the halo distribution.
In particular,
we consider the halo power spectrum, $P_{\hh}$, the halo-mass cross-spectrum,
$P_{\hm}$, as well as all the possible bispectra $B_{\hhh}$, $B_{\hhm}$ and $B_{\hmm}$. 
Our main results can be summarized as follows:
\begin{enumerate}
\item
In a $\Lambda$CDM model at $z=0$,
tree-level SPT does not accurately model non-linearities in the momenta
of the matter distribution on spatial scales of the order of $10-30 \Mpc$.
\item
The simple second-order LEB fits very well all halo spectra and bispectra when either $N$-body simulations or tree-level 
SPT are used in the modelling of the clustering amplitudes for the matter 
distribution.  However, the bias parameters derived from the models
based on SPT are heavily biased with respect to 
the case when non-linearities are accurately modelled.
This might explain why studies that interpreted 
different statistics of the galaxy distirbution based on SPT reached 
inconsistent
conclusions regarding the non-linear bias of optically selected galaxies 
\citep[e.g.][]{Verdeetal2002,Gaztanagaetal05}.
\item The LEB models applied to counts in cells requires an
optimal smoothing scale of several $\Mpc$ to match the halo statistics
from the simulations. For our halos,
this corresponds to a few Lagrangian radii
but is also of the same order of the spatial scales under analysis.
\item 
Comparing the parameter 
constraints for the fully non-linear LEB obtained 
from the different spectra, we find some inconsistencies.
In particular, the non-linear bias parameter extracted from the cross-spectrum 
$P_{\hm}$ 
is incompatible with the results from all the other statistics.  
The main difference is that $P_{\hm}$ strongly favours a positive value 
for $b_2$ that is very close to zero, whereas the posterior distributions 
derived from all other spectra prefer a negative $b_2$ in the range
$-0.3\lesssim b_2\lesssim -0.2$. General agreement, instead, 
is found for the linear bias parameter, $b_1$.
\item
Non-trivial shot-noise corrections in $P_{\hh}$ might be invoked to reconcile
the bias-parameters extracted from $P_{\hm}$ and $P_{\hh}$. However this
cannot explain the differences between the constraints from 
the shot-noise free statistics $P_{\hm}$ and $B_{\hmm}$.
This suggests that further complexity should be added to 
second-order LEB in order to match all halo statistics.
\item
Analysis of the cross-correlation coefficients defined for the two-point and 
three-point 
statistics reveal further subtle inconsistencies contained in the LEB 
truncated to second order,
suggesting it is too simple a model to describe halo bias with high accuracy.
\end{enumerate}

A final remark is in order.
Our numerical study is based on simulations with a fixed background cosmology and
focuses on retrieving the bias parameters when the cosmological
parameters are perfectly known.
However, this is not the case for actual galaxy surveys where bias and cosmology
are generally estimated simultaneously.
To transform our method into a resourceful tool for data analysis, 
we will need to explore how the shapes and 
amplitudes of the moments of the non-linear matter density field 
depend on the unknown cosmological parameters without having to run an exorbitant 
amount of $N$-body simulations \citep[e.g.][]{Angulo-White-2010}
-- a topic we shall explore in future work.


\section*{Acknowledgements}

JEP thanks Xun Shi and Andr\'{e}s Balaguera-Antol\'{i}nez for useful discussions. We thank V. Springel for
making public {\tt GADGET-2} and for providing his {\tt B-FoF} halo
finder, and R.~Scoccimarro for making public his {\tt 2LPT} code. JEP
and CP were partially supported by funding provided through the SFB-Transregio
33 “The Dark Universe” by the Deutsche Forschungsgemeinschaft. RES acknowledges support from ERC Advanced Grant 246797 “GALFORMOD”.



\bibliographystyle{mn2e}
\bibliography{refs}


\appendix

\onecolumn


\section{Spectral relationships}

\subsection{Relationship between ${\mathcal P}_{(l,m)}$ and the $n$-point multispectra}\label{app:one}

We now derive the relation between the functions ${\mathcal
  P}_{(l,m)}$ and the multi-point matter spectra.

To begin, the functions ${\mathcal P}_{(l,m)}$ are defined:
\ba
\left<\Delta^{(l)}(\bk_1|R)\Delta^{(m)}(\bk_2|R)\right> & \equiv & 
(2\pi)^3 \delta^D(\bk_{1}+\bk_{2}) {\mathcal P}_{(l,m)}(\bk_1)
\\
& = & \int 
\prod_{i=1}^{l} \left\{ \frac{\dq_i}{(2\pi)^3}\right\}
\prod_{j=1}^{m} \left\{\frac{\dpp_j}{(2\pi)^3}\right\}
(2\pi)^3\delta^{D}(\bk_1-\bq_{1\dots l})
(2\pi)^3\delta^{D}(\bk_2-\bp_{1\dots m}) \times \nn \\
& & 
\times\, 
\left<\frac{}{}
\tilde{\delta}(\bq_1|R)\dots\tilde{\delta}(\bq_l|R)
\tilde{\delta}(\bp_1|R)\dots\tilde{\delta}(\bp_m|R)
\right> \label{eq:ln}
\ea
The ensemble average of the $l+m$ density fields may be evaluated to
give: 
\be 
\left<
\tilde{\delta}(\bq_1|R)\dots\tilde{\delta}(\bq_l|R)
\tilde{\delta}(\bp_1|R)\dots\tilde{\delta}(\bp_m|R)
\right> \equiv (2\pi)^3
\delta^D(\bq_{1\dots l}+\bp_{1\dots m}){\mathcal P}_{(l+m)}(\bq_1,\dots,\bq_l,\bp_1,\dots,\bp_m) \ .
\ee
On inserting the above definition into \Eqn{eq:ln} we find
\ba
\left<\Delta^{(l)}(\bk_1|R)\Delta^{(m)}(\bk_2|R)\right>
& \equiv & \int 
\prod_{i=1}^{l} \left\{ \frac{\dq_i}{(2\pi)^3}\right\}
\prod_{j=1}^{m} \left\{\frac{\dpp_j}{(2\pi)^3}\right\}
(2\pi)^3\delta^{D}(\bk_1-\bq_{1\dots l})
(2\pi)^3\delta^{D}(\bk_2-\bp_{1\dots m}) \times \nn \\
& & \times\, (2\pi)^3
\delta^D(\bq_{1\dots l}+\bp_{1\dots m}){\mathcal P}_{(l+m)}(\bq_1,\dots,\bq_l,\bp_1,\dots,\bp_m)
\ea
Integrating over the first two Dirac delta functions in yields:
\ba
\left<\Delta^{(l)}(\bk_1|R)\Delta^{(m)}(\bk_2|R)\right>
& \equiv & (2\pi)^3
\delta^D(\bk_{1}+\bk_{2}) \int 
\prod_{i=1}^{l-1} \left\{ \frac{\dq_i}{(2\pi)^3}\right\}
\prod_{j=1}^{m-1} \left\{\frac{\dpp_j}{(2\pi)^3}\right\} \times \nn \\
& & \times\, {\mathcal P}_{(l+m)}(\bq_1,\dots,\bq_{l-1},\bk_1-\bq_{1\dots (l-1)},\bp_1,\dots,\bp_{m-1},
\bk_2-\bp_{1\dots(m-1)})
\ea
Hence,
\be
{\mathcal P}_{(l,m)}(\bk_1)
= \int 
\prod_{i=1}^{l-1} \left\{ \frac{\dq_i}{(2\pi)^3}\right\}
\prod_{j=1}^{m-1} \left\{\frac{\dpp_j}{(2\pi)^3}\right\}
{\mathcal P}_{(l+m)}(\bq_1,\dots,\bq_{l-1},\bk_1-\bq_{1\dots (l-1)},\bp_1,\dots,\bp_{m-1},
\bk_2-\bp_{1\dots(m-1)})
\ee
Lastly, we may change integration variables in the following way:
$\tilde{\bq}_{2}\rightarrow \bq_2-\bq_1$, $\tilde{\bq}_3\rightarrow
\bq_3-\tilde{\bq}_2$, \dots, upon which the above expression may be
written as:
\be 
{\mathcal P}_{(l,m)}(\bk_1) =
\int 
\prod_{i=1}^{l-1} \left\{ \frac{\dq_i}{(2\pi)^3}\right\}
\prod_{j=1}^{m-1} \left\{\frac{\dpp_j}{(2\pi)^3}\right\}
{\mathcal P}_{(l+m)}(\bq_1,\bq_2-\bq_1,\dots,\bk_1-\bq_{l-1},\bp_1,\bp_2-\bp_1,\dots,\bk_2-\bp_{m-1})
\label{eq:Plm}
\ee
Terms up to and including the quadspectrum may be written:
\ba 
{\mathcal P}_{(1,1)}(\bk_1) & = & {\mathcal P}_{(2)}(\bk_1,\bk_2) = {\mathcal P}(\bk_1) \\
{\mathcal P}_{(2,1)}(\bk_1) & = & \int \frac{\dq_1}{(2\pi)^3} {\mathcal P}_{(3)}(\bq_1,\bk_1-\bq_1,\bk_2)
= \int \frac{\dq_1}{(2\pi)^3} {\mathcal B}(\bq_1,\bk_1-\bq_1,\bk_2) \\
{\mathcal P}_{(3,1)}(\bk_1) & = & \int \frac{\dq_1}{(2\pi)^3}\frac{\dq_2}{(2\pi)^3}
{\mathcal P}_{(4)}(\bq_1,\bq_2-\bq_1,\bk_1-\bq_2,\bk_2) \\
{\mathcal P}_{(2,2)}(\bk_1) & = & \int \frac{\dq_1}{(2\pi)^3}\frac{\dpp_1}{(2\pi)^3}  
{\mathcal P}_{(4)}(\bq_1,\bk_1-\bq_1,\bp_1,\bk_2-\bp_1) \\
{\mathcal P}_{(4,1)}(\bk_1) & = & \int \frac{\dq_1}{(2\pi)^3}\frac{\dq_2}{(2\pi)^3}
\frac{\dq_3}{(2\pi)^3} 
{\mathcal P}_{(5)}(\bq_1,\bq_2-\bq_1,\bq_3-\bq_2,\bk_1-\bq_3,\bk_2) \\
{\mathcal P}_{(3,2)}(\bk_1) & = & \int \frac{\dq_1}{(2\pi)^3}\frac{\dq_2}{(2\pi)^3}
\frac{\dpp_1}{(2\pi)^3} 
{\mathcal P}_{(5)}(\bq_1,\bq_2-\bq_1,\bk_1-\bq_2,\bp_1,\bk_2-\bp_1) 
\ea


\subsection{Relationship between ${\mathcal B}_{(l,m,n)}$ and the $n$-point multispectra}\label{app:two}

In a similar fashion, we may now derive the relation between the
functions ${\mathcal B}_{(l,m,n)}$ and the multi-point matter spectra.

To begin, the functions ${\mathcal B}_{(l,m,n)}$ are defined:
\ba
\left<\Delta^{(l)}(\bk_1|R)\Delta^{(m)}(\bk_2|R)\Delta^{(n)}(\bk_3|R)\right>
& \equiv & \int 
\prod_{i=1}^{l} \left\{ \frac{\dq_i}{(2\pi)^3}\right\}
\prod_{j=1}^{m} \left\{\frac{\dpp_j}{(2\pi)^3}\right\}
\prod_{k=1}^{n} \left\{\frac{\dss_k}{(2\pi)^3}\right\}
(2\pi)^3\delta^{D}(\bk_1-\bq_{1\dots l})(2\pi)^3 \times
\nn \\
& & \hspace{-3.5cm}\times
\delta^{D}(\bk_2-\bp_{1\dots m})
(2\pi)^3\delta^{D}(\bk_3-\bs_{1\dots n})
\left<\frac{}{}
\tilde{\delta}(\bq_1|R)\dots\tilde{\delta}(\bq_l|R)
\tilde{\delta}(\bp_1|R)\dots\tilde{\delta}(\bp_m|R)
\tilde{\delta}(\bs_1|R)\dots\tilde{\delta}(\bs_n|R)
\right> \label{eq:lmn}
\ea
The ensemble average of the $l+m+n$ density fields may be evaluated to give: 
\ba 
\left<
\tilde{\delta}(\bq_1|R)\dots\tilde{\delta}(\bq_l|R)
\tilde{\delta}(\bp_1|R)\dots\tilde{\delta}(\bp_m|R)
\tilde{\delta}(\bs_1|R)\dots\tilde{\delta}(\bs_n|R)
\right> 
& \equiv & (2\pi)^3
\delta^D(\bq_{1\dots l}+\bp_{1\dots m}+\bs_{1\dots n}) \times\nn \\
& & \times {\mathcal P}_{(l+m+n)}
(\bq_1,\dots,\bq_l,\bp_1,\dots,\bp_m,\bs_1,\dots,\bs_n) \ .
\ea
On inserting the above definition into \Eqn{eq:lmn} we find
\ba
\left<\Delta^{(l)}(\bk_1|R)\Delta^{(m)}(\bk_2|R)\Delta^{(n)}(\bk_3|R)\right>
& \equiv & 
\int 
\prod_{i=1}^{l} \left\{\frac{\dq_i}{(2\pi)^3}\right\}
\prod_{j=1}^{m} \left\{\frac{\dpp_j}{(2\pi)^3}\right\}
\prod_{k=1}^{n} \left\{\frac{\dss_k}{(2\pi)^3}\right\}
(2\pi)^{12}
\delta^{D}(\bk_1-\bq_{1\dots l})
\delta^{D}(\bk_2-\bp_{1\dots m}) \times
\nn \\
& & \hspace{-1.4cm}\times
\delta^{D}(\bk_3-\bs_{1\dots n})\delta^D(\bq_{1\dots l}+\bp_{1\dots m}+\bs_{1\dots n})
{\mathcal P}_{(l+m+n)}(\bq_1,\dots,\bq_l,\bp_1,\dots,\bp_m,\bs_1,\dots,\bs_n)
\ea
On integrating over the first three Dirac delta functions gives:
\ba
\left<\Delta^{(l)}(\bk_1|R)\Delta^{(m)}(\bk_2|R)\Delta^{(n)}(\bk_3|R)\right>
& \equiv & (2\pi)^{3}\delta^D(\bk_1+\bk_2+\bk_3)
\int 
\prod_{i=1}^{l} \left\{\frac{\dq_i}{(2\pi)^3}\right\}
\prod_{j=1}^{m} \left\{\frac{\dpp_j}{(2\pi)^3}\right\}
\prod_{k=1}^{n} \left\{\frac{\dss_k}{(2\pi)^3}\right\} \times
\nn \\
& & \times
{\mathcal P}_{(l+m+n)}(\bq_1,\dots,\bk_1-\bq_{1\dots l-1},\bp_1,\dots,\bk_2-\bp_{1\dots m-1},\bs_1,\dots,\bk_3-\bs_{1\dots n-1})
\ea
As for the power spectrum, we may change integration variables in the
following way, $\tilde{\bq}_{2}\rightarrow \bq_2-\bq_1$,
$\tilde{\bq}_3\rightarrow \bq_3-\tilde{\bq}_2$, \dots. After which we
find,
\ba 
{\mathcal B}_{(l,m,n)}(\bk_1,\bk_2,\bk_3) & = & 
\int 
\prod_{i=1}^{l-1} \left\{ \frac{\dq_i}{(2\pi)^3}\right\}
\prod_{j=1}^{m-1} \left\{\frac{\dpp_j}{(2\pi)^3}\right\}
\prod_{k=1}^{n-1} \left\{\frac{\dss_k}{(2\pi)^3}\right\} 
\nn \\
&  & 
\times {\mathcal P}_{(l+m+n)}(\bq_1,\bq_2-\bq_1,\dots,\bk_1-\bq_{l-1},\bp_1,\bp_2-\bp_1,\dots,\bk_2-\bp_{m-1},
\bs_1,\bs_2-\bs_1,\dots,\bk_3-\bs_{n-1}) \ .
\ea
%


\subsection{Proof of the symmetry of ${\mathcal P}_{(l,m)}$}\label{app:Plmsym}

We now prove that the spectra ${\mathcal P}_{(l,m)}$ are symmetric in
their indices $m$ and $l$: 
\be 
{\mathcal P}_{(l,m)} = {\mathcal P}_{(m,l)} \label{eq:lm=ml} .
\ee

Consider \Eqn{eq:Plm}, on relabelling the variables $\bp_i=\bq_{l+i-1}$,
and writing $P_{(l+m)}=P_{(m+l)}$, we find,
\be 
{\mathcal P}_{(l,m)}(\bk_1) =
\int 
\left\{ \frac{\dq_1}{(2\pi)^3}\right\}\dots \left\{ \frac{\dq_{l+m-2}}{(2\pi)^3}\right\}
{\mathcal P}_{(m+l)}(\bq_1,\dots,\bk_1-\bq_{l-1},\bq_l,\dots,\bk_2-\bq_{l+m-2})
\ee
On changing the integration variables to $\bp_i=-\bq_i$, we find 
\be 
{\mathcal P}_{(l,m)}(\bk_1) =
\int 
\frac{\dpp_1}{(2\pi)^3}\dots \frac{\dpp_{l+m-2}}{(2\pi)^3}
{\mathcal P}_{(m+l)}(-\bp_1,\dots,\bk_1+\bp_{l-1},-\bp_l,\dots,\bk_2+\bp_{l+m-2}) \ .
\ee
Parity invariance of the $n$-point correlation functions means that
${\mathcal P}_{n}(\bp_1,\dots,\bp_n)={\mathcal
  P}_{n}(-\bp_1,\dots,-\bp_n)$ \citep[for a proof
  see][]{Smithetal2008b}. Under parity invariance, we find 
\be 
{\mathcal P}_{(l,m)}(\bk_1) =
\int 
\left\{ \frac{\dpp_1}{(2\pi)^3}\right\} \dots \left\{ \frac{\dpp_{l+m-2}}{(2\pi)^3}\right\}
{\mathcal P}_{(m+l)}(\bp_1,\dots,-\bk_1-\bp_{l-1},\bp_l,\dots,-\bk_2-\bp_{l+m-2}) \ .
\ee
We may switch $\bk_1=-\bk_2$ and $\bk_2=-\bk_1$,
\be 
{\mathcal P}_{(l,m)}(\bk_1) =
\int 
\left\{ \frac{\dpp_1}{(2\pi)^3}\right\} \dots \left\{ \frac{\dpp_{l+m-2}}{(2\pi)^3}\right\}
{\mathcal P}_{(m+l)}(\bp_1,\dots,\bk_2-\bp_{l-1},\bp_l,\dots,\bk_1-\bp_{l+m-2}) \ .
\ee
Next we may rearrange the arguments of the $n$-point spectra, since it
is totally symmetric under exchange symmetry: ${\mathcal
  P}_{n}(\bp_1,\dots,\bp_n)={\mathcal
  P}_{n}(\bp_i,\dots,\bp_1,\dots,\bp_n)$, whereupon
\be 
{\mathcal P}_{(l,m)}(\bk_1) =
\int 
\left\{ \frac{\dpp_1}{(2\pi)^3}\right\} \dots \left\{ \frac{\dpp_{l+m-2}}{(2\pi)^3}\right\}
{\mathcal P}_{(m+l)}(\bp_l,\dots,\bk_1-\bp_{l+m-2},\bp_1,\dots,\bk_2-\bp_{l-1}) \ .
\ee
Finally on changing variables $\bp_{l+i-1}=\bq_i$, we obtain
\be 
{\mathcal P}_{(l,m)}(\bk_1) =
\int 
\prod_{i=1}^{m-1} \left\{ \frac{\dq_i}{(2\pi)^3}\right\}
\prod_{j=1}^{l-1} \left\{\frac{\dpp_j}{(2\pi)^3}\right\}
{\mathcal P}_{(m+l)}(\bq_1,\dots,\bk_1-\bq_{m-1},\bp_1,\dots,\bk_2-\bp_{l-1}) = {\mathcal P}_{(m,l)}(\bk_1) \ ,
\ee
and this completes the proof of \Eqn{eq:lm=ml}.



\section{Modelling $B^{\rm (s)}_{(2,1,1)}$ with perturbation theory}\label{app:pt}

In order to better understand what drives the amplitude and functional
form of the $B^{\rm (s)}_{(l,m,n)}$ we have attempted to model the
signal with standard perturbation theory techniques. Rather than
modelling all of the spectra, we have focused on the lowest order
non-trivial term $B_{(2,1,1)}$.

To begin, consider again \Eqn{eq:P4}, this came from:  
\be
\left<\Delta^{(2)}(\bk_1|R)\Delta^{(1)}(\bk_2|R)\Delta^{(1)}(\bk_3|R)\right> 
= 
\int \frac{\dq_1}{(2\pi)^3} \frac{\dq_2}{(2\pi)^3} (2\pi)^3
\delta^{D}(\bk_1-\bq_1-\bq_2) 
\left<
\tilde{\delta}(\bq_1|R)
\tilde{\delta}(\bq_2|R)
\tilde{\delta}(\bk_2|R)
\tilde{\delta}(\bk_3|R)
\right> \label{eq:B211a}
\ee
The above ensemble averaged product can be broken into connected and 
disconnected terms. Hence:
\ba
\left<\tilde{\delta}(\bq_1|R)
\dots
\tilde{\delta}(\bq_4|R)
\right> & = & \left<
\tilde{\delta}(\bq_1|R)
\dots
\tilde{\delta}(\bq_4|R)
\right>_c 
+ 
\left<\tilde{\delta}(\bq_1|R)\tilde{\delta}(\bq_2|R)\right>
\left<\tilde{\delta}(\bq_3|R)\tilde{\delta}(\bq_4|R)\right> \nn \\ 
& & +
\left<\tilde{\delta}(\bq_1|R)\tilde{\delta}(\bq_3|R)\right> 
\left<\tilde{\delta}(\bq_2|R)\tilde{\delta}(\bq_4|R)\right>
+ 
\left<\tilde{\delta}(\bq_1|R)\tilde{\delta}(\bq_4|R)\right>
\left<\tilde{\delta}(\bq_2|R)\tilde{\delta}(\bq_3|R)\right>
\ea
If we consider the disconnected terms, these may be written in terms of
power spectra (c.f.~\Eqn{eq:Pk}). We note that the first disconnected
term is vanishing unless $\bk_1$ is the null vector. The remaining 
terms are:
\ba
\left<\tilde{\delta}(\bq_1|R)
\dots
\tilde{\delta}(\bq_4|R)
\right>  &  = &
(2\pi)^3\delta^{D}(\bq_{1\dots4})
{\mathcal T}(\bq_1,\bq_2,\bq_3,\bq_4)\nn \\
& & +
(2\pi)^3{\mathcal P}(q_1) P(q_2) 
\left[
\delta^{D}(\bq_1+\bq_3)\delta^{D}(\bq_2+\bq_4)+
\delta^{D}(\bq_1+\bq_4)\delta^{D}(\bq_2+\bq_3)
\right]
\ea
On inserting the above expression into \Eqn{eq:B211a} and computing the 
integral over $\bq_2$, we find:
\be
{\mathcal B}^{\rm (s)}_{(2,1,1)}
= 
\frac{2}{3}\left[
{\mathcal P}(\bk_2){\mathcal P}(\bk_3) + 2\,{\rm cyc}
\right] 
+\frac{1}{3}\int \frac{\dq_1}{(2\pi)^3}  
\left[{\mathcal T}(\bq_1,\bk_1-\bq_1,\bk_2,\bk_3) 
+{2\,\cyc}\right] \ ,
\ee
where we have factored out the Dirac delta. The power spectrum and
trispectrum may be evaluated using standard Eulerian perturbation
theory \citep[for a review see][]{Bernardeauetal2002}.  The important
results that we will need are that

At one loop level we have 
\ba
{\mathcal B}^{\rm (s)}_{(2,1,1)}
& = & 
\frac{2}{3}\left[
{\mathcal P}_{(0)}(\bk_2){\mathcal P}_{(0)}(\bk_3) + 2\,{\rm cyc}
\right] +
\frac{2}{3}\left[
 {\mathcal P}_{(0)}(\bk_2){\mathcal P}_{(1\ell)}(\bk_3) 
+{\mathcal P}_{(1\ell)}(\bk_2){\mathcal P}_{(0)}(\bk_3) 
+ 2\,{\rm cyc}
\right] \nn \\
& & +\frac{1}{3}\int \frac{\dq_1}{(2\pi)^3}  
\left[{\mathcal T}(\bq_1,\bk_1-\bq_1,\bk_2,\bk_3) 
+{2\,\cyc}\right] \ .
\ea

The function ${\mathcal P}_{(0)}$ denotes the smoothed linear matter power
spectrum and ${\mathcal P}_{(1\ell)}$ denotes the `1-loop'
correction. The 1-Loop term may be written as the 
\be
\mathcal{P}_{1\ell}(\bk) =  \mathcal{P}_{22}(\bk) + \mathcal{P}_{13}(\bk) 
\ee
where the loop integrals are \dots
\ba 
\mathcal{P}_{13}(k) & = & \frac{\mathcal{P}_{11}(k)k^3}{252(2\pi)^2}\int_{0}^{\infty} dx\, x^2 
\mathcal{P}_{11}(xk)\left\{-42x^{2}+100-\frac{158}{x^{2}}+\frac{12}{x^{4}}+
\frac{3}{x}(1-x^2)^3(7x^2+2)\log\left[\frac{x+1}{|x-1|}\right]\right\} ; \\
\mathcal{P}_{22}(k) & = & 2 \int_0^{\infty} \frac{dq}{(2\pi)^2} q^2 \mathcal{P}_{11}(q)
\int_{-1}^{1} d\mu \mathcal{P}_{11}(k\psi(x,\mu))
\left\{\frac{5}{7}+\frac{1}{2}\frac{\mu-x}{\psi(x,\mu)}
\left[\frac{x}{\psi(x,\mu)}+\frac{\psi(x,\mu)}{x}\right]
+\frac{2}{7}\left[\frac{\mu-x}{\psi(x,\mu)}\right]^2\right\}^2\ ,
\ea
where $x=q/k$ and where $\psi^2(x,\mu)=1+x^2-2x\mu$.

The connected tree-level contribution to the trispectrum is given by:
\be \mathcal{T}(\bq_1,\bq_2,\bq_3,\bq_4) = 
4\mathcal{T}_{2211}(\bq_1,\bq_2,\bq_3,\bq_4)+
6\mathcal{T}_{3111}(\bq_1,\bq_2,\bq_3,\bq_4)
\ee
where the two types of term are:
\ba
  \label{eq:Tafull}
  \mathcal{T}_{2211}(\bq_1,\bq_2,\bq_3,\bq_4)  & = & 
  \mathcal{P}_1 \mathcal{P}_2 \left[\frac{}{} 
  \mathcal{P}_{13} F_2( \bq_1,- \bq_{13}) F_2( \bq_2, \bq_{13}) + 
  \mathcal{P}_{14} F_2( \bq_1,- \bq_{14}) F_2( \bq_2, \bq_{14})  
  \right] \nn\\
  & & + 
  \mathcal{P}_1 \mathcal{P}_3 \left[\frac{}{} 
  \mathcal{P}_{12} F_2( \bq_1,- \bq_{12}) F_2( \bq_3, \bq_{12}) + 
  \mathcal{P}_{14} F_2( \bq_1,- \bq_{14}) F_2( \bq_3, \bq_{14})  
  \right]  \nn \\
  & & +
  \mathcal{P}_1 \mathcal{P}_4 \left[\frac{}{}  
  \mathcal{P}_{12} F_2( \bq_1,- \bq_{12}) F_2( \bq_4, \bq_{12}) + 
  \mathcal{P}_{13} F_2( \bq_1,- \bq_{13}) F_2( \bq_4, \bq_{13})  
  \right]  \nn \\
  & & + 
  \mathcal{P}_2 \mathcal{P}_3 \left[\frac{}{}  
  \mathcal{P}_{21} F_2( \bq_2,- \bq_{21}) F_2( \bq_3, \bq_{21}) + 
  \mathcal{P}_{24} F_2( \bq_2,- \bq_{24}) F_2( \bq_3, \bq_{24})  
  \right]  \nn \\
  & & + 
  \mathcal{P}_2 \mathcal{P}_4 \left[\frac{}{}  
  \mathcal{P}_{21} F_2( \bq_2,- \bq_{21}) F_2( \bq_4, \bq_{21}) + 
  \mathcal{P}_{23} F_2( \bq_2,- \bq_{23}) F_2( \bq_4, \bq_{23})  
  \right]  \nn \\
  & & + 
  \mathcal{P}_3 \mathcal{P}_4 \left[\frac{}{}  
  \mathcal{P}_{31} F_2( \bq_3,- \bq_{31}) F_2( \bq_4, \bq_{31}) + 
  \mathcal{P}_{32} F_2( \bq_3,- \bq_{32}) F_2( \bq_4, \bq_{32})  
  \right] \,, 
\ea
and
\be
\label{eq:tbfull}
  \mathcal{T}_{3111}(\bq_1,\bq_2,\bq_3,\bq_4) =  
  F_3( \bq_1, \bq_2, \bq_3) \mathcal{P}_1 \mathcal{P}_2 \mathcal{P}_3 
+ F_3( \bq_2, \bq_3, \bq_4) \mathcal{P}_2 \mathcal{P}_3 \mathcal{P}_4  
+ F_3( \bq_3, \bq_4, \bq_1) \mathcal{P}_3 \mathcal{P}_4 \mathcal{P}_1 
+ F_3( \bq_4, \bq_1, \bq_2) \mathcal{P}_4 \mathcal{P}_1 \mathcal{P}_2 \,,
\ee
where $\mathcal{P}_{i}\equiv \mathcal{P}_{\rm{lin}}(\bq_{i})$, $\mathcal{P}_{ij}\equiv
\mathcal{P}_{\rm{lin}}(|\bq_{i}+ \bq_{j}|)$ and $ \bq_{ij}\equiv \bq_{i}+
\bq_{j}$.
The calculation of the second-order coupling functions is straightforward. The
result is
\ba
F_2^{\rm (s)}(\bq_1,\bq_2) & = &  \frac{5}{14}\left[\alpha(\bq_1,\bq_2)+\alpha(\bq_2,\bq_1)\right]
+\frac{2}{7}\beta(\bq_1,\bq_2) \ ; \\
G_2^{\rm (s)}(\bq_1,\bq_2) & = &  \frac{3}{14}\left[\alpha(\bq_1,\bq_2)+\alpha(\bq_2,\bq_1)\right]
+\frac{4}{7}\beta(\bq_1,\bq_2)\ ; \\
F_3^{\rm (s)}({\bq_1},{\bq_2},{\bq_3}) & = &
\frac{7}{54}
\left[\frac{}{}\alpha({\bq_1}, \bq_{23})F_2^{\rm (s)}({\bq_2},{\bq_3})+
\alpha({\bq_2}, \bq_{13})F_2^{\rm (s)}({\bq_1},{\bq_3}) 
  +\alpha({\bq_3}, \bq_{12})F_2^{\rm (s)}({\bq_1},{\bq_2})\right]\nn\\
  & & + \frac{4}{54}
\left[\frac{}{}\beta({\bq_1}, \bq_{23})G_2^{\rm (s)}({\bq_2},{\bq_3})+
\beta({\bq_2}, \bq_{13})G_2^{\rm (s)}({\bq_1},{\bq_3}) 
 +\beta({\bq_3}, \bq_{12})G_2^{\rm (s)}({\bq_1},{\bq_2})\right] \nn \\
 & & 
+\frac{7}{54}
\left[\frac{}{}\alpha( \bq_{12},{\bq_3})G_2^{\rm (s)}({\bq_1},{\bq_2})+
\alpha( \bq_{13},{\bq_2})G_2^{\rm (s)}({\bq_1},{\bq_3}) 
 +\alpha( \bq_{23},{\bq_1})G_2^{\rm (s)}({\bq_2},{\bq_3})\right] \,.
\label{eq:F3}
\ea
where we introduced the two fundamental mode coupling functions
\be
  \label{eq:alpha}
  \alpha( \bq_{1}, \bq_{2}) = 
  \frac{( \bq_{1}+ \bq_{2})\cdot  \bq_{1}}{\bq_{1}^{2}} \ ; \hspace{1cm}
  \beta({\bq_1},{\bq_2})  = 
  \frac{|{\bq_1}+{\bq_2}|^2({\bq_1}\cdot{\bq_2})}{2\bq_1^2 \bq_2^2}\,.
\ee
We then proceeded with integrating as in equation
(\ref{eq:P4}) taking care to cut the integral off at scales that
exceed twice the filter radius.

\end{document}